\documentclass[%
 reprint,
superscriptaddress,
 amsmath,amssymb,
 aps,
prx
]{revtex4-2}

\usepackage{braket}
\usepackage{graphicx}% Include figure files
\usepackage{dcolumn}% Align table columns on decimal point
\usepackage{bm}% bold math
\usepackage{changes}
\usepackage{xcolor}
\usepackage{comment}
\usepackage{physics}
\usepackage{bbm}
\usepackage{chngcntr}
\usepackage[colorlinks=true,linkcolor=blue,citecolor=red]{hyperref}

\newcommand{\about}{{\sim}}

\let\vec\mathbf
\newcommand{\vk}{\vec{k}}
\newcommand{\cre}[2]{{#1}^{\dagger}_{#2}}
\newcommand{\des}[2]{{#1}_{#2}}
\newcommand{\normord}[1]{:#1:}
\newcommand{\expec}[1]{\left\langle #1 \right\rangle}
\newcommand{\tMF}{\text{MF}}
\newcommand{\ies}[1]{i_{#1} \eta_{#1} \sigma_{#1}}
\newcommand{\iesUp}[1]{i_{#1} \eta_{#1} \uparrow}
\newcommand{\iesDn}[1]{i_{#1} \eta_{#1} \downarrow}
\newcommand{\iesDnMin}[1]{i_{#1} - \downarrow}
\newcommand{\iesC}[1]{i_{#1}, \eta_{#1}, \sigma_{#1}}
\newcommand{\iesCUp}[1]{i_{#1}, \eta_{#1}}
\newcommand{\iesCDn}[1]{i_{#1}, \eta_{#1}}
\newcommand{\one}{\mathbbm{1}}

\newcommand{\UN}[1]{\mathrm{U} \left( #1 \right)}
\newcommand{\SUN}[1]{\mathrm{SU} \left( #1 \right)}
\begin{document}

\preprint{APS/123-QED}

\title{Dynamical correlations and order in magic-angle twisted bilayer graphene}% Force line breaks with \\

\author{Gautam~Rai}
\thanks{These authors contributed equally to this work.}
\affiliation{I. Institute of Theoretical Physics, University of Hamburg, Notkestrasse 9, 22607 Hamburg, Germany}

\author{Lorenzo~Crippa}
\thanks{These authors contributed equally to this work.}
\affiliation{Institut f\"ur Theoretische Physik und Astrophysik and W\"urzburg-Dresden Cluster of Excellence ct.qmat, Universit\"at W\"urzburg, 97074 W\"urzburg, Germany}

\author{Dumitru~C\u{a}lug\u{a}ru}
\affiliation{Department of Physics, Princeton University, Princeton, New Jersey 08544, USA}

\author{Haoyu~Hu}
\affiliation{Donostia International Physics Center, P. Manuel de Lardizabal 4, 20018 Donostia-San Sebastian, Spain}

\author{Francesca Paoletti}
\affiliation{Institut f\"ur Theoretische Physik und Astrophysik and W\"urzburg-Dresden Cluster of Excellence ct.qmat, Universit\"at W\"urzburg, 97074 W\"urzburg, Germany}

\author{Luca~de'~Medici}
\affiliation{LPEM, ESPCI Paris, PSL Research University, CNRS, Sorbonne Universit\'e, 75005 Paris France}

\author{Antoine~Georges}
\affiliation{Coll\`ege de France, Universit\'e PSL, 11 place Marcelin Berthelot, 75005 Paris, France}
\affiliation{Center for Computational Quantum Physics, Flatiron Institute, 162 Fifth Avenue, New York, New York 10010, USA}
\affiliation{Centre de Physique Théorique, Ecole Polytechnique, CNRS, Institut Polytechnique de Paris, 91128 Palaiseau Cedex, France}
\affiliation{DQMP, Université de Genève, 24 quai Ernest Ansermet, CH-1211 Genève, Suisse}

\author{B.~Andrei~Bernevig}
\affiliation{Department of Physics, Princeton University, Princeton, New Jersey 08544, USA}
\affiliation{Donostia International Physics Center, P. Manuel de Lardizabal 4, 20018 Donostia-San Sebastian, Spain}
\affiliation{IKERBASQUE, Basque Foundation for Science, Bilbao, Spain}

\author{Roser Valent\'\i}
\affiliation{Institut f\"ur Theoretische Physik, Goethe Universit\"at Frankfurt, Max-von-Laue-Strasse 1, 60438 Frankfurt am Main, Germany}

\author{Giorgio~Sangiovanni}
\email{sangiovanni@physik.uni-wuerzburg.de}
\affiliation{Institut f\"ur Theoretische Physik und Astrophysik and W\"urzburg-Dresden Cluster of Excellence ct.qmat, Universit\"at W\"urzburg, 97074 W\"urzburg, Germany}

\author{Tim~Wehling}
\email{tim.wehling@uni-hamburg.de}
\affiliation{I. Institute of Theoretical Physics, University of Hamburg, Notkestrasse 9, 22607 Hamburg, Germany}
\affiliation{The Hamburg Centre for Ultrafast Imaging, 22761 Hamburg, Germany}

\date{\today}

\begin{abstract}
%New
The interplay of dynamical correlations and electronic ordering is pivotal in shaping phase diagrams of correlated quantum materials. In magic angle twisted bilayer graphene, transport, thermodynamic and spectroscopic experiments pinpoint at a competition between distinct low-energy states with and without electronic order, as well as between localized and delocalized charge carriers. In this study, we utilize Dynamical Mean Field Theory (DMFT) on the topological heavy Fermion (THF) model of twisted bilayer graphene to investigate the emergence of electronic correlations and long-range order in the absence of strain. We contrast moment formation, Kondo screening and ordering on a temperature basis and explain the nature of emergent correlated states based on three central phenomena: (i) the formation of local spin and valley isospin moments around 100K, (ii) the ordering of the local isospin moments around 10K preempting Kondo screening, and (iii) a cascadic redistribution of charge between localized and delocalized electronic states upon doping. At integer fillings, we find that low energy spectral weight is depleted in the symmetric phase, while we find insulating states with gaps enhanced by exchange coupling in the zero-strain ordered phases. Doping away from integer filling results in distinct metallic states: a ``bad metal" above the ordering temperature, where scattering off the disordered local moments suppresses electronic coherence, and a ``good metal" in the ordered states with coherence of quasiparticles facilitated by isospin order. This finding reveals coherence from order as the microscopic mechanism behind the Pomeranchuk effect observed experimentally in [Nature \textbf{592}, 214 (2021), Nature \textbf{592}, 220 (2021)]. Upon doping, there is a periodic charge reshuffling between localized and delocalized electronic orbitals leading to cascades of doping-induced Lifshitz transitions, local spectral weight redistributions and periodic variations of the electronic compressibility ranging from nearly incompressible to negative. Our findings highlight the essential role of charge transfer, hybridization and ordering in shaping the electronic excitations and thermodynamic properties in twisted bilayer graphene and provide a unified understanding of the most puzzling aspects of scanning tunneling spectroscopy, transport, and compressibility experiments.

\end{abstract}

\maketitle
%\tableofcontents

\section{Introduction}
When two layers of graphene are stacked on top of each other with a relative twist angle of 1.1 degrees (the magic angle), the emergent long-wavelength moir\'{e} pattern gives rise to a band structure with extremely flat bands at the charge neutrality point~\protect\cite{SUA10, LOP07, bistritzerMoireBandsTwisted2011, REN21}. Electronic interaction effects are enhanced in the flat bands, leading to a rich low-temperature phase diagram where superconducting~\protect\cite{caoUnconventionalSuperconductivityMagicangle2018, luSuperconductorsOrbitalMagnets2019, ohEvidenceUnconventionalSuperconductivity2021, yankowitzTuningSuperconductivityTwisted2019}, insulating \protect\cite{caoCorrelatedInsulatorBehaviour2018}, correlated metallic \protect\cite{polshynLargeLinearintemperatureResistivity2019, jaouiQuantumCriticalBehaviour2022}, and exotic magnetic phases \protect\cite{nuckollsStronglyCorrelatedChern2020, serlinIntrinsicQuantizedAnomalous2020, sharpeEmergentFerromagnetismThreequarters2019, luSuperconductorsOrbitalMagnets2019,xieFractionalChernInsulators2021, che20a} have been observed in experiment by manipulating the charge carrier density. There has been a large theoretical effort towards understanding the insulating~\protect\cite{songMagicAngleTwistedBilayer2022, KEN18, WU19, HUA19, CLA19, CHR20, SAI20,
THO18, RAD18, OCH18, PAD18, XU18b, LIU19,RAD19, kangStrongCouplingPhases2019, CEA20, KHA20, VAF20, EUG20, ZHA20,SOE20, BUL20,XIE20a, brillauxAnalyticalRenormalizationGroup2022a,KAN20a, CHA20, LED20, CHI20B, ABO20, REP20a, XU21, PAR21a, kwanDomainWallCompetition2021, kwanExcitonBandTopology2021, chenRealizationTopologicalMott2021, ZHA21, KAN21, KUM21, DA21, WU21a, LIA21, LIU21,XIE21, LIU21a, kwanKekuleSpiralOrder2021, HOF22, xiePhaseDiagramTwisted2023, wagnerGlobalPhaseDiagram2022, zhangPolynomialSignProblem2023, CHR22, 
DA19, CHA21, SHE21, POT21, calugaruSpectroscopyTwistedBilayer2022,
WU20b, SEO19, REP20, BUL20a, PhysRevLett.129.076401,huangEvolutionQuantumAnomalous2024, LIU12} and superconducting~\protect\cite{GUI18, WU18, GUO18, GUI18, LIU18, PEL18, XU18, ISO18, YOU19, LIA19, GON19, ROY19, JUL20, KON20, chatterjeeSkyrmionSuperconductivityDMRG2022, CHI20a, KHA21, LEW21, WAN21, PAR21b, VEN18, DOD18, PO18,
YUA18, KEN18, WU19, HUA19, CLA19, CHR20, SAI20, BAL20, BER21b, FER21, 
WU19a, 
XIE20,
FER20, blasonLocalKekuleDistortion2022, angeliValleyJahnTellerEffect2019} phases, ferromagnetism \protect\cite{PIX19, SEO19, WU20b, REP20, BUL20a, kwanDomainWallCompetition2021, liuAnomalousHallEffect2020}, 
%transport \protect\cite{PAD20}, 
the topological properties \protect\cite{XIE20, 
ZOU18, PO18c, HEJ19a, HEJ19a, AHN19, PO19, SON19,
LIU19a, SON21, PhysRevLett.125.236804, WAN21a}, 
and on constructing suitable models~\protect\cite{songMagicAngleTwistedBilayer2022,
DAI16, JAI16, UCH14, WIJ15,
moonOpticalAbsorptionTwisted2013, KAN18, carrExactContinuumModel2019, fangAngleDependentInitioLowEnergy2019, TAR19, EFI18, PO19, LIU19a, koshinoMaximallyLocalizedWannier2018, ZHA19, HUA20a,WIL20,FU20,PAD20, VAF21,BER21,BER21a,  HEJ21,CAO21a, chouKondoLatticeModel2023, DAV22,SHI22a, wangUnusualMagnetotransportTwisted2023, LAU23, liTopologicalMixedValence2023, KEN21}. Simultaneously, there has been extensive experimental work~\protect\cite{AND21, caoStrangeMetalMagicAngle2020,STE20,XIE21a,KER19,JIA19,wongCascadeElectronicTransitions2020,dasSymmetrybrokenChernInsulators2021,choiCorrelationdrivenTopologicalPhases2021,CHO21, CHO21a, parkFlavourHundCoupling2021,LU21, rozenEntropicEvidencePomeranchuk2021,saitoIsospinPomeranchukEffect2021,DAS22,Seifert2020,Otteneder2020,LIS21,PhysRevResearch.3.013153,HES21,PhysRevMaterials.6.024003,GRO22, CHO19, TSC21, KER19, XIE19, CAL23, LIU21e, SAI21, CAO21, Yu2022}. For integer fillings, static mean-field approaches such as Hartree-Fock \protect\cite{songMagicAngleTwistedBilayer2022, kwanKekuleSpiralOrder2021, wagnerGlobalPhaseDiagram2022, BUL20, XIE20a, LIU21a, LIU21} revealed
a number of candidate ordered states related by the approximate symmetries of the system that are very close in energy. Depending on filling and the strain and relaxation properties of the sample, these include the time-reversal symmetry breaking Kramer's inter-valley coherent (K-IVC) state, valley-polarized states (VP)  \protect\cite{songMagicAngleTwistedBilayer2022} and incommensurate Kekul\'{e} spiral (IKS) states \protect\cite{kwanKekuleSpiralOrder2021, wagnerGlobalPhaseDiagram2022, XU18b} accompanied by a time-reversal symmetry preserving inter-valley coherent (T-IVC) order. The near-degeneracy of the ordered states suggests that the true state is sample-dependent owing to perturbations such as substrate effects, defects, and most importantly strain, in line with recent experiments~\protect\cite{nuckollsQuantumTexturesManybody2023, calugaruSpectroscopyTwistedBilayer2022}.

While the static mean-field approach is suitable for studying the integer-filled case deep in the ordered regime, understanding the temperature-dependent phase diagram, fractional dopings and the competition of fluctuating local moments and Kondo screening in the symmetric state with isospin polarization in the ordered state requires a dynamic treatment of local correlations. A suitable many-body method for this is dynamical mean-field theory (DMFT) \protect\cite{georgesDynamicalMeanfieldTheory1996, metznerCorrelatedLatticeFermions1989,Georges_DMFT_1992}. Previous applications of DMFT have successfully investigated the symmetric phase using the Wannier-construction--based multi-orbital projector models~\protect\cite{haule_mott-semiconducting_2019,dattaHeavyQuasiparticlesCascades2023a} and the topological heavy Fermion model~\protect\cite{huSymmetricKondoLattice2023a}, shedding light on Kondo physics and Fermi-surface resetting cascade transitions~\protect\cite{wongCascadeElectronicTransitions2020}. Concurrently with the present work, Zhou et al.~\protect\cite{zhouKondoPhaseTwisted2024} have used DMFT to study an effective single-valley model which imposes one valley to be fully occupied and frozen at all temperatures and all fillings, while the other valley is occupied upon doping without any additional symmetry-breaking. These studies focused on fully~\protect\cite{haule_mott-semiconducting_2019, dattaHeavyQuasiparticlesCascades2023a, huSymmetricKondoLattice2023a} (or partially~\protect\cite{zhouKondoPhaseTwisted2024}) symmetric states. Therefore, the interplay between dynamic correlations and spontaneous symmetry-breaking remains an outstanding question that needs to be addressed. 

In this paper, we apply DMFT on a set of symmetry-broken states of unstrained magic-angle twisted bilayer graphene (TBLG)---treating static and dynamic effects on the same footing, and compare the resulting spectral and thermodynamic observables with calculations in the symmetric state. In the symmetric (or interchangeably disordered) state we explicitly suppress long-range order. Our calculations provide a unified understanding of scanning tunneling spectroscopy \protect\cite{wongCascadeElectronicTransitions2020}  and compressibility experiments \protect\cite{TOM19, rozenEntropicEvidencePomeranchuk2021, saitoIsospinPomeranchukEffect2021, pierceUnconventionalSequenceCorrelated2021}, and some puzzling aspects of transport \protect\cite{saitoIsospinPomeranchukEffect2021, shenCorrelatedStatesTwisted2020, caoStrangeMetalMagicAngle2020} experiments based on three effects: (i) the formation of local isospin moments around 100~K, (ii) a cascadic charge reshuffling between localized and delocalized electronic states, and (iii) the ordering of local isospin moments around 10K. With our novel charge self-consistent Hartree-Fock+DMFT approach (described in Sec.~\ref{sec:methods}), we are able to contrast the temperature scales of moment formation, ordering and Kondo screening, give quantitative estimates of the ordering temperature on a DMFT level, and provide a \emph{microscopic} {theoretical understanding} of the temperature- and doping-dependent electronic phase diagram of TBLG. Our approach gives us access to the interacting spectral functions and Fermi surfaces, which link directly to scanning tunneling \protect\cite{wongCascadeElectronicTransitions2020, nuckollsQuantumTexturesManybody2023, nuckollsStronglyCorrelatedChern2020, ohEvidenceUnconventionalSuperconductivity2021} and quantum twist angle microscopies \protect\cite{inbarQuantumTwistingMicroscope2023} and lay the basis for an understanding of (magneto)transport experiments and superconductivity in MATBLG. %We expect that our charge self-consistent Hartree-Fock+DMFT will find fruitful application to a host of correlated materials in the years to come.

In Sec.~\ref{sec:results}, we apply our findings to a range of observations in the extensive experimental literature on the non-superconducting state of TBLG: we provide a microscopic understanding of transport in and around the correlated insulating states, and identify \emph{coherence from order} as the microscopic mechanism behind the \emph{isospin Pomeranchuk effect}. We link cascades in the local density in scanning tunneling spectroscopy and compressibility experiments with Lifshitz transitions, and we provide a proper mathematical treatment of the negative compressibility regions observed in this system. Our main results are summarized below.

 At integer fillings in the symmetric state, there is depletion of spectral weight at the Fermi level. In the zero-strain ordered state, there is a hard gap, and the spectral functions are comparable to Hartree-Fock calculations \protect\cite{songMagicAngleTwistedBilayer2022}. We find ordering temperatures of $\about 15$K  at $\nu = 0, -1$, and $\about 10$K at $\nu=-2$ (where $\nu$ measures the number of charge carriers per moir\'e unit cell with respect to the charge neutrality point). These ordering temperatures are about an order of magnitude lower than the Hartree-Fock prediction, indicating strong suppression of long-range order by local dynamic fluctuations. Upon raising the temperature from $\about 10$~K to $\about 100$--$200$~K, our DMFT calculations reveal that charge fluctuations are progressively restored --- in other words charge fluctuations are frozen out around the Hartree-Fock ordering temperatures. Above $\about 3$~K we find that the local spin susceptibility is approximately inversely proportional to the temperature. Taken together, the freezing of charge fluctuations and the spin-susceptibility point towards a local moment regime spanning the temperature range between the DMFT ordering temperatures ($\about 10$~K) and Hartree-Fock ordering temperatures ($\about 100$~K).
 
 %We find that charge fluctuations remain frozen up to $\about 100$--$200$~K (roughly around the Hartree-Fock ordering temperatures) above which,  we find a slow crossover to \emph{free-orbital} physics. At integer fillings, we find a low-temperature Curie-Weiss behavior of the local spin susceptibility \protect\cite{huSymmetricKondoLattice2023a}. Upon raising the temperature, charge fluctuations are progressively restored and the local moments eventually melt around the Hartree-Fock scale of $\about 100$--$200$~K, leading to the existence of a local moment regime spanning the temperature range between the DMFT ordering temperatures ($\about 10$~K) and Hartree-Fock ordering temperatures ($\about 100$~K).
 
 We further extend our symmetry-broken calculations to fractional fillings. At small doping away from the insulator, we find a doping-induced insulator-to-metal transition involving the population of coherent light charge carriers. We characterize the Fermi surface across a range of fillings, and find that the topology, the orbital character, and the coherence of the Fermi surface depends on the filling, and the absence or presence of long-range order. We identify the changes in the Fermi surface with a sequence of Lifshitz transitions associated with a redistribution of charge between localized and delocalized orbitals. These Lifshitz transitions manifest themselves as the filling-induced cascade transitions seen in compressibility \protect\cite{zondinerCascadePhaseTransitions2020, rozenEntropicEvidencePomeranchuk2021, saitoIsospinPomeranchukEffect2021, pierceUnconventionalSequenceCorrelated2021} and spectroscopic measurements \protect\cite{wongCascadeElectronicTransitions2020}. 

Entropy and transport measurements have observed the isospin Pomeranchuk effect in TBLG \protect\cite{rozenEntropicEvidencePomeranchuk2021, saitoIsospinPomeranchukEffect2021, shenCorrelatedStatesTwisted2020, caoStrangeMetalMagicAngle2020}, in a phenomenological analogy to helium-3 \protect\cite{pomeranchukTheoryLiquid3He1950}. The Pomeranchuk effect has previously been discussed in the solid state context in the Hubbard model \protect\cite{georgesNumericalSolutionHubbard1992, georgesPhysicalPropertiesHalffilled1993}, and in cold atoms \protect\cite{wernerInteractionInducedAdiabaticCooling2005,Taka_2012}. 
At certain filling fractions, raising the temperature induces a transition from a metallic state (resistivity $\approx 10^{-1}$~k$\Omega$) to a near-insulating state (resistivity $\approx 10^1$~k$\Omega$) at around $5$--$10$~K \protect\cite{saitoIsospinPomeranchukEffect2021, shenCorrelatedStatesTwisted2020}. Our approach allows us to distinguish between two distinct metallic states in TBLG---an order-induced coherent good metal below the DMFT ordering temperature, and an incoherent bad metal. Our results therefore point to a microscopic mechanism underlying the observations of ``Pomeranchuk physics" in transport and thermodynamics experiments, with good agreement of the temperature scale.

The doping dependencies of the electronic spectra and the existence of disordered moments at temperatures above $\about 10$~K is closely linked to the dependence of the chemical potential $\mu$ on doping $\nu$ in the symmetric phase. Our calculations unveil a balancing mechanism between the filling of the correlated and uncorrelated subspaces, by which the former is progressively occupied with increasing total filling, and the latter is cyclically filled and depleted. We discuss the correlated nature of the system at integer and fractional fillings, showing how, coherently with what is known about the physics of the periodic Anderson model, Mott-like behavior emerges at integer values of the \emph{total} filling. We compare our DMFT results with the exact solution of the topological heavy Fermion model \protect\cite{songMagicAngleTwistedBilayer2022} (see also Sec.~\ref{sec:model}) in the zero-hybridization limit \protect\cite{huKondoLatticeModel2023c} and explain the observed features in the experimentally measurable \textit{charge compressibility}. This is found to exhibit a saw-tooth behavior and negative values in extended ranges of dopings, once the geometrical capacitance contribution is subtracted out analogously to experiments~\protect\cite{Eisenstein1992,Riley2015,Schakel_2001}. We find that the sawtooth features in the compressibility wash out and the nearly incompressible states fade away at $\about 100$~K, which is related to the onset of charge fluctuations. Our observation is consistent with with temperature-dependent experimental data \protect\cite{saitoIsospinPomeranchukEffect2021}, and underlines the role of fluctuating local moments in the \emph{cascade transitions}. 
\begin{figure*}
    \centering
    \includegraphics{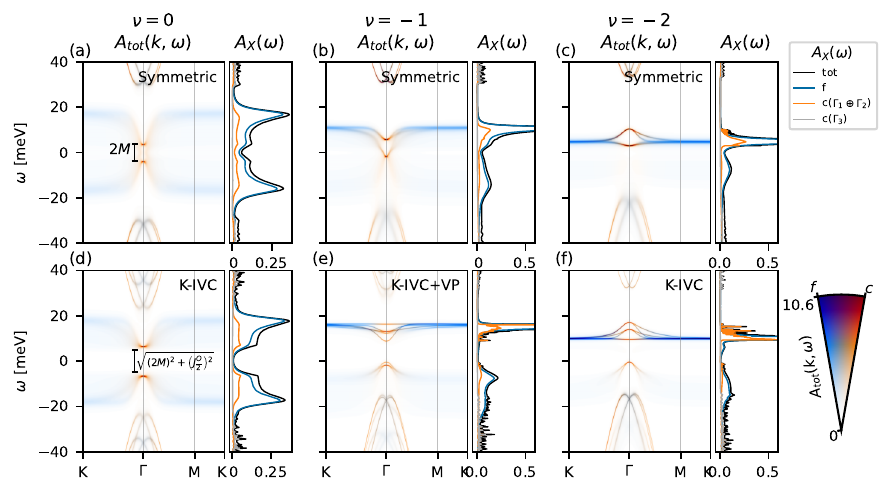}
    \caption{The momentum-resolved $A(k,\omega)$ and momentum-integrated $A(\omega)$ spectral functions at integer filling in the symmetric (top panels) and symmetry-broken phases (bottom panels) at the fillings $\nu=0,-1,-2$ (from left to right). In the color plot for $A_{tot}(k,w)$, the hue represents the orbital character (blue for $f$ vs red for $c$) of the spectral weight. The different lines in the $A(\omega)$ plot denote total ($X = \text{tot}$) or orbital-projected ($X = f, c(\ldots)$) spectral functions. The data is at $7.7$~K.}
    \label{fig:Akw_int}
\end{figure*}
\section{Model}\label{sec:model}

We use the topological heavy Fermion (THF) model from Ref.~\protect\cite{songMagicAngleTwistedBilayer2022} to describe the electronic structure of TBLG. This model, derived from the microscopic interacting Bistritzer-MacDonald model \protect\cite{bistritzerMoireBandsTwisted2011}, connects a set of completely localized $f$-orbitals to highly dispersive $c$-orbitals that carry the topology. Per spin and valley ($\sigma\in\{\uparrow, \downarrow\}$, $\eta\in\{+,-\}$ respectively), there are two $f$-orbitals ($\alpha\in\{1,2\}$), and four $c$-orbitals (two of each forming the $\Gamma_3$ ($a\in\{1,2\}$) and $\Gamma_1\oplus\Gamma_2$ ($a\in\{3,4\}$) representations). The $f$-orbitals make up most of the flat bands; the exception is at the $\Gamma$ point in the moir\'{e} Brillouin zone, where the flat band character changes to that of the $\Gamma_1 \oplus \Gamma_2$ $c$-orbitals through $f$-$c$ hybridization. The THF Hamiltonian can be written as
\begin{align}
    \hat{H}_{THF} = \underbrace{\hat{H}_c + \hat{H}_{cf}}_{\hat{H}_0} + \hat{H}_{U} + \hat{H}_W + \hat{H}_V + \hat{H}_J.
\end{align}
The terms comprising the non-interacting Hamiltonian $\hat{H}_0$ are $\hat{H}_c$, which contains the dispersion of the $c$-orbitals, and $\hat{H}_{cf}$, which contains the hybridization between the two subspaces. $\hat{H}_0$ defines two important energy scales: the splitting of the $\Gamma_1 \oplus \Gamma_2$ $c$-subspace, $M = 3.7$~meV, which sets the bandwidth of the flat bands, and the $f$-$c$ hybridization term at $\Gamma$, $\gamma=-24.8$~meV, which sets the gap between the flat bands and the high-energy bands. The four terms in the interacting part of the Hamiltonian are $\hat{H}_U$ ($\hat{H}_V$), the density-density interaction in the $f$($c$)-subspace, $\hat{H}_W$, the density-density interaction between $c$- and $f$-states, and $\hat{H}_J$, the exchange interaction between the $f$- and $c$-subspaces. They are calculated by performing double-gated Coulomb integrals (see Supplementary material~\protect\cite{SM} or \protect\cite{songMagicAngleTwistedBilayer2022} for the definition of each term). Following typical experimental setups, we use a gate distance of 10~nm for main results of this manuscript. We've checked that the results are robust within a range of reasonable gate distances and consequently, a reasonable range of interactions (see Supplementary Material~\protect\cite{SM}).

In our calculation, we treat $\hat{H}_U$ with DMFT, taking its local many-body effects into account, and the remaining interaction terms via static mean-field decoupling. We perform two sets of calculations: (a) allowing for symmetry-broken states where symmetry-breaking in the first iteration is guided by Hartree-Fock results from \protect\cite{songMagicAngleTwistedBilayer2022}; (b) in a fully symmetric state. For later convenience, we define $n_f$, $n_c$, and $n = n_f+n_c$ to be the number of $f$-, $c$-, and total electrons per moir\'e unit cell in the system. The corresponding fillings with respect to the charge neutrality point are given by $\nu_f = n_f-4$, $\nu_c = n_c-8$, and $\nu = n-12$. Due to the exact particle-hole symmetry of the THF model (which may be broken by additional terms not included in this study), the physics at positive and negative $\nu$ is related by a particle-hole transformation, and we will limit our discussions to $\nu\leq 0$. In this paper, isospin refers to a generalized spin consisting of electron spin, valley, and orbital degrees of freedom. The corresponding local moments are referred to interchangeably as isospin moments, local moments, and local isospin moments.

\section{Methods}\label{sec:methods}

We split the total Hamiltonian into a static and a dynamic part,
\begin{align}
    \hat{H}_{stat} =& \hat{H}_c + \hat{H}_{fc} + \hat{H}_W^{MF} + \hat{H}_V^{MF} + \hat{H}_J^{MF} \nonumber\\&- 3.5 U \sum_{\alpha\eta\sigma} f^\dagger_{\alpha\eta\sigma}f_{\alpha\eta\sigma},\\
    \hat{H}_{dyn} =&  \frac{U}{2}\sum_{(\alpha\eta\sigma)\neq (\alpha'\eta' \sigma')}f^\dagger_{\alpha\eta\sigma}f_{\alpha\eta\sigma}f^\dagger_{\alpha'\eta'\sigma'}f_{\alpha'\eta'\sigma'}.
\end{align}
The superscript $MF$ represents a static mean-field decoupled interaction term (Hartree+Fock for $\hat{H}^{MF}_W$ and $\hat{H}^{MF}_J$ and Hartree for $\hat{H}^{MF}_V$). 

$\hat{H}_{stat}$ plays the role of the lattice Hamiltonian in the DMFT calculation. It must be self-consistently determined as the mean-field decoupled interaction terms depend on the system's density matrix $\rho$. $\hat{H}_{dyn}$ acts on the $f$-subspace only and induces a frequency-dependent self-energy in the $f$-subspace. We solve the impurity problem with two continuous-time quantum Monte Carlo (CT-QMC) hybridization expansion solvers (TRIQS-cthyb \protect\cite{parcolletTRIQSToolboxResearch2015, sethTRIQSCTHYBContinuousTime2016, aichhornTRIQSDFTToolsTRIQS2016} and w2dynamics \protect\cite{parraghConservedQuantitiesInvariant2012, wallerbergerW2dynamicsLocalOne2019}). We converge two self-consistency loops at once: the DMFT self-consistency condition for the self-energy $\Sigma$ of the $f$-subspace, and the Hartree-Fock mean-field condition for the total density matrix $\rho$. Details of the calculations, including the CT-QMC parameters and a comparison of TRIQS and w2dynamics results, are given in the Supplementary Material~\protect\cite{SM} (see also references \protect\cite{gubernatisQuantumMonteCarlo1991,krabergerMaxEnt2023,kaufmann2021anacont} therein). 

The results shown in this paper are obtained for the lowest energy ordered phases predicted by a Hartree-Fock analysis \protect\cite{songMagicAngleTwistedBilayer2022} of the THF model in the absence of strain. That is, at the charge neutrality point ($\nu=0$), both spin sectors are half-filled with K-IVC order, upon hole doping once $(\nu=-1)$, one spin sector is K-IVC ordered, while the other spin sector is valley polarized, and at half-filling $(\nu=-2)$, one spin sector has K-IVC order, while the other has no long-range order. We emphasize that these states and their flat- and chiral-$U(4)$ related counterparts valley-polarized (VP), and intervalley-coherent (T-IVC) states are very close in energy. The true ordered state is therefore sensitive to defects, substrate effects, strain, and other sample-dependent perturbations. These effects can be incorporated with DMFT on the THF model with additional terms, and will be the subject of a future publication. Our goal is to make universal statements about the interplay of correlations and ordering in this system, which likely does not depend on which of the several competing low-temperature ordered phases the system is in.

For the symmetry-broken calculations at integer fillings, we bias the system towards the chosen symmetry-broken solution by applying a weak polarizing field for the first few DMFT iterations, and then turning it off for the remainder until self-consistency is reached. (See Supplemental Material~\protect\cite{SM} for the definition of the polarizing field). In the ordered phase, we approach fractional fillings around each integer filling by gradually doping the integer-filling self-consistent solution in small increments.

\section{Results}\label{sec:results}
\begin{figure}
    \centering
    \includegraphics{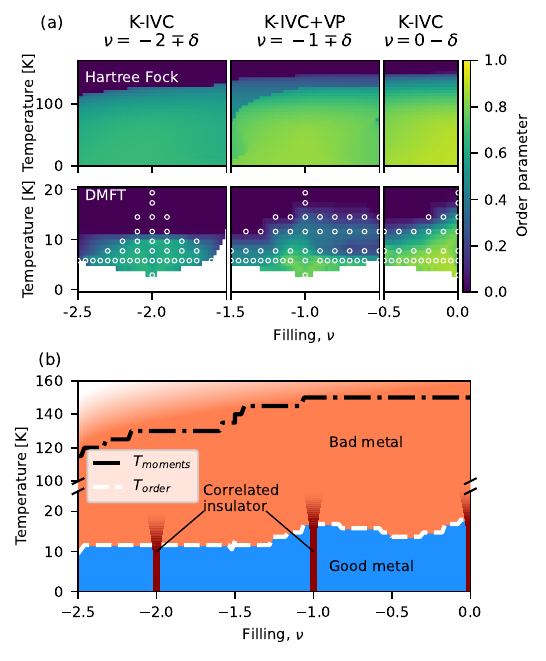}
    \caption{(a) The magnitude of the symmetry-breaking order parameter at various fillings and temperatures. In the bottom panel, white circles represent numerical data points from DMFT simulations at and around the $\nu=-2$ (K-IVC), $\nu=-1$ (K-IVC+VP), and $\nu = 0$ (K-IVC) parent states in the left, middle, and right panels respectively. The space between circles is filled by linear interpolation. The top panel shows the same quantity from a Hartree-Fock simulation (note the different vertical scale). In the dark blue regions, the self-consistency loop flows to the symmetric state, indicating that the symmetry-broken solution does not exist. (b) A schematic filling-temperature phase diagram: thermally activated charge fluctuations are frozen below the Hartree-Fock ordering temperature ($T_{\text{moments}}$), allowing local moments to progressively form in the orange region. These local moments order below the DMFT ordering temperature ($T_{\text{order}}$) in the blue region. The DMFT ordering temperature marks the boundary between a bad metal and an order-facilitated good metal at fractional fillings as discussed in Sec.~\ref{sec:fermiology} and Sec.~\ref{sec:integer-fillings} in the main text. The correlated insulators at integer fillings emerging below $T_{\text{order}}$ (dark red) fade into the bad metal above $T_{\text{order}}$ forming regions with most strongly suppressed quasiparticle weights, c.f. Fig.~\ref{fig:decoupled_comparison}b).}
    \label{fig:stability}
\end{figure}
\subsection{(Nearly) insulating states at integer filling}
Fig.~\ref{fig:Akw_int} shows the DMFT spectral functions in the symmetric and the symmetry-broken phases at integer fillings $\nu=0,-1,-2$. In all cases, the local Hubbard term in $\hat{H}_{dyn}$ shifts the flat-band spectral weight of the $f$-subspace away from the Fermi level to form lower and upper Hubbard bands. The remaining low-energy excitations have residual $f$- and $\Gamma_1\oplus \Gamma_2$ $c$-orbital character, while the $\Gamma_3$ $c$-spectral weight remains pushed away to higher energies ($\about \gamma = 24$~meV) by the $f$-$c$ hybridization term. We point out two generic differences between the spectral functions in the symmetric ((a)-(c)) and symmetry-broken ((d)-(f)) phases. First, the Hubbard bands are relatively sharp in the symmetry-broken phase compared to those in the symmetric phase, indicating that dynamic correlations are weaker in the symmetry-broken phases \protect\cite{sangiovanniStaticDynamicalMeanfield2006}. Second, while low-energy spectral weight is depleted also in the symmetric state, symmetry-breaking supports a robust insulating gap in the absence of strain. This is best seen comparing the momentum-integrated spectral functions in the right panels of Fig.~\ref{fig:Akw_int}(a) and (d). In the symmetric case, spectral weight is reduced at the Fermi level but does not vanish. In the ordered case, the spectral weight vanishes and there is a robust gap. This can be understood in the language of the THF model. At the $\Gamma$ point, the bare (with $f$-$c$ hybridization turned off) dispersion of the THF model has two contributions at zero energy: per spin and valley, there is a pair of $f$-orbitals with completely flat bands at zero, and a pair of $\Gamma_3$ $c$-orbitals that contribute a pair of particle-hole related parabolic bands touching at zero. $f$-$c$ hybridization moves both of these contributions away from zero to the high-energy bands at the $\Gamma$~point. Precisely at the $\Gamma$ point, the flat bands are entirely of $\Gamma_1\oplus\Gamma_2$ $c$-character. Generically, the $c$-electron spectral weight is affected by interactions in the $f$-sector directly by the $f$-$c$ exchange interaction ($H_J$), and indirectly through hybridization effects. The former directly gaps out the $c$-electron spectral weight near the Fermi level. This is evident in the energy difference between the bright spots in the spectral functions at the $\Gamma$ point at the charge neutrality point (CNP) in Fig.~\ref{fig:Akw_int}(a) and Fig.~\ref{fig:Akw_int}(d). In the symmetric phase (a), the $\Gamma_1 \oplus \Gamma_2$ $c$-orbitals retain the $2M$ splitting of the non-interacting Hamiltonian. In the symmetry-broken case (d), the splitting gains an additional contribution from the $f$-$c$ exchange term $\hat{H}_J$ and is given by $\sqrt{(2M)^2 + (OJ/2)^2}$, where $O$ is the off-diagonal inter-valley term in the $f$-subspace density matrix. Note that we treat $\hat{H}_J$ with Hartree-Fock, which neglects any dynamical renormalization effects. Deep in the ordered state, these renormalization effects are expected to be weak. In the symmetric state, it has been shown by poor man's scaling~\protect\cite{chouScalingTheoryIntrinsic2023} that the renormalized $J$ is always reduced at low temperatures. Fortuitously, the Hartree-Fock decoupled $\hat{H}_J$ is negligible in the symmetric state---the Fock term vanishes due to its dependence on off-diagonal terms in the density matrix, and the Hartree term is a small ($\about 
\frac{\nu}{8}J$) effective $c$-$f$ double-counting term.

What is the nature of the (nearly) insulating states at integer fillings? Given the close relation between the THF and the periodic Anderson model (PAM), we attempt a classification of insulating states of the THF model in terms of the phenomenology of the periodic Anderson model. In PAM-like models, different types of insulators including band insulators, Kondo insulators, Mott insulators, and charge transfer insulators have been established~\protect\cite{imada_metal-insulator_1998,amaricciMottTransitionsPartially2017}. 

At integer $\nu_f$, the limit of $U\to \infty$ corresponds to removing all $f$-states from the THF Hamiltonian. However, without $c$-$f$ coupling (and thus also without $f$-states at all) the THF model is metallic~\protect\cite{songMagicAngleTwistedBilayer2022} and not gapped. Clearly, the absence of a gap at $U\to \infty$ in the symmetric state, rules out TBLG at integer filling being a genuine Mott insulator, in agreement with the interpretations of \protect\cite{haule_mott-semiconducting_2019,dattaHeavyQuasiparticlesCascades2023a}. At the experimentally relevant temperatures of a few K, a Kondo insulator at integer fillings is unlikely, since the $f$-electron moments are not yet fully screened down to $\about 1$~K~\protect\cite{huSymmetricKondoLattice2023a}.

The absence of a hard Mott gap can be traced back to a finite $t_{fc}$ in our THF model, similarly to what has been discussed in Ref.~\protect\cite{demediciMottTransitionKondo2005}: even at strong interaction strengths, as long as the $f$-electrons can hop into a non-interacting band crossing the Fermi level, a Mott insulator with a clean gap is prevented. The heavy Fermi-liquid that replaces it has typically a very low coherence temperature. In accordance with these general expectations (and despite the differences between our model and those of Ref.~\protect\cite{demediciMottTransitionKondo2005}) we find (see section~\ref{sec:integer-fillings}) that the quasiparticle weight never vanishes in the symmetric state, even close to integer fillings where it is heavily suppressed, displaying the commensurability effects characteristic of incipient Mott phases. Still, in order to see a clean gap, long-range ordering is needed, as we find in our calculations for broken-symmetry states.

The presence of $f$- and $c$- spectral weight near the Fermi level is reminiscent of ``p-metals''~\protect\cite{amaricciMottTransitionsPartially2017} or charge transfer (CT) insulators~\protect\cite{imada_metal-insulator_1998}. In typical transition metal oxide-based CT insulators like NiO or cuprates \protect\cite{imada_metal-insulator_1998}, the electronic gap is bounded by transition metal $3d$ and oxygen $2p$-spectral weight from above and below, respectively. The insulating states at $\nu=-2$ with dominantly $c$-spectral weight below the Fermi level and $f$-spectral weight above (c.f. Figs. \ref{fig:Akw_int} (c) and (f) ) resemble this CT scenario. Also the insulating states $\nu=0$ and $\nu=-1$ are similar to CT insulators, yet with one decisive difference to the usual CT and $p$-metal case: in TBLG at $\nu=0$ and $\nu=-1$ we have delocalized $c$-bands dispersing in between the $f$-type Hubbard bands from above \textit{and} below. Also differently from the transition metal oxide CT cases, the spectral depletion regions near the Fermi level of TBLG in the symmetric state rely on hybridization between delocalized ($c$) and localized ($f$) states. In the ordered states of TBLG, it is the exchange interaction between $c$ and $f$ orbitals leading to mass terms \protect\cite{huKondoLatticeModel2023c} which markedly enhance the gaps in the $c$-sector and thus also the total gaps.

\subsection{Fluctuations and the stability of the ordered solution}Next, we study how doping or raising the temperature affects the ordered states at integer fillings. Fig.~\ref{fig:stability}(a) shows the order parameter of the self-consistent solution in a doping-temperature plane generated by gradually doping a particular ordered state at an integer filling. We define the order parameter in the symmetry-broken phase at arbitrary filling by the matrix inner product with the traceless part of the corresponding parent state density matrix (see Supplementary Material \protect\cite{SM} and references~\protect\cite{calugaru_ipt,PUL80} therein). At high enough temperatures, the system flows to the disordered phase under the self-consistency loop. We find the threshold temperature to be $\about 15$~K at $\nu=0, -1$, and $\about 10$~K at $\nu=-2$. The ordering temperature predicted by Hartree-Fock is an order of magnitude higher ($\about 100$-$150$~K; see upper panel of Fig.~\ref{fig:stability}(a)). Our DMFT simulations show that long-range order is suppressed by local dynamic fluctuations down to about $10$~K. Spatial fluctuations, which are neglected in single-site DMFT, are expected to reduce the ordering temperature further. Just below the DMFT ordering temperature, doping away from an integer solution also leads to a disordered solution, resulting in dome-shaped ordered regions as seen in the lower panel of Fig.~\ref{fig:stability}(a). At low enough temperatures, the ordered solution continued from either neighboring integer filling may co-exist at a given fractional filling.

The Hartree-Fock ordering temperature encodes the onset of thermally activated charge fluctuations. Above this temperature thermal smearing results in the Hartree-Fock equations converging to the symmetric unpolarized solution. Below this temperature near integer fillings, one charge sector predominantly contributes to the many-body configurations, allowing the formation of local moments (see Supplementary Material \protect\cite{SM} for the numerical analysis of the sector statistics). Since Hartree-Fock neglects dynamic fluctuations, moments order immediately upon formation. Therefore, the temperature range between the Hartree-Fock and DMFT ordering temperatures is the regime of fluctuating local moments. Correspondingly, in this temperature range, the local spin susceptibility follows the Curie-Weiss law (see~\protect\cite{huSymmetricKondoLattice2023a} and Supplementary Material~\protect\cite{SM}). Note that near half-integer filling an intermediate valence regime with quantum charge fluctuations between two (but not more) neighboring sectors is realized below the Hartree-Fock ordering temperature preventing the identification of well-defined local moments \protect\cite{huSymmetricKondoLattice2023a}, while also additional charge sectors get activated above this temperature scale.

Anticipating the results of Sec.~\ref{sec:MIT} and Sec.~\ref{sec:fermiology}, where we discuss the metallic states below and above the ordering temperature, we construct a schematic phase diagram in Fig.~\ref{fig:stability}(b). Below the ordering temperature of $\about 10$~K, we find an insulator at integer fillings, and a good metal at fractional fillings. The state above the ordering temperature is discussed in Sec.~\ref{sec:fermiology}, where we will show that the spectral weight at the Fermi level is generically incoherent, indicating a bad metal. Our results point to a \emph{coherence from order} at low temperatures, which we expand on in Sec.~\ref{sec:coherence}.

\begin{figure*}
    \centering
    \includegraphics[width=\textwidth]{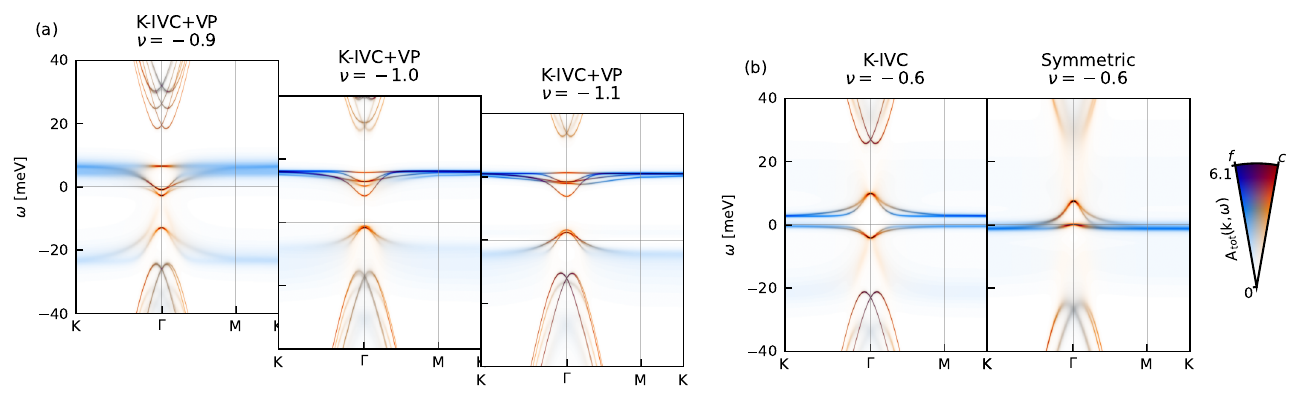}
    \caption{Spectral functions at $5.8~K$ (a) at and close to $\nu=-1$ in the ordered phase (b) at $\nu=-0.6$ in the ordered and symmetric phases. In (a), the electron and hole doped panels are shifted such that the Hubbard bands match up with the middle panel. Lightly doping away from an integer-filled insulator moves the Fermi level into the coherent $c$-spectral weight around the $\Gamma$-point leading to a coherent metal. Upon further doping as in (b), the Fermi level hits the $f$-part of the flat bands, leading to coherent and incoherent spectral weight at the Fermi level.}
    \label{fig:MIT}
\end{figure*}

\subsection{Doping-induced insulator-to-metal transition}\label{sec:MIT}

Away from integer filling, the normal state of TBLG is metallic. The nature of the metallic state depends on the filling and temperature. In fact, we find distinct behaviors upon doping with electron or holes away from the insulators at $\nu=0$ or $\nu=-1$ on one hand and $\nu=-2$ on the other hand. In Fig.~\ref{fig:MIT}(a), we show how the spectral function changes as we slightly dope the system away from $\nu=-1$ at $T=5.8$~K in the ordered state with electrons or holes. In this case, we find that the Fermi level moves into the dispersive part of the low-energy bands around the $\Gamma$ point: the active bands are coherent with delocalized $c$-character. Here, we expect Fermi liquid-like behavior. Metallic behavior derived from $c$-orbitals on both sides of the insulator points back to the peculiar property of the charge-transfer insulator at $\nu=-1$ with $c$-bands dispersing both above and below. In contrast, the behavior at $\nu=-2$ is asymmetric with respect to doping. Consider the spectral function in Fig.~\ref{fig:Akw_int}~(f). The $c$-part of the upper band is flattened. While hole doping at $\nu=-2$ would lead to metallic behavior just like at $\nu=-1$, electron doping would start to immediately occupy the $f$-orbitals. This is much more reminiscent of a conventional charge transfer insulator, with localized carriers on one side and delocalized carriers on the other side of the gap.

Upon further doping away from $\nu=-1$ towards $\nu=0$, the Fermi level eventually also hits the localized $f$-part of the flat bands. This occurs generically once between every two successive integer fillings, and is shown in Fig.~\ref{fig:MIT}(b) for $\nu=-0.6$. The Supplementary Material~\protect\cite{SM} includes a movie showing the evolution of the spectral function as the filling is varied (\url{https://youtu.be/cw3K0YEsPU0}). The ordered phase (in the left panel) has a splitting in the $f$-subspace that is absent in the symmetric phase (in the right panel). This is a consequence of a feedback of the ferromagnetic exchange interaction $\hat{H}_J$, which is only active in the ordered phase. Due to the isospin order in the $f$-sector, $\hat{H_J}$ behaves like a polarizing field in the $c$-sector, inducing analogous isospin order in the $c$-sector. The induced isospin order in the $c$-sector in turn causes a small polarizing field in the $f$-sector, resulting in the splitting seen in the left panel of Fig.~\ref{fig:MIT}(b).

In both the symmetric and ordered state, hitting the localized $f$ part of the spectrum induces a charge reshuffling between the localized and delocalized subspaces, resulting in the sawtooth pattern of orbital-resolved filling seen in \protect\cite{huSymmetricKondoLattice2023a, kangCascadesLightHeavy2021, huKondoLatticeModel2023c}. This is precisely the region of fillings where experiments see negative compressibilities \protect\cite{saitoIsospinPomeranchukEffect2021, zondinerCascadePhaseTransitions2020}. We discuss the orbital-resolved fillings along with the compressibility further in Sec.~\ref{sec:compressibility}. These filling regions are also associated with a sequence of Lifshitz transitions. We identify these Lifshitz transitions with the experimentally observed cascade transitions \protect\cite{wongCascadeElectronicTransitions2020, zondinerCascadePhaseTransitions2020} in Sec.~\ref{sec:fermiology}.

\begin{figure*}
    \centering
    \includegraphics[width=\textwidth]{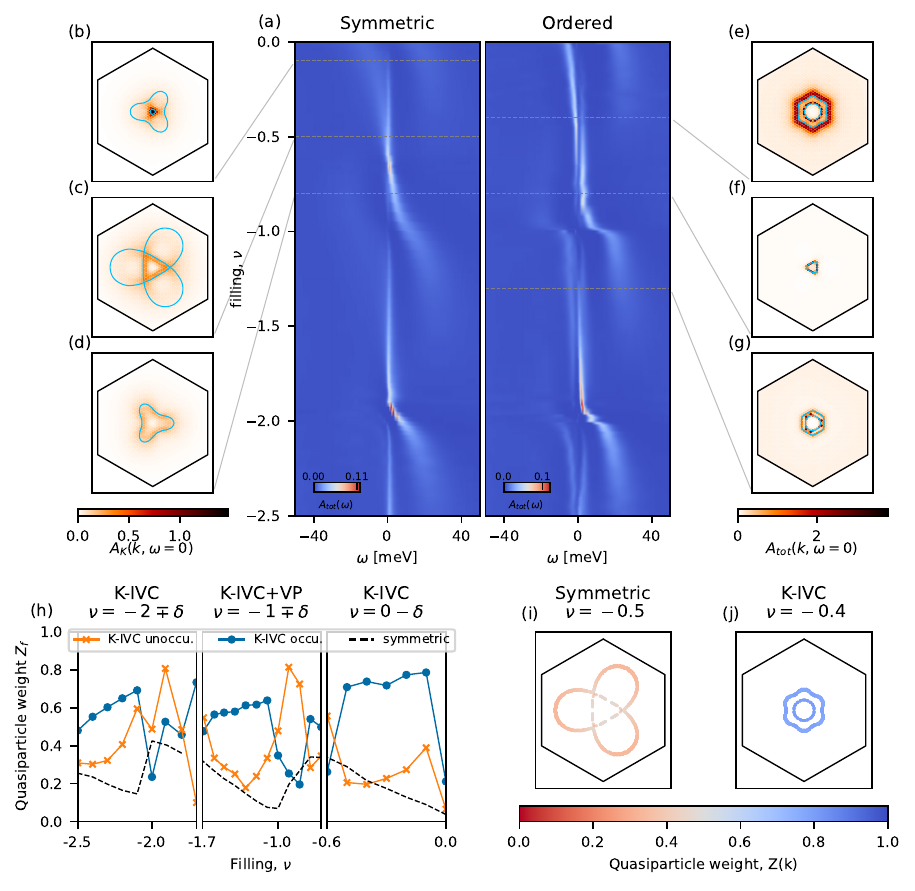}
    \caption{(a) The momentum-integrated spectral function $A_{tot}(\omega)$ $5.8~K$ in the symmetric phase (left) and in the ordered phase (right) as a function of hole doping. The electron-doped side (with respect to CNP) is related to the hole-doped side by a partice-hole transformation. (b)-(d) The $K$-valley--projected zero-energy spectral function in the first moir\'{e} Brillouin zone in the symmetric phase at select fillings. (e)-(g) The full zero-energy spectral function in the first moir\'{e} Brillouin zone in the ordered state at select fillings. The light blue lines mark the zeros of the effective Hamiltonian as identified by a quasiparticle analysis (see main text). (h) A comparison of the $f$-orbital quasiparticle weight $Z_f$ in the symmetric and ordered states at the same temperature $T=5.8$~K. In the ordered case, only the spin-up K-IVC occupied and unoccupied orbitals are shown for clarity. (i)-(j) The quasiparticle weight projected onto the quasiparticle basis along the zeros of the quasiparticle hamiltonian. All data is at $T=5.8$~K. }
    \label{fig:Fermiology}
\end{figure*}

\subsection{Cascade transitions and signatures of order}\label{sec:cascade}In Fig.~\ref{fig:Fermiology}(a), we show the momentum-integrated spectral function as a function of filling in the symmetric and symmetry-broken phases. In both cases, there is a reconstruction of the low-energy spectral features upon changing the doping by an integer value. These are the \emph{cascade transitions} that have been seen experimentally with scanning tunnel spectroscopy (STS) \protect\cite{wongCascadeElectronicTransitions2020, nuckollsStronglyCorrelatedChern2020} and which have been similarly found in the DMFT study of Datta et al.~\protect\cite{dattaHeavyQuasiparticlesCascades2023a} for the symmetric state. In the symmetric state, at integer fillings, the spectral function has a two-peak structure, with a lower and an upper peak that are similarly far away from zero, but with their relative spectral weight depending on which integer filling the system is tuned to (for instance, compare $\nu=-2$ and $\nu=-1$ in Fig.~\ref{fig:Fermiology}). Upon hole doping away from an integer filling, the two peak structure shifts to higher energies, the lower peak merges with a zero-energy resonance, and the upper peak fades away. Upon further hole doping, as the system approaches the next integer filling, the zero energy resonance shifts to higher energies becoming the new upper peak, and a new lower peak emerges.

The symmetry-broken state behaves the same way except that there is fine low-energy structure, owing to isospin order. The details of the fine structure will depend on the particular ordered state. In particular, the zero-energy peak at fractional fillings, made up primarily of $f$-spectral weight, is split by a feedback effect from the exchange interaction $\hat{H}_J$. This splitting is present as long as there is isospin order. See Fig.~\ref{fig:stability} for the sets of filling and temperature values where this fine structure may be present. Note that our calculation does not include spatial fluctuations, which might further suppress ordering temperatures.

\subsection{Fermiology and Lifshitz transitions}\label{sec:fermiology}We perform a quasiparticle analysis at the Fermi level to better understand the Fermi surface reconstruction and the Lifshitz transitions underlying the cascades. The exact shape of the Fermi surface will depend on perturbations such as strain, but some general features such as the existence and loose integer-periodic nature of the Lifshitz transitions will remain. We numerically find the zeros, $\hat{H}_{qp}(k)\ket{\psi^{i}(k)} = 0$, of the quasiparticle Hamiltonian
\begin{equation}\label{eq:Hqp}
   \hat{H}_{qp}(k)\,=\,-\,\hat{Z}^\frac{1}{2}\mathfrak
     {Re}\left[G^{-1}(\omega+i\epsilon, k)\right]_{\omega\to 0}\hat{Z}^{\frac{1}{2}}, 
\end{equation}

where $\hat{Z}=\left(\left.1 - \partial_{i\omega}\mathfrak{Im}\hat{\Sigma}_{tot}(i\omega)\right|_{i\omega\to 0}\right)^{-1}$ is the matrix quasiparticle weight. $\hat{\Sigma}_{tot}$ is the $f$-subspace self-energy promoted to the full $f\oplus c$ space by padding with zeros in the $c$-sector (the $c$-sector is treated on a mean-field level and the static contribution to the self-energy is included in the mean-field terms in $\hat{H}_{stat}$), the superscript $i$ on $\ket{\psi^i(k)}$ labels the zeros in case there is degeneracy at any $k$-point. 
The zeros of $\hat{H}_{qp}$ point at potential quasiparticles at the Fermi level. In particular, the quasiparticle weight $\hat{Z}$ evaluated at the zeros gives information on the coherence of spectral weight at the Fermi level in the form of a $k$-dependent quasiparticle weight \begin{equation}
\frac{1}{Z^{i}(k)} = \bra{\psi^i(k)}\hat{Z}^{-1}\ket{\psi^i(k)} 
\label{eq:Zk}
\end{equation}along a potential Fermi contour. For an uncorrelated Fermi liquid, $Z^i(k)=1$, and $\ket{\psi^i(k)}$ would give the quasiparticle eigenstates at the Fermi surface. Technical details on the root-finding algorithm used to find $\ket{\psi^i(k)}$ and the rest of the quasiparticle analysis are provided in the Supplementary Material~\protect\cite{SM}.

The blue lines in Fig.~\ref{fig:Fermiology}(b)-(f), which we henceforth refer to as zero energy lines (ZEL), mark the location of the zeros $\ket{\psi^i(k)}$ in $k$-space overlaid on the spectral function at the Fermi level in the first moir\'{e} Brillouin zone at select fillings. For clarity, the symmetric phase data in Fig.~\ref{fig:Fermiology}(b)-(d) are projected to the $K$-valley only. The $K'$ valley contribution is related by a $C_2$ rotation. We find that the topology of the ZEL changes with doping. In our zero-strain calculation, between the CNP and $\nu=-1$, we see three regimes. Upon lightly doping away from CNP, (b), The $K$-projected ZEL consists of two concentric contours. The three lobes of the outer contour jutting away from the $\Gamma$ point have low quasiparticle weight, and correspondingly smeared out spectral weight. The inner ring and the inner sections of the outer ring are primarily of $c$-character and more coherent, and form a Fermi contour. Halfway to $\nu=-1$ (c), the Fermi level hits the flat $f$-part of the flat band (see also Fig.~\ref{fig:MIT}(b)). This coincides with the concentric ZEL connecting to form a trefoil knot. This ZEL has three points where it intersects itself corresponding to potential van Hove singularities at the Fermi level. As in the previous case, however, the outer lobes of the ZEL have low quasiparticle weight and are primarily $f$-character. Finally, approaching $\nu=-1$, the ZEL consists of a single closed contour, as seen in (d). We reiterate that the other valley contributes additional $60$ degree rotated copies of these Fermi surfaces restoring $C_6$ symmetry.

In the ordered case Fig.~\ref{fig:Fermiology}(e)-(g), the zeros are not valley-decoupled as the charge carriers occupy inter-valley coherent orbitals. The color map therefore shows the full spectral weight with both valleys included. Generically, away from integer fillings, the zeros are coherent with high quasiparticle weight. The symmetry and topology of the Fermi surface depends on the properties of the active orbital at the chosen filling. Doping with holes away from CNP, the Fermi surface consists of K-IVC orbitals getting depleted, resulting in a Fermi surface that is $6$-fold symmetric. Near $\nu=-1$, there are two occupied K-IVC orbitals in one spin sector, and an occupied valley-polarized orbital in the other spin sector. Electron (hole) doping corresponds to occupying (depleting) the unoccupied (occupied) valley-polarized orbital, resulting in a Fermi surface that is $3$-fold symmetric. We emphasize that perturbations such as strain will change the details of the ordered state and the shape and symmetries of the Fermi surface. However, there will still be Lifshitz transitions between integer fillings when the Fermi level hits the incoherent $f$-band.

\subsection{Pomeranchuk physics and coherence from order}\label{sec:coherence}An important takeaway from Fig.~\ref{fig:Fermiology} is the different nature of the metallic state in the symmetric and ordered phases. Generically, the spectral weight at the Fermi level in the ordered state originates from much more coherent excitations than in the disordered state. This can be seen from Fig.~\ref{fig:Fermiology}(h), where we show the quasiparticle weight in the orbital basis over a range of fillings, and from Fig.~\ref{fig:Fermiology}(i)-(j), where we show the quasiparticle weight in the quasiparticle basis $Z(k)$ from Eq. (\ref{eq:Zk}) at chosen fillings in both the ordered and symmetric phase \emph{at the same temperature}. The quasiparticle weight is clearly higher in the ordered phase than in the symmetric phase, indicating that quasiparticles are more coherent in the ordered state than in the symmetric state \emph{at the same temperature}. Accordingly, the transport and quasiparticle scattering rates (see Supplementary Material~\protect\cite{SM}) are lower in the ordered phase for most fillings indicating that there is less scattering than in the symmetric phase, with higher quasiparticle lifetime. 
Thus, generically at fractional fillings, we expect a good metal in the ordered state, and a bad metal in the disordered state. A quantitative analysis of transport is the subject of an ongoing work. With an ordering temperature of about $10$--$15$~K, this interpretation is consistent with transport experiments \protect\cite{shenCorrelatedStatesTwisted2020, caoStrangeMetalMagicAngle2020, rozenEntropicEvidencePomeranchuk2021, saitoIsospinPomeranchukEffect2021}, which observe a sharp drop in resistivity at temperatures below $5$--$10$~K for a range of fillings around the correlated insulating phases. Note that single-site DMFT neglects all spatial fluctuations, which would reduce the ordering temperatures further. With this analysis, we are able to provide a microscopic explanation for the isospin Pomeranchuk effect in TBLG \protect\cite{shenCorrelatedStatesTwisted2020, caoStrangeMetalMagicAngle2020, rozenEntropicEvidencePomeranchuk2021, saitoIsospinPomeranchukEffect2021}. The bad metal at high temperatures is the result of an incoherent metal stabilized by the isospin entropy of pre-formed local moments, in analogy to solid helium-3 in the original Pomeranchuk effect \protect\cite{pomeranchukTheoryLiquid3He1950}. The low-temperature Fermi liquid is a coherent metal induced by spontaneous symmetry-breaking. The spectral function in the ordered phase is composed of bands of coherent quasiparticles, which become occupied upon doping away from the insulating states in the close vicinity of integer filling. In this regard, in contrast to the original Pomeranchuk effect, it is coherence facilitated by order that is responsible for the resistivity drop.

Taken together, our DMFT study reveals the following similarities and differences between the Pomeranchuk physics in He-3 and TBLG. In both systems, disordered fluctuating (iso)spin moments give rise to a high-entropy high temperature phase. Both, in He-3 and TBLG, the (iso)spin entropy is suppressed in the low-temperature state. Both, the low-temperature state of TBLG (at non-integer filling) and He-3, can have $T$-linear Fermi liquid derived contributions to the entropy. Yet, there is a decisive difference between the low-temperature states of TBLG and He-3: In TBLG, the entropy suppression is due to ordering of the (iso)spin moments, while He-3 realizes a Fermi liquid without long range spin-order. In other words, there are no local moments in He-3 at low $T$, while the isospin moments still exist in TBLG albeit ordered. Hence, the Pomeranchuk physics of TBLG is similar to the coherence from order physics of metallic ferromagnets like SrRuO$_3$~\protect\cite{Allen_SrRuO3_1996} or relatives of Fe-based superconductors~\protect\cite{Edelmann_PRB2017}.

\noindent
\subsection{Importance of integer total fillings}\label{sec:integer-fillings} 
\begin{figure*}
    \centering
    \includegraphics[width=\textwidth]{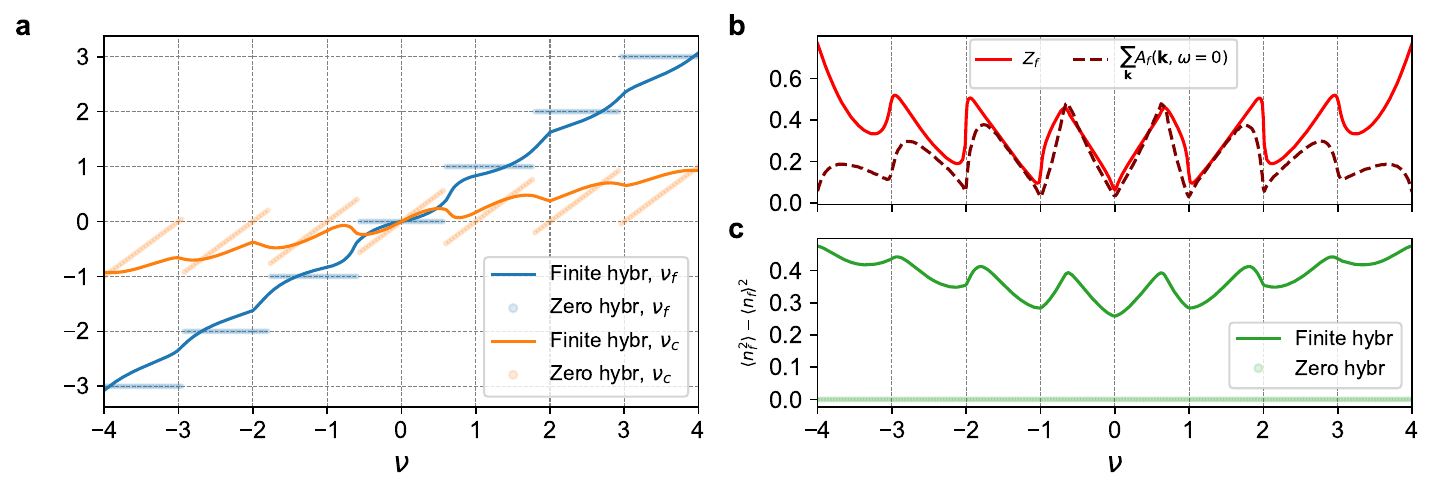}
    \caption{
    (a) Partial filling of the $f$- and $c$-orbitals as a function of the total filling $\nu$ in the symmetric phase. The dark solid lines show the DMFT results of the full model with hybridization. The light dotted curves are results from the zero-hybridization model solved exactly in \protect\cite{huKondoLatticeModel2023c}. (b) Quasiparticle weight for the $f$-orbitals, calculated at 11.6 $K$ and the local $f$-spectral weight at the Fermi level (c) Charge fluctuations of the $f$-orbitals $\langle n_{f}^{2}\rangle -\langle n_{f}\rangle^{2}$, as a function of the total filling $\nu$. Both quantities in (b) and (c) are indicators of the importance in DMFT of the integer values of $\nu$, in contrast to the zero-hybridization solution.
    }
    \label{fig:decoupled_comparison}
\end{figure*}

Fig.~\ref{fig:decoupled_comparison} reports the filling of the $f$- and $c$-orbitals, as a function of the total filling $\nu$. It is instructive to compare the DMFT solution of the full model with hybridized $f$- and $c$-orbitals to the solution of the zero-hybridization model from \protect\cite{huKondoLatticeModel2023c}. This is illustrated in Fig.~\ref{fig:decoupled_comparison}(a) where the solid lines refer to the solution with finite hybridization solved with DMFT and the light circles to the zero-hybridization case. It is clear that the model with hybridization obeys a similar energetic balance to the one without hybridization. The $c$-orbitals are periodically filled up and emptied, upon increasing $\nu$. In DMFT, this yields the characteristic, albeit smoothened out, sawtooth behavior. These overall trends displayed by $\nu_f$ and $\nu_c$ are a consequence of the strong correlations due to the $f$-$f$-Coulomb terms together with $f$-$c$ and $c$-$c$ interactions terms that are present both in the zero-hybridization solution of Ref.~\protect\cite{huKondoLatticeModel2023c} and in our DMFT calculation.

There is a further, more fundamental difference between finite and zero hybridization, namely the special role played in the former---and not in the latter---by the integer values of the total filling $\nu$.
In the zero hybridization model, $\nu=-3$, $\nu=-2$, $\nu=-1$ and so on, have no special meaning apart from trivially imposing $\nu_c=0$. On the contrary, the DMFT solution with finite hybridization bears a clear witness of the integer values of the total $\nu$, as evident from Fig.~\ref{fig:decoupled_comparison}(b)-(c).
The quasiparticle weight of the $f$-orbitals and the local $f$-spectral weight at the Fermi level, calculated from QMC via the low-T estimator $\sum_{\mathbf{k}}A_{f}(\mathbf{k},\omega=0)=\frac{\beta}{\pi}G_{f}^{\mathrm{QMC}}(\tau=\beta/2)$, display particularly strong variations approaching integer values of $\nu$. At the same time, the fluctuations of the $f$-occupation $\langle n_f^2 \rangle - \langle n_f \rangle^2$ are suppressed near integer values of $\nu$, in contrast to the zero-hybridization case, for which this quantity is flat and vanishes. In the Supplementary Material \protect\cite{SM}, we show that scattering rate has an analogous strongly filling-dependent behavior. Everywhere apart from the precise fillings at which the cusps in these two quantities occur, the behavior evidenced by these two indicators of the many-body nature of the $f$-electrons reveals a crucial property of the THF model with its full heavy-light fermion hybridization.

Periodic Anderson models ubiquitously show Mott-like behaviors at integer values of the total filling, rather than at integer fillings of the correlated subspace \protect\cite{Georges1993,amaricciMottTransitionsPartially2017}. This is independent of whether or not Mott insulating phases are fully realized and is to some extent counterintuitive. One may indeed naively think that the strongest propensity to Mottness is present at integer values of $\nu_f$, not of $\nu$. This ``commensurability'' physics as a function of total filling is obviously possible only when the hybridization puts the correlated and itinerant subspaces in communication and is captured by DMFT. Indeed, there are two main features that render DMFT especially suited to the task at hand: first, being by construction in the thermodynamic limit it is able to treat integer and fractional fillings on an equal footing. Second, DMFT has been proven to be able to solve models possessing degrees of freedom with different correlation strength in the low-energy subspace (as are $f$- and $c$-orbitals in our case) taking the charge fluctuations occurring between the two fully into account.

\begin{figure}
    \centering
    \includegraphics[width=\columnwidth]{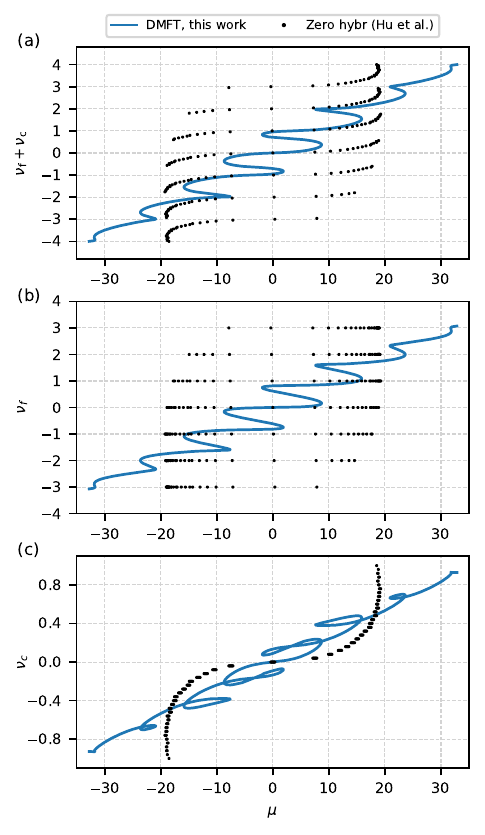}
    \caption{Filling with respect to CNP as a function of the chemical potential: (a) total filling, (b) $f$-electron filling, and (c) $c$-electron filling at T=11.6 K. The solid blue curves refer to DMFT simulations tuning the chemical potential to obtain a target (total) filling. The black dots show results for the zero hybridization model solved in \protect\cite{huKondoLatticeModel2023c}. }
    \label{fig:nvsmu}
\end{figure}
\begin{figure*}
    \centering
    \includegraphics[width=\textwidth]{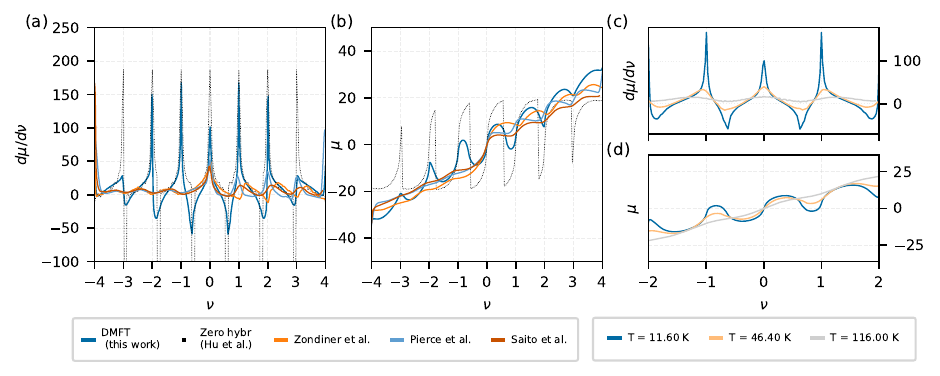}
    \caption{(a) Inverse compressibility obtained from the fixed-occupation DMFT simulations for $T=11.6 K$, compared with the experimental data from Zondiner et al. \protect\cite{zondinerCascadePhaseTransitions2020} ($B_{||}=0$, $T=4K$, and $\theta=1.13^\circ$), Pierce et al. \protect\cite{pierceUnconventionalSequenceCorrelated2021} ($B_{||}=0$, %$T=?K$ \LC{(not clear from the paper)} and 
    $\theta=1.06^\circ$), and Saito et al. \protect\cite{saitoIsospinPomeranchukEffect2021} ($B_{||}=0$, $T=12K$ and $\theta=1.06^\circ$), and the zero-hybridization model from \protect\cite{huKondoLatticeModel2023c}. (b) The chemical potential corresponding to the inverse compressibilities in (a). (c) Inverse compressibility and (d) chemical potential as a function of doping for three different temperatures.}
    \label{fig:expcomparison}
\end{figure*}

\noindent
\subsection{Compressibility}\label{sec:compressibility}

In Fig.~\ref{fig:nvsmu} we show the total, $f$- and $c$-fillings as a function of the \emph{intrinsic} chemical potential $\mu$. A discussion on the precise definition of the \emph{intrinsic} chemical potential and on how the mean-field interaction terms are operationally accounted for in our charge self-consistent DMFT calculation can be found in the Supplementary Material~\protect\cite{SM} (see also references~\protect\cite{Anisimov1991,Solovyev1994,Czyzik1994} therein); in short, this amounts to separating intrinsic from geometric contributions to the thermodynamic potential \protect\cite{koppCalculationCapacitancesConductors2009, liVeryLargeCapacitance2011} and to the chemical potential. What follows uses the \emph{intrinsic} chemical potential $\mu$ unless otherwise stated.

The blue solid lines in Fig.~\ref{fig:nvsmu} report the total and orbital-resolved fillings as a function of $\mu$. These are to be compared with the black dots, which report the same for the zero-hybridization model. 
In both cases, the overall behavior of the filling is a consequence of an energetic balance between the orbital species. The $f$-$c$ hybridization constitutes the additional feature captured by our DMFT study with respect to the zero-hybridization case described in~\protect\cite{huKondoLatticeModel2023c}, and causes a breakdown of the rigid-band picture.
The role of the $f$-$c$ hybridization is to provide a ``smoothing'' effect on the cross-talk between the two subspaces. Note that, differently from the standard periodic Anderson models considered in literature, the THF model~\protect\cite{songMagicAngleTwistedBilayer2022} features a specific momentum, and hence real-space, dependence of the hybridization term $H^{fc}(\mathbf{k})$~.

The capricious ups and downs of $\nu_c$ shown in Fig.~\ref{fig:nvsmu}(c) have to be contrasted to the progressive filling of the uncoupled reservoir of itinerant electrons in the zero-hybridization case: in both cases, when an additional flat $f$-electron band passes through the Fermi level, it provides a large, rapid increase in occupation, which is compensated by emptying the dispersive $c$-bands. This behavior is discontinuous in the decoupled case, with $\nu_{c}$ travelling multiple times from $\approx -0.85$ to $\approx +0.85$ along the same path marked by the black dots in Fig.~\ref{fig:nvsmu}(c). 
By contrast, the presence of $f$-$c$ hybridization and broadening in the DMFT solution forces both $\nu_{f}$ and $\nu_{c}$ to vary continuously. This results in the behaviors highlighted by the blue curves in Fig.~\ref{fig:nvsmu}(b) and (c) respectively: the $f$-electron occupation monotonically increases with respect the total $\nu$ (see also Fig.~\ref{fig:decoupled_comparison}(a)); when pictured upon varying $\mu$, it shows a series of continously connected plateaux. These occur trivially at integer values in the zero-hybridization model, while the DMFT solution with hybridization displays plateaux that are not pinned to integer $\nu_f$, except obviously for $\nu_f=0$ because of particle-hole symmetry.
Due to the previously mentioned occupation balancing mechanism, $\nu_{c}$ has instead to necessarily decrease in some intervals upon increasing the total $\nu$, resulting in the peculiar ``loops'' of Fig.~\ref{fig:nvsmu}(c).
A movie showing the evolution of the momentum-resolved spectral function across the filling range is included in the Supplementary Material~\protect\cite{SM}(\url{https://youtu.be/cw3K0YEsPU0}). It clearly shows the mutual transfer of low-energy spectral weight between $f$ and $c$. 

For comparison to experiments, we extract the electronic compressibility of the model and  plot its inverse in Fig.~\ref{fig:expcomparison}(a). One can see how the full resets of $\mu$ in the zero-hybridization case result in extremely pronounced negative spikes. In DMFT, these negative regions are much less prominent and the intensity varies between the charge neutrality point and the full/empty fillings. However, the minima of the DMFT data coincide with the negative spikes of the zero-hybridization model, and the positive spikes at integer fillings are consistent between the two methods. {While the maxima in the compressibilities found here and by the DMFT study of Datta et al.~\protect\cite{dattaHeavyQuasiparticlesCascades2023a} are in qualitative agreement, there are no negative compressibilities reported in Ref.~\protect\cite{dattaHeavyQuasiparticlesCascades2023a}. We assign this discrepancy to geometric capacitance terms, which may not have been fully subtracted in Ref.~\protect\cite{dattaHeavyQuasiparticlesCascades2023a}.}

The comparison of our DMFT results to experiments, shown in panels (a) and (b) of  Fig.~\ref{fig:expcomparison} reveals an overall reasonable agreement between DMFT and experiments from \protect\cite{zondinerCascadePhaseTransitions2020,pierceUnconventionalSequenceCorrelated2021, saitoIsospinPomeranchukEffect2021}. Yet, even if DMFT suppresses the negative compressibilities with respect to zero hybridization, these regions are still overestimating the experimentally observed ones (compare the blue to the green, red NS purple curves in Fig.~\ref{fig:expcomparison}). Among the different experimental curves, showing some discrepancies between one another, it is interesting to notice how the position of the DMFT positive peaks at integer fillings match rather well those of~\protect\cite{pierceUnconventionalSequenceCorrelated2021}, while the negative peaks and the overall behavior of the chemical potential are more similar to that of~\protect\cite{zondinerCascadePhaseTransitions2020}. Since the THF model is by design particle-hole symmetric, it doesn't account for the asymmetries experimentally measured for electron and hole doping. Our results are a better match for positive doping than negative doping. A calculation including strain and additional particle-hole symmetry breaking terms will be the subject of a future study.

In Fig.~\ref{fig:expcomparison}(c-d), we show the temperature-dependence of the inverse compressibility and the chemical potential. The peaks in the inverse compressibility (step features in the chemical potential) get progressively thermally broadened and are almost completely washed out at $\about 100$~K. This compares nicely with experimental reports on the temperature-dependence of the inverse compressibility \protect\cite{saitoIsospinPomeranchukEffect2021}, where the features also wash out at $\about 100$~K. This is the temperature scale at which thermally activated charge fluctuations become appreciable and the Hartree-Fock solution is no longer isospin-polarized. In the supplementary material~\protect\cite{SM}, we confirm by calculating the temperature-dependent sector statistics (the weight of the different charge sectors in the many body configurations) that this is also the temperature scale where there is no longer one predominant charge sector. In other words, the experimentally observed structures of the inverse  compressibility and its temperature dependence pinpoint to the thermal activation of charge fluctuations and the dissolution of local moment physics around $\sim 100$~K.

\subsection{Significance and origin of the negative compressibilities.} All chemical potential terms discussed so far referred to the intrinsic chemical potential, while only the total thermodynamic potential determines instabilities of the system. The total thermodynamic potential has to include also the geometric contributions, which are present in both, the theoretical model ~\protect\cite{songMagicAngleTwistedBilayer2022} and all experimental realizations of TBLG~\protect\cite{zondinerCascadePhaseTransitions2020,saitoIsospinPomeranchukEffect2021,rozenEntropicEvidencePomeranchuk2021,wongCascadeElectronicTransitions2020}: the charge required to dope TBLG away from charge neutrality is taken from gate electrodes. The geometric contribution to the total thermodynamic potential is the classical electrostatic energy that builds up upon charge transfer from the gate to TBLG. The intrinsic chemical potential does not involve the contribution from electrostatic potential associated with the charge transfer between gate and TBLG and thus, in turn, a negative compressibility referring only to a derivative involving the \emph{intrinsic} chemical potential $dn/d\mu$ does neither imply necessarily an instability of the system nor a tendency to phase separate.

In the THF model of Ref.~\protect\cite{songMagicAngleTwistedBilayer2022}, the TBLG system is supposed to sit in the middle of two gate electrodes, providing an electric potential dependent on the gate separation. As detailed in the Supplementary Material~\protect\cite{SM}, this entails a large geometric capacitance term, which has to be taken into account when determining the \emph{total} TBLG compressibility. In the Supplementary Material~\protect\cite{SM}, we show the same data of Figs.~\ref{fig:nvsmu} and ~\ref{fig:expcomparison}, plotted without the subtraction of the geometrical capacitance contribution, i.e. referring to the \emph{full} TBLG system, which necessarily includes the gates where the doping charges are taken from. There, most of the negative compressibility regions are gone, though a small region survives close to $\nu \pm 0.6$. As detailed in the Appendix, this effect is related to the form of the double-gate  screened Coulomb integrals considered in the original THF model, which sets an inter-gate distance $\xi=10$nm~\protect\cite{songMagicAngleTwistedBilayer2022}. The geometric contribution depends linearly on the distance between the capacitor plates, which suggests that an interaction term corresponding to a larger separation would have removed even the remaining $\nu \pm 0.6$ negative compressibility region. Similar geometric effects are also ubiquitously present in experimental realizations~\protect\cite{zondinerCascadePhaseTransitions2020,saitoIsospinPomeranchukEffect2021,rozenEntropicEvidencePomeranchuk2021,wongCascadeElectronicTransitions2020}, where TBLG is typically placed in single or multigate structures in order to achieve gate controlled charge doping. In the experimentally realized setups, assuming the electrodes/back-gate to be ideal metals, which do not feature quantum capacitance effects, the only contribution to the negative compressibility comes from exchange/correlation contributions on the TBLG layer~\protect\cite{Kopp2009}, which are accounted for in DMFT locally to all orders.

% \LC{It is further to note that the negative values and divergences in compressibility are subject to the thermodynamics of the system, and tend to wash out at higher temperatures together with the local moments. As an illustration of this, we show in Fig.~\ref{fig:expcomparison}(c-d) the behavior of $d\mu/d\nu$ and $\mu$ at various values of temperature [MISSING DATA...]: the negative compressibility regions, divergences and the localization of maxima and minima are completely overhauled already at $T \sim 100$ K. More details on the effect of temperature on local moments can be found in the Supplementary Material~\protect\cite{SM}.}

\section{Conclusions}
Our study provides a unified understanding of dynamic correlations and spontaneous symmetry-breaking in TBLG, allowing us to reconcile a set of complementary experiments spanning a wide range of temperatures. Strong electronic interactions give rise to the emergence of local isospin moments at temperatures on the order of $\about 100$~K, which order around $\about 10$~K. This result suggests that ordering precedes full Kondo screening which manifests at lower temperatures \protect\cite{huSymmetricKondoLattice2023a}. Furthermore, our comparison between DMFT and Hartree-Fock ordering temperatures reveals a noteworthy difference of one order of magnitude, indicating the significant influence of local quantum fluctuations on the temperature-dependent phase diagram of TBLG. Note that the true ordering temperatures will be lower than what is predicted by single-site DMFT, as it neglects spatial fluctuations. 

Once the local moments have formed, regardless of whether they order or not, exchange-correlation effects in the localized $f$-states lead to ``integer-periodic" variations in the compressibility ranging from nearly incompressible to negative values as found in capacitance experiments \protect\cite{saitoIsospinPomeranchukEffect2021, zondinerCascadePhaseTransitions2020, pierceUnconventionalSequenceCorrelated2021}. Concomitantly, there is a periodic
redistribution of charge between the $f$- and the $c$-states upon doping. This charge reshuffling is responsible for the doping-induced cascade transitions seen in scanning tunneling spectroscopy \protect\cite{wongCascadeElectronicTransitions2020} and explains why the cascades first appear at a temperature of $\about 100$~K~\protect\cite{saitoIsospinPomeranchukEffect2021}.

Our study focuses on finite temperatures $T\gtrsim 5$~K and highlights the special role of total integer fillings and the appearence of (nearly) incompressible insulating states there. While there is a depletion of low-energy spectral weight already in the symmetric phase, exchange interactions generically support a hard gap at integer fillings in the ordered phase in the absence of strain.

Regarding the nature of the metallic states in TBLG,  we find disordered local moments which cause scattering and predominantly incoherent low-energy electronic spectral weight in the temperature range of approximately $10$~K$<T<100$~K. Ordering of the isospin moments leads to coherence and the appearance of well-defined quasiparticles for $T\lesssim 10$~K. This order-induced coherence stands behind the isospin Pomeranchuk effect observed in transport experiments. Above the ordering temperature but below the temperature of moment formation, the isospin entropy of the local moments stabilizes an incoherent ``bad metal" phase, which manifests in incoherent spectral weight at the Fermi level. Below the ordering temperature, the Fermi surface is composed of delocalized coherent quasiparticle excitations implying Fermi liquid-like behavior. This change from incoherent to coherent spectral weight at the Fermi level can explain the generic resistivity drop seen in many experiments below $\sim 10$~K~\protect\cite{luSuperconductorsOrbitalMagnets2019,caoStrangeMetalMagicAngle2020,rozenEntropicEvidencePomeranchuk2021,saitoIsospinPomeranchukEffect2021,parkFlavourHundCoupling2021}. In contrast to the original Pomeranchuk effect in He-3, the suppression of entropy and coherence in TBLG at low temperatures come from the ordering of local moments and not from 
their disappearance.  
%fate of local moments. 
The coherence from order physics demonstrated here for TLBG reveals similarities between TBLG and materials such as metallic ferromagnets like SrRuO$_3$~\protect\cite{Allen_SrRuO3_1996} and compounds related to Fe-based superconductors~\protect\cite{Edelmann_PRB2017}.

We note that the single-site DMFT neglects non-local fluctuations and any dynamical renormalization effects on $\hat{H}_J$. Both of these approximations will lead to an overestimate of the ordering temperature, so the true ordering temperature is likely to be a bit lower. The renormalization of $\hat{H}_J$ is expected to be weak except possibly far away from integer fillings, which is also where the filling-dependent Kondo temperature is expected to find its local maxima (see \protect\cite{zhouKondoPhaseTwisted2024} and Supplementary Material~\protect\cite{SM}). These effects, however, are likely secondary to strain and particle-hole symmetry breaking lattice effects \protect\cite{herzog-arbeitmanHeavyFermionsEfficient2024}, which have been shown to have a huge effect on the low-temperature ordered states \protect\cite{kwanKekuleSpiralOrder2021, wagnerGlobalPhaseDiagram2022, blasonLocalKekuleDistortion2022, angeliValleyJahnTellerEffect2019}. The inclusion of these effects is a natural extension of our work.

Our results show that ordering affects electronic spectra and even Fermi surface topologies in metallic states. Thus, we expect that further symmetry-breaking, i.e. particularly superconductivity at lower temperatures $T\about 1$~K, will be impacted by this order-facilitated coherence.

\begin{acknowledgments}
We thank Dimitri Efetov, Piers Coleman, Andrew Millis, Alexei M. Tsvelik, Sankar Das Sarma, Xi Dai, Andrey Chubukov, Francisco Guinea, 
Maria J. Calderón, Leni Bascones, Ali Yazdani, Shahal Ilani, Pablo Jarillo-Herrero,  Silke Paschen, Elio K\"onig, and Georg Rohringer for useful discussions.
G.R. and T.W. gratefully acknowledge funding and support from the European Commission via the Graphene
Flagship Core Project 3 (grant agreement ID: 881603), and the Deutsche Forschungsgemeinschaft (DFG, German Research Foundation) through the cluster of excellence ``CUI: Advanced Imaging of Matter" of the Deutsche Forschungsgemeinschaft  (DFG EXC 2056, Project ID 390715994), and the DFG priority program SPP2244 (Project ID 443274199). 
G.R, L.C, T.W., G.S. and R.V. thank the DFG for funding through the research unit QUAST FOR
5249 (project ID: 449872909; projects P4 and P5).
G.R. gratefully acknowledges the computing time granted by the Resource Allocation Board and provided on the supercomputer Lise and Emmy at NHR@ZIB and NHR@Göttingen as part of the NHR infrastructure (project ID hhp00061).
L.C and G.S. were supported by the W\"urzburg-Dresden Cluster of Excellence on Complexity and Topology in Quantum Matter –ct.qmat Project-ID 390858490-EXC 2147, and gratefully acknowledge the Gauss Centre for Supercomputing e.V. (www.gauss-centre.eu) for funding this project by providing computing time on the GCS Supercomputer SuperMUC at Leibniz Supercomputing Centre (www.lrz.de).
L.C. gratefully acknowledges the scientific support and HPC resources provided by the Erlangen National High Performance Computing Center (NHR@FAU) of the Friedrich-Alexander-Universität Erlangen-Nürnberg (FAU) under the NHR project b158cb. NHR funding is provided by federal and Bavarian state authorities. NHR@FAU hardware is partially funded by the German Research Foundation (DFG) – 440719683.
R.V. research was supported in part by the National Science Foundation under Grants No. NSF PHY-1748958 and PHY-2309135. 
L.d.M. is supported by the European Commission through the ERC-CoG2016, StrongCoPhy4Energy, GA No724177.
B. A. B.'s work was primarily supported by the DOE Grant No. DE-SC0016239, the Simons Investigator Grant No. 404513. 
H. H. was supported by the European Research Council (ERC) under the European Union’s
Horizon 2020 research and innovation program (Grant Agreement No. 101020833). 
D. C. was supported by the DOE Grant No. DE-SC0016239 by  the Gordon and Betty Moore Foundation through Grant No. GBMF8685 towards the Princeton theory program, the Gordon and Betty Moore Foundation’s EPiQS Initiative (Grant No. GBMF11070), Office of Naval Research (ONR Grant No. N00014-20-1-2303), BSF Israel US foundation No. 2018226 and NSF-MERSEC DMR. 
The Flatiron Institute is a division of the Simons Foundation. 
\end{acknowledgments}

\bibliography{refs_for_arxiv}
\clearpage

 % \appendix
\onecolumngrid
\begin{center}
    \Large\textbf{Supplementary Material for ``Dynamical correlations and order in magic-angle twisted bilayer graphene"}
\end{center}
 \setcounter{page}{1}
% \counterwithin{figure}{section}
% \include{SI-for-arxiv-Aug8}

%%%%%%%%%%%%%%%%%%%%

%%%%%%%%%%%%%%%%%%%%%
%SMSMSMSMSMSMS
%%%%%%%%%%%%%%%%%%%%%%

\setcounter{section}{0}
\setcounter{figure}{0}
\renewcommand{\thefigure}{S\arabic{figure}}
\renewcommand{\thesection}{S\arabic{section}}
% \renewcommand{\thesubsection}{\alph{subsection}}

%%%%%%%%%%%%%%%%%%
%%%%%%%%%%%%%%%%%%
\let\vec\mathbf

%%%%%%%%%%%%%%%%%%
%%%%%%%%%%%%%%%%%%

% \title{Supplemental material for\\Dynamical correlations and order in magic-angle twisted bilayer graphene}% Force line breaks with \\

% \setcounter{page}{1}
% \counterwithin{figure}{section}
\section{Non-interacting Hamiltonian}\label{app:H0}
The topological heavy Fermion model is derived and described in \protect\cite{songMagicAngleTwistedBilayer2022}. For completeness, we rewrite the single particle Hamiltonian:
\begin{align}
    \hat{H}_0=& \overbrace{\sum_{|\mathbf{k}|<\Lambda_c}\sum_{aa'\eta\sigma}H_{aa'}^{(c,\eta)}(\mathbf{k})c^\dagger_{\mathbf{k}a\eta\sigma}c_{\mathbf{k}a'\eta\sigma}}^{\hat{H}_c}\nonumber\\
    &+\underbrace{\frac{1}{\sqrt{N_M}}\sum_{|\mathbf{k}|<\Lambda_c, \mathbf{R}}\sum_{a\alpha\eta\sigma}\left[e^{i\mathbf{k}\cdot\mathbf{R} - \frac{|\mathbf{k}|^2\lambda^2}{2}}H_{a\alpha}^{(fc,\eta)}(\mathbf{k})f^\dagger_ {\mathbf{R}\alpha\eta\sigma}c_{\mathbf{k}a\eta\sigma} + h.c.\right]}_{\hat{H}_{cf}}.
    \label{app:eqn:sp_ham_thf}
\end{align}
The fist term is the dispersion of the $c$-sector, and $\Lambda_c$ is a momentum cutoff. The second term is the coupling of the dispersionless $f$-sector with the $c$-sector. $N_M$ is the number of moir\'e unit cells, $\mathbf{R}$ labels the moir\'e lattice vectors, $\lambda = 0.3375 a_M$ is a damping factor where $a_M$ is the moir\'e lattice constant. The spin and valley degrees of freedom are decoupled, and at each momentum vector $\mathbf{k}$, the Hamiltonian takes the form of $6\times 6$ matrix per spin and per valley. It can be represented compactly by $H^{(c,\eta)}(\mathbf{k})$ and $H^{(fc,\eta)}(\mathbf{k})$ as shown below
\begin{align}
    H^{(c,\eta)}(\mathbf{k}) = \begin{pmatrix}
        0_{2\times 2} & v_\star(\eta k_x \sigma_0 + i k_y \sigma_z)\\
        v_\star(\eta k_x \sigma_0 - i k_y \sigma_z) & M\sigma_x
    \end{pmatrix}, \label{app:eqn:H_c}
\end{align}
$\sigma_{0,x,y,z}$ are Pauli matrices in orbital space, $M = 3.697$~meV defines the bandwidth of the flat bands, $v_\star=-4.303$~eV~\AA{} is a fitting parameter. 
\begin{align}
H^{(fc,\eta)}(\mathbf{k}) = \begin{pmatrix}
    \gamma\sigma_0 + v'_\star(\eta k_x\sigma_x + k_y\sigma_y) & 0_{2\times 2}
\end{pmatrix}, \label{app:eqn:H_fc}
\end{align}
$\gamma = -24.75$~meV sets the gap between the flat band and the high energy bands, and $v'_\star=1.622$~eV \AA{} is a fitting parameter. 

\section{Interaction Hamiltonian}\label{app:H_int}
The interaction Hamiltonian of the THF model is derived in detail in the SI of \protect\cite{songMagicAngleTwistedBilayer2022}. We summarize their calculation here and bring their result to the form we use for our DMFT calculations. The interaction Hamiltonian is computed by performing the Coulomb integrals in the Bistritzer-Macdonald basis
\begin{align}
    \hat{H}_I =\frac{1}{2}\int d^2 \mathbf{r}_1 d^2 \mathbf{r}_2 V(\mathbf{r}_1 - \mathbf{r}_2) : \hat{\rho}(\mathbf{r}_1)::\hat{\rho}(\mathbf{r}_2):,
\end{align}
where $V(\mathbf{r}_1 - \mathbf{r}_2)$ is the double-gate-screened Coulomb interaction, and $:\hat{\rho}:$ is the normal-ordered density operator. Of all the terms generated, the most important ones are identified to be the density-density interaction in the $f$($c$)-subspace, $\hat{H}_U$($\hat{H}_V$), the density-density interaction mixing $c$- and $f$-subspaces, $\hat{H}_W$, and the exchange interaction between the $f$- and $c$-subspaces $\hat{H}_J$
\begin{align}
    \hat{H}_I \approx \hat{H}_U + \hat{H}_W + \hat{H}_V + \hat{H}_J.
\end{align}
We treat $\hat{H}_U$ with DMFT, $\hat{H}_W$ and $\hat{H}_J$ with Hartree+Fock, and $\hat{H}_V$ with Hartree. Importantly, we treat every term, including the ones in the delocalized $c$-subspace, on at least a Hartree level. This means that we capture any relative static energy shifts between the different orbitals upon doping. We calculate the Hartree+Fock mean-fields self-consistently making this essentially a charge self-consistent DMFT calculation. 

In the following, it is convenient to define the normal-ordered operator with the colon symbol
\begin{align}
    :f^\dagger_{\mathbf{R}\alpha_1\eta_1\sigma_1} f_{\mathbf{R}\alpha_2\eta_2\sigma_2}: &= f^\dagger_{\mathbf{R}\alpha_1\eta_1\sigma_1} f_{\mathbf{R}\alpha_2\eta_2\sigma_2} - \frac{1}{2}\delta_{\alpha_1\eta_1\sigma_1;\alpha_2\eta_2\sigma_2},\\
    :c^\dagger_{\mathbf{k_1}a_1\eta_1\sigma_1} c_{\mathbf{k_2}a_2\eta_2\sigma_2}: &= c^\dagger_{\mathbf{k_1}a_1\eta_1\sigma_1} c_{\mathbf{k_2}a_2\eta_2\sigma_2} - \frac{1}{2}\delta_{\mathbf{k_1}\alpha_1\eta_1\sigma_1;\mathbf{k_2}\alpha_2\eta_2\sigma_2},
\end{align}
and the following density matrices
\begin{align}
    O^f_{\alpha_1\eta_1\sigma_1;\alpha_2\eta_2\sigma_2} &= \frac{1}{N_M} \sum_\mathbf{R} \bra{\Psi}:f^\dagger_{\mathbf{R}\alpha_1\eta_1\sigma_1} f_{\mathbf{R}\alpha_2\eta_2\sigma_2}:\ket{\Psi},\\
    O^{c''}_{\alpha_1\eta_1\sigma_1;\alpha_2\eta_2\sigma_2} &= \frac{1}{N_M} \sum_{|\mathbf{k}|<|\Lambda_c} \bra{\Psi}:c^\dagger_{\mathbf{k_1}a_1\eta_1\sigma_1} c_{\mathbf{k_2}a_2\eta_2\sigma_2}:\ket{\Psi},\,\,\,\,a_1,a_2\in[3,4]\\
    O^{cf}_{a\eta_1\sigma_1;\alpha_2\eta_2\sigma_2} &= \frac{1}{\sqrt{N_M}} \sum_{|\mathbf{k}|<|\Lambda_c} e^{-i\mathbf{k}\cdot\mathbf{R}}\bra{\Psi}c^\dagger_{\mathbf{k}a\eta_1\sigma_1} f_{\mathbf{R}\alpha\eta_2\sigma_2}\ket{\Psi}.
\end{align}
$O_f$ is the density matrix in the $f$-sector, $O^{c''}$ is the density matrix in the $\Gamma_1\oplus\Gamma_2$ $c$-sector, and $O^{cf}$ is the $c$-$f$ off-diagonal block.
\subsection{$\hat{H}_{U}$: $f$-$f$ density-density term (on-site)}
\begin{align}
    \hat{H}_{U} =&  \frac{U}{2}\sum_\mathbf{R}\sum_{(\alpha\eta\sigma)\neq (\alpha'\eta' \sigma')}
    \left(f^\dagger_{\mathbf{
    R}\alpha\eta\sigma} f_{\mathbf{
    R}\alpha\eta\sigma} -\frac{1}{2}\right)
    \left(f^\dagger_{\mathbf{R}\alpha'\eta'\sigma'}f_{\mathbf{
    R}\alpha'\eta'\sigma'}-\frac{1}{2}\right),
\end{align}
with $U=57.95$~meV. We treat this term dynamically. In order to perform our charge self-consistent DMFT simulations, it is useful to separate out the bilinear term that enforces charge neutrality. Collecting the bilinear terms in $\hat{H}_U$ and collapsing the sum we obtain
\begin{align}
    \hat{H}_U = \frac{U}{2}\sum_{\mathbf{R}}\sum_{(\alpha\eta\sigma)\neq (\alpha'\eta' \sigma')}f^\dagger_{\mathbf{R}\alpha\eta\sigma}f_{\mathbf{R}\alpha\eta\sigma}f^\dagger_{\mathbf{R}\alpha'\eta'\sigma'}f_{\mathbf{R}\alpha'\eta'\sigma'}
    - 3.5 U \sum_{\mathbf{R}}\sum_{\alpha\eta\sigma} f^\dagger_{\mathbf{R}\alpha\eta\sigma}f_{\mathbf{R}\alpha\eta\sigma} + \mathcal{O}(1).
\end{align}
We call the first term $\hat{H}_{dyn}$, and it enters the DMFT calculation as the interaction Hamiltonian of the impurity problem. The second term forms part of the static Hamiltonian $\hat{H}_{stat}$ along with the remaining three interaction terms decoupled via static mean fields, and the non-interacting terms. 
\subsection{$\hat{H}_V$: $c$-$c$ density-density term}
We treat the density-density term within the $c$-subspace on the Hartree level. Since the primary effect of $\hat{H}_V$ is to shift the relative energies of the $c$- and $f$-subspaces, and all the symmetry-breaking happens in the $f$-subspace, we neglect the Fock contribution (in accordance with \protect\cite{songMagicAngleTwistedBilayer2022}). The Hartree contribution is a static shift in the $c$-subspace depending only on the $c$-electron filling $\nu_c$
\begin{align}
    \hat{H}_V^{MF} = V \nu_c \sum_{|\mathbf{k}|<\Lambda,a\eta\sigma} c^\dagger_{\mathbf{k}a\eta\sigma}c_{\mathbf{k}a\eta\sigma} + \mathcal{O}(1),
\end{align}
where $V = 48.33$~meV and $\nu_c$ is the filling of the $c$-electrons with respect to charge neutrality. Refer to SI of \protect\cite{songMagicAngleTwistedBilayer2022} for derivation.

\subsection{$\hat{H}_W$: $c$-$f$ density-density term}
The Hartree term involves only the orbital-resolved occupations, contributing a static shift in the $f$-($c$-)sector depending on the occupation of the $c$-($f$-)sector:
\begin{align}
    \sum_{|\mathbf{k}|<\Lambda_c}\sum_{a\eta\sigma}W_a\nu_f:c^\dagger_{\mathbf{k}a\eta\sigma}c_{\mathbf{k}a\eta\sigma}: +     
\sum_{\mathbf{R}}\sum_{a\alpha\eta\sigma}W_a\nu_{c,a}:f^\dagger_{\mathbf{R}\alpha\eta\sigma}f_{\mathbf{R}\alpha\eta\sigma}: + \mathcal{O}(1).
\end{align}
The Fock term involves the $c$-$f$ density matrix:
\begin{align}
    -\frac{1}{\sqrt{N_M}}\sum_{\mathbf{R}}\sum_\mathbf{|\mathbf{k}|<\Lambda_c} \sum_{\alpha\alpha'\eta\eta'\sigma\sigma'} W_a(O^{cf}_{a\eta'\sigma',\alpha\eta\sigma}e^{i\mathbf{k}\cdot\mathbf{R}}f^\dagger_{\mathbf{R}\alpha\eta\sigma}c_{\mathbf{k}a\eta'\sigma'} + h.c.) + \mathcal{O}(1).
\end{align}
Thee coupling to the $\Gamma_3$ $c$-sector is $W_1=W_2=44.03$~meV, and the coupling to the $\Gamma_1\oplus\Gamma_2$ sector is $W_3=W_4=50.20$~meV. See SI of \protect\cite{huSymmetricKondoLattice2023a} for derivation.

\subsection{$\hat{H}_J$: $c$-$f$ exchange term}
$\hat{H}_J$ is a ferromagnetic exchange coupling with $J=16.38$~meV. The Hartree term takes the form
\begin{align}
    & \frac{J}{2} \sum_\mathbf{R} \sum_{\alpha\alpha'\eta\eta'\sigma\sigma'}:f^\dagger_{\mathbf{R},\alpha\eta\sigma}f_{\mathbf{R},\alpha'\eta'\sigma'}:O^{c''}_{\alpha'\eta'\sigma',\alpha\eta\sigma} \nonumber\\&+     \frac{J}{2} \sum_\mathbf{|\mathbf{k}|<\Lambda_c} \sum_{\alpha\alpha'\eta\eta'\sigma\sigma'}:c^\dagger_{\mathbf{k},\alpha\eta\sigma}c_{\mathbf{k'},\alpha'\eta'\sigma'}:O^{f}_{\alpha'\eta'\sigma',\alpha\eta\sigma} + \mathcal{O}(1)
\end{align}
For the Fock term, it is useful to define the mean fields $V_3$, $V_4$ (following the nomenclature in \protect\cite{huSymmetricKondoLattice2023a})
\begin{align}
    V_3 &= \sum_{\mathbf{R}}\sum_{|\mathbf{k}|<\Lambda_c}\sum_{\alpha\eta\sigma}\frac{e^{i\mathbf{k}\cdot\mathbf{R}}\delta_{1,\eta(-1)^{\alpha+1}}}{N_M\sqrt{N_M}} \bra{\psi}f^\dagger_{\mathbf{R}\alpha\eta\sigma}c_{\mathbf{k}\alpha+2\eta\sigma}\ket{\psi}\\
    V_4 &= \sum_{\mathbf{R}}\sum_{|\mathbf{k}|<\Lambda_c}\sum_{\alpha\eta\sigma}\frac{e^{i\mathbf{k}\cdot\mathbf{R}}\delta_{-1,\eta(-1)^{\alpha+1}}}{N_M\sqrt{N_M}} \bra{\psi}f^\dagger_{\mathbf{R}\alpha\eta\sigma}c_{\mathbf{k}\alpha+2\eta\sigma}\ket{\psi}
\end{align}
The Fock term is
\begin{align}
    J \sum_{\mathbf{R}}\sum_{|\mathbf{k}|<\Lambda_c}\sum_{\alpha\eta\sigma}\left\{\frac{e^{-i\mathbf{k}\cdot\mathbf{R}}}{\sqrt{N_M}}\left[\delta_{1,\eta(-1)^{\alpha+1}}f^\dagger_{\mathbf{R}\alpha\eta\sigma}c_{\mathbf{k}\alpha+2\eta\sigma}V_3^*
+\delta_{-1,\eta(-1)^{\alpha+1}}f^\dagger_{\mathbf{R}\alpha\eta\sigma}c_{\mathbf{k}\alpha+2\eta\sigma}V_4^*
    \right] + h.c.\right\} + \mathcal{O}(1)
\end{align}
\subsection{Summary}
All together, we arrive at the Hamiltonian $\hat{H}_{stat} + \hat{H}_{dyn}$ for the ``charge self-consistent" DMFT calculation with
\begin{align}
    \hat{H}_{dyn} =&\frac{U}{2}\sum_{\mathbf{R}}\sum_{(\alpha\eta\sigma)\neq (\alpha'\eta' \sigma')}f^\dagger_{\mathbf{R}\alpha\eta\sigma}f_{\mathbf{R}\alpha\eta\sigma}f^\dagger_{\mathbf{R}\alpha'\eta'\sigma'}f_{\mathbf{R}\alpha'\eta'\sigma'}\\
    \hat{H}_{stat} =& \hat{H}_0  - 3.5 U \sum_{\mathbf{R}}\sum_{\alpha\eta\sigma} f^\dagger_{\mathbf{R}\alpha\eta\sigma}f_{\mathbf{R}\alpha\eta\sigma} + V \nu_c \sum_{|\mathbf{k}|<\Lambda,a\eta\sigma} c^\dagger_{\mathbf{k}a\eta\sigma}c_{\mathbf{k}a\eta\sigma} \nonumber\\
    &+   \sum_{|\mathbf{k}|<\Lambda_c}\sum_{a\eta\sigma}W_a\nu_f:c^\dagger_{\mathbf{k}a\eta\sigma}c_{\mathbf{k}a\eta\sigma}: +     
\sum_{\mathbf{R}}\sum_{a\alpha\eta\sigma}W_a\nu_{c,a}:f^\dagger_{\mathbf{R}\alpha\eta\sigma}f_{\mathbf{R}\alpha\eta\sigma}: \nonumber\\
&+   -\frac{1}{\sqrt{N_M}}\sum_{\mathbf{R}}\sum_\mathbf{|\mathbf{k}|<\Lambda_c} \sum_{\alpha\alpha'\eta\eta'\sigma\sigma'} W_a(O^{cf}_{a\eta'\sigma',\alpha\eta\sigma}e^{i\mathbf{k}\cdot\mathbf{R}}f^\dagger_{\mathbf{R}\alpha\eta\sigma}c_{\mathbf{k}a\eta'\sigma'} + h.c.) \nonumber\\
&+ \frac{J}{2} \sum_\mathbf{R} \sum_{\alpha\alpha'\eta\eta'\sigma\sigma'}(\eta\eta' + (-1)^{\alpha+\alpha'}):f^\dagger_{\mathbf{R},\alpha\eta\sigma}f_{\mathbf{R},\alpha'\eta'\sigma'}:O^{c''}_{\alpha'\eta'\sigma',\alpha\eta\sigma} \nonumber\\
&+     \frac{J}{2} \sum_\mathbf{|\mathbf{k}|<\Lambda_c} \sum_{\alpha\alpha'\eta\eta'\sigma\sigma'}(\eta\eta' + (-1)^{\alpha+\alpha'}):c^\dagger_{\mathbf{k},\alpha\eta\sigma}c_{\mathbf{k'},\alpha'\eta'\sigma'}:O^{f}_{\alpha'\eta'\sigma',\alpha\eta\sigma}\nonumber\\
  &+  J \sum_{\mathbf{R}}\sum_{|\mathbf{k}|<\Lambda_c}\sum_{\alpha\eta\sigma}\left\{\frac{e^{-i\mathbf{k}\cdot\mathbf{R}}}{\sqrt{N_M}}\left[\delta_{1,\eta(-1)^{\alpha+1}}f^\dagger_{\mathbf{R}\alpha\eta\sigma}c_{\mathbf{k}\alpha+2\eta\sigma}V_3^*
+\delta_{-1,\eta(-1)^{\alpha+1}}f^\dagger_{\mathbf{R}\alpha\eta\sigma}c_{\mathbf{k}\alpha+2\eta\sigma}V_4^*
    \right] + h.c.\right\}\nonumber\\
   &+ \mathcal{O}(1).
\end{align}
In addition to the self-energy $\Sigma$, we have a second self-consistently computed quantity---the full $f\oplus c$ density matrix of the system $\rho$. They are both needed to fix $\hat{H}_{stat}$.

\section{DMFT Implementation}\label{app:implementation}
In practice, we have two iteration-dependent quantities: the density matrix of the system $\rho^i$ and the $f$-projected self-energy $\Sigma^i$, where the label $i$ refers to the iteration number. The dynamic self-energy is zero in the $c$-subspace, so we can always promote $\Sigma^i$ to the full $c\oplus f$ space by padding with zeros. The static part of the $c$-sector self-energy is already included in the mean-field terms in $\hat{H}_{stat}$ and we assume that there is no frequency-dependent component. In each iteration, $\rho^i$ fixes $\hat{H}_{stat}$, which together with $\Sigma^i$ gives us the local lattice Green function. We project the local lattice Green function to the $f$-subspace and compute the Weiss field, which we then input to the CT-QMC impurity solver in addition to $\hat{H}_{dyn}$. The impurity solver gives us a new self-energy $\Sigma^{i+1}$ and a new interacting impurity Green function from which we compute the new density matrix $\rho^{i+1}$. This step is repeated until the density matrix and the self-energy converge. We choose a static self-energy for the initial guess. The exact value is inconsequential as long as it is reasonable, i.e. as long as the $f$-bands lie close to the Fermi level between the upper and lower $c$-bands. We use the initial choice of the density matrix to specify a particular symmetry-broken phase or the symmetric phase.

\begin{figure}
    \centering
    \includegraphics{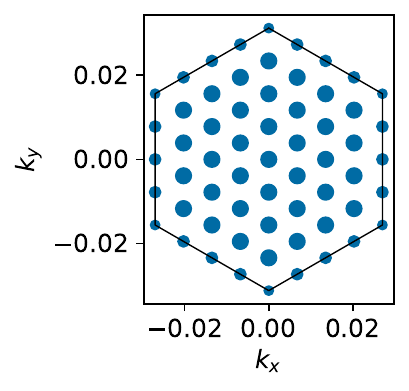}
    \caption{Regular centered sampling of the first moir\'e Brillouin zone. Edges and corners are weighted $1/2$ and $1/3$ respectively. In this example 
    \texttt{sample\_len=4}.}
    \label{fig:BZ_sampling}
\end{figure}

In DMFT, we consider the local Hamiltonian. We consider a single moir\'e lattice site labelled by $\mathbf{R}=(0,0)$, so that $N_M=1$. We sum momenta over the first moir\'e Brillouin zone (FMBZ), meaning that all references to $|\mathbf{k}|<\Lambda_c$ are rather $\mathbf{k}\in FMBZ$. We sample the FMBZ with a regular centered grid. The resolution is parametrized by a number \texttt{sample\_len}, which counts the number of steps between $\Gamma$ and $K$. Points on the edges and corners are degenerate, and therefore weighted $1/2$ and $1/3$ respectively. See Fig.~\ref{fig:BZ_sampling} for an example grid.  

Note that for calculations in the symmetric phase, we simplified the interaction Hamiltonian further by neglecting the Fock part of $\hat{H}_W$ and setting the $W$-coupling with both $c$-sectors to be equivalent $W_a \to \frac{W_1+W_3}{2} = 47.12$~meV. The $\hat{H}_J$ contribution to the symmetric state is zero. This allows the static self-consistency loop to be absorbed into a double counting term as described in the SI of \protect\cite{huSymmetricKondoLattice2023a} which is relatively seamless to incorporate in DMFT codes. We have checked for a range of parameter values that the differences with the full calculation are solely quantitative and small.

As an extra check to ensure the accuracy of our calculations, we implemented our model twice with two different DMFT codes independently, and compared test cases. GR worked with the TRIQS suite of packages, and LC worked with w2dynamics.

\subsection{TRIQS runs}
\subsubsection{DMFT parameters}
For summation within the DMFT runs, we used a grid with \texttt{sample\_len=9}, which has 271 points. We considered our system converged when the maximum element of the difference of the current and previous density matrix reached order $10^{-3}$. This took between 50-200 iterations. At most fillings and temperatures, we used a mixing factor of 0.9, i.e. the new self energy and density matrix were nine parts the output of the impurity solver and one part the previous self-energy or density matrix. In regions where convergence was difficult (where the Fermi level hits the flat band, and negative compressibility is seen), we used lower mixing factors. For most fillings and temperatures, we updated the self-energy and density matrix every iteration. In regions of difficult convergence, we updated the density matrix only every 2, 3, or 5 iterations.

\subsubsection{QMC parameters}
In the convergence runs, we used the QMC solver with $1.2\times 10^7$ total cycles (measurements) distributed on 192 cores, with 5000 warmup cycles. Each cycle consisted of 1000 QMC moves. We measured in the imaginary time basis, and performed a tail fit with $4$ moments. The tail fitting window, defined by the lower (\texttt{fit\_min\_n})  and upper Matsubara frequency index (\texttt{fit\_max\_n}), depends strongly on temperature and we chose it heuristically to be \texttt{fit\_min\_n}~=~$\lfloor 160T^{-0.9577746167096538}) \rfloor$ and \texttt{fit\_max\_n}~=~$\lfloor209 T^{-0.767031728744396}\rfloor$

\subsubsection{Analytic continuation}\label{app:continuation}
We use the maximum entropy method \protect\cite{gubernatisQuantumMonteCarlo1991} as implemented in triqs\_maxent package \protect\cite{krabergerMaxEnt2023} to obtain the spectral functions on the real axis. We first analytically continue the imaginary-frequency self-energy obtained from the CT-QMC solver to a real-frequency self-energy. Then, we use the real-frequency self-energy to compute the Green functions, spectral functions, etc on the real axis. We always perform the analytic continuation step in the natural basis defined by the chosen symmetry-broken phase (see Sec.~\ref{app:breaking_symmetries}). In this basis, the self-energy is diagonal, which allows us to avoid the complications inherent in analytically continuing matrix quantities.

The self-energy continuation method uses an intermediate auxiliary object $G_{aux}(\omega)$. We use a hyperbolic mesh for the auxiliary spectral function $A_{aux}(\omega)$, such that we have an energy resolution of about $0.01$, $0.2$, and $1$~meV at $\omega = 0, 1$ and $20$~meV respectively.

\subsection{w2dynamics runs}
The $k$-space Hamiltonian grid for the \textit{w2dynamics} runs was independently generated following the same weighting philosophy described above. To account for the larger weight of the bulk points with respect to the edges and corners of the BZ, an oversampling strategy was adopted considering duplicated $k$-points in the number of $6$ for the bulk, $3$ for the edges and $2$ for the corners, for a total of 2071 $k$-points.
\subsubsection{DMFT parameters}
As a criterion for convergence of the DMFT simulations we considered the step-by-step variation of local observables such as orbital/spin/valley - resolved occupations with a threshold of $10^{-4}$. Since the compressibility runs were obtained at high temperature ($\approx 11 K$), i.e. in the disordered phase, convergence was rather quick and never required more than 50 DMFT loops. In order to properly follow the compressibility curve in the negative slope branches, we run every simulation at increasing values of $\nu$ starting from the converged self-energy of the previous step. For the same reason, the mixing factor was set at $0.5$.  The spacing between doping values was in general $0.025$, but it was decreased in more challenging intervals such as the negative slope branches of the compressibility curves and around the divergence points of the inverse compressibility. The total number of points for the fixed-$\nu$ simulations is 325, and  for the fixed-$\mu$ simulations it is 218.

\subsubsection{QMC parameters}

For the QMC runs with $T = 11.6 K$ we used 144 CPU cores. At each DMFT loop a sufficient number of measurements was found to be ${\texttt{Nmeas}} = 7.5\cdot10^{4}$ per core, with a number of steps ${\texttt{Ncorr}}=2000$ per core between successive measurements to avoid autocorrelation effects. ${\texttt{Nwarmup}}=2\cdot 10^{6}$ steps per core were run before each measurement begun.
We measured in imaginary time domain and used the Legendre polynomial basis, with maximum order $\texttt{NLegMax}=35$.

\subsubsection{Analytic continuation}\label{app:continuation_w2dyn}
Analytical continuation was performed starting from the last DMFT run using the MaxEnt method from the \textit{ana\_cont} software suite~\protect\cite{kaufmann2021anacont}. For the local spectral function $A(\omega)$ plots, we performed analytic continuation of the Matsubara-summed local Green's function, extracted from the last DMFT loop of each simulation. For the momentum-resolved spectral function plots, we analytically continue the self-energy, and use the resulting real-frequency $\Sigma(\omega)$ in the calculation of $A(\mathbf{k},\omega)$. The self-energy has no off-diagonal components in the symmetric phase. We used a regular mesh of $1001$ real-axis frequency points in the interval $[-250:+250]$ meV for $A(\omega)$ and $[-300:+300]$ meV for $\Sigma(\omega)$ and no preblur.

\subsection{TRIQS vs. w2dynamics comparison}
\begin{figure}
    \centering
    \includegraphics{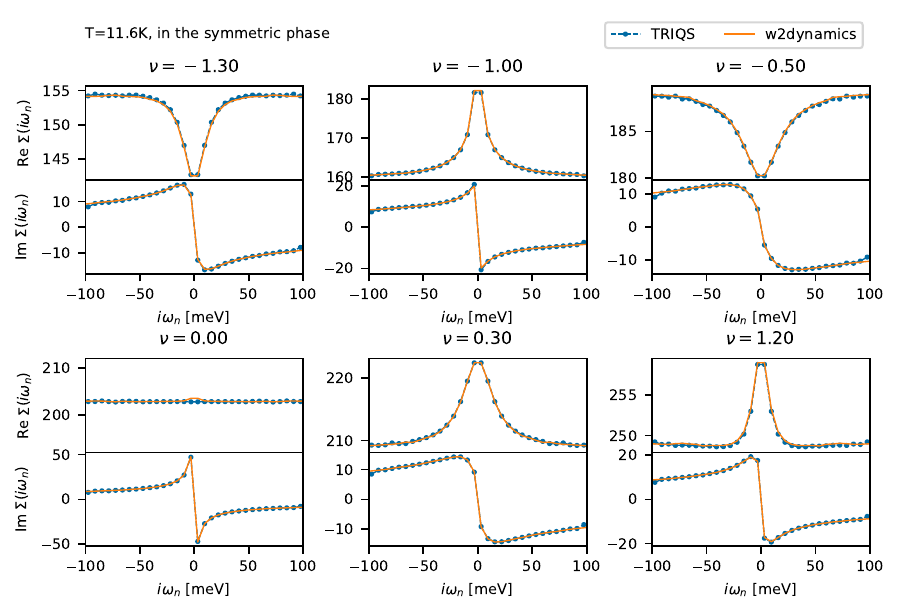}
    \caption{A comparison of the self-energies obtained from TRIQS and w2dynamics at various fillings at 11.6 K in the symmetric phase.}
    \label{fig:w2-triqs-benchmark}
\end{figure}

Fig.~\ref{fig:w2-triqs-benchmark} compares the converged self-energies obtained from each code at various fillings at 11.6 K. The self-energies are robust not only to the details of the implementation of two QMC solvers, but also to the independently made choices of DMFT and QMC parameters. For example, in the TRIQS runs, the QMC measurements were made in the imaginary time basis, while in the w2dynamics runs, the measurements were made in the Legendre basis. 
Data from TRIQS simulations was used to generate the following figures: 1, 2, 3, 4a (ordered state data only), 4b--j, S3, and S4. Data from w2dynamics simulations was used to generate the following figures: 4a (symmetric phase data only), 5, 6, S6, and S7.

\subsection{Robustness to variation in interaction parameters}

In \protect\cite{calugaruTwistedBilayerGraphene2023}, some of authors of the present work analytically derive the THF model interacting and non-interacting parameters for a range of several physical parameters, including the twist angle and the gate distance. Changing the twist angle directly affects the non-interacting dispersion. The relevant parameter in this discussion is the energy scale of the hybridization between the localized and delocalized sectors, $\gamma$, which controls the gap between the flat bands and the remote bands. Varying the gate distance affects the strength of all the interaction parameters: when the gates are closer to the TBLG sample, the interaction is screened more strongly. Heuristically, heavy Fermion phenomenology is expected to survive as long as the flat bands remain flat ($M=U$) and hybridization $\gamma$ as discussed in \protect\cite{calugaruTwistedBilayerGraphene2023}.

\begin{figure}
    \centering
    \includegraphics[width=0.7\textwidth]{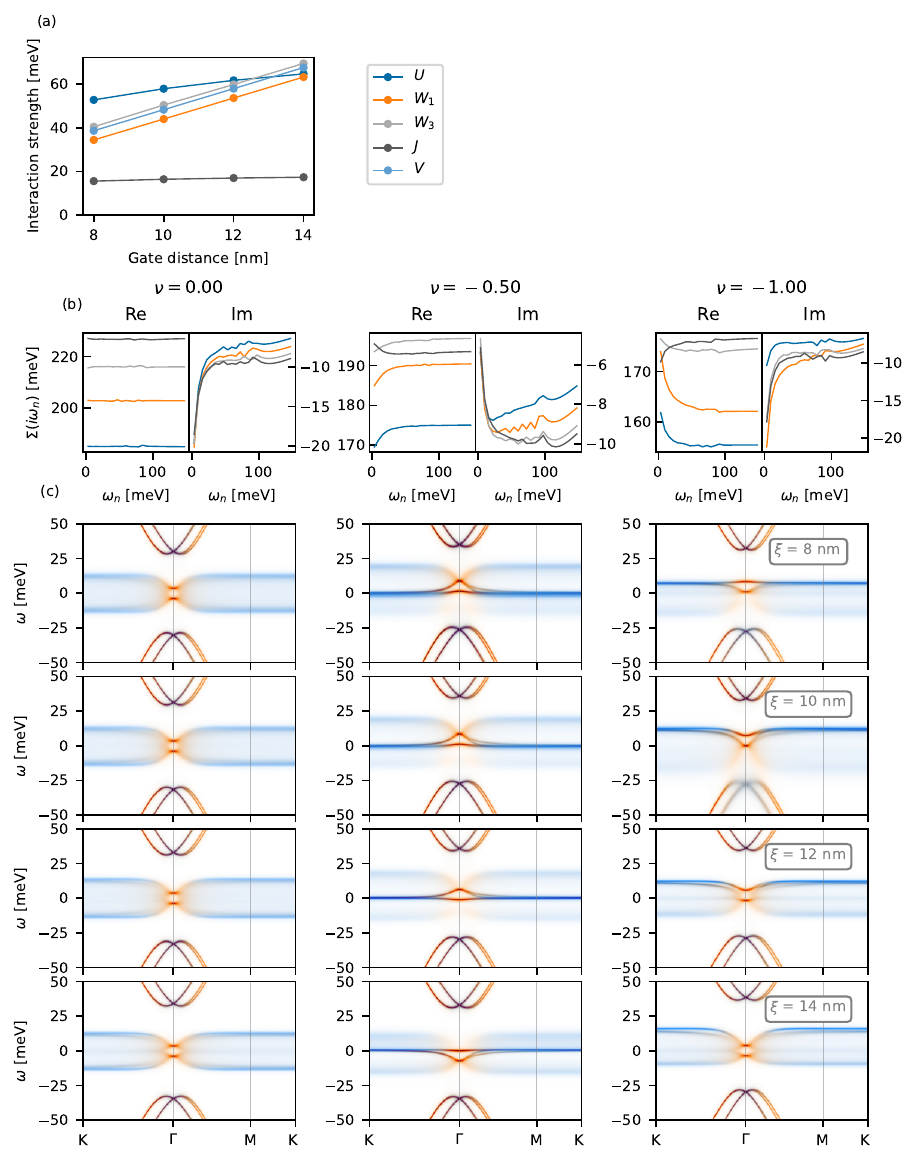}
    \caption{(a) Interaction parameters as a function of gate distance for twiste angle $\theta=1.05$ deg. (b) DMFT self-consistent Matsubara self-energy compared for different values of the gate distance at fillings $\nu=0, -0.5, -1$ at temperature $T=11.6$K. (c) DMFT spectral functions at fillings $\nu=0, -0.5, -1$ from left to right, for gate distances $\xi=8, 10, 12, 14$~nm from top to bottom.}
    \label{fig:robustness}
\end{figure}

In Fig. \ref{fig:robustness}, we compare the self-energies and spectral functions at three representative fillings ($\nu=0$, $\nu=-0.5$, and $\nu=-1$) for four different gate distances ($\xi=8, 10, 12, 14$  nm) at 1.05 deg in the symmetric phase. Fig. \ref{fig:robustness} shows the strength of the interaction parameters at these values of the gate distance. %At this twist angle, $\gamma = 24$~meV, which means that in all these cases $\gamma/U<0.5$. 
Note first that the imaginary part of the self-energy is qualitatively the same and quantitatively similar at each filling regardless of the gate distance. 
%The real part of the self-energy has a static part that can differ by $\sim 10$~meV, which reflects the magnitude of the change in the various density-density terms (U, V, W) as the gate distance is varied. 
Also, the qualitative features of the spectral functions remain the same as the gate distance is varied: ph-symmetric(-asymmetric) Hubbard bands at $\nu=-0, (-1)$ with c spectral weight around Gamma at the Fermi level; f-spectral weight pinned to the Fermi level at $\nu=-0.5$.

{With this, we conclude that three crucial features that govern the phenomenology—the formation of local moments, the ordering of local moments, and the redistribution of charge between the f- and c-sectors, are all robust to reasonable changes in the values of the interaction parameters, where \emph{reasonable} has an experimentally relevant characterization in terms of the gate distance. The precise temperature scale of local moment formation and ordering may be affected, but we expect the order of magnitude, 100K and 10K respectively to be robust. Likewise the precise filling where the c-f reshuffling events occur (controlled to first order by $U-(W_1+W_3)/2$) will change. This has an experimentally measurable effect in that it affects the precise filling where the local minima in the compressibility occurs. And indeed, there is a lot of variation in experimental data in this regard (c.f. Fig.~7).}

\section{Breaking symmetries}\label{app:breaking_symmetries}

We specify a chosen symmetry-broken state through the \emph{parent state} density matrix, $\rho^0$. In the parent states, all doped charge goes to the $f$-subspace, i.e. the $c$-subspace is always half-filled. In the $f$-subspace, we choose the up spin sector to be in the K-IVC phase, and the down spin sector to be in the K-IVC, valley-polarized, or disordered phases at fillings $\nu= 0, -1, -2$.
\begin{equation}
\begin{aligned}\label{eq:parent-state-rho}
    \rho^{0}_{f, \uparrow}(\nu) &= \frac{1}{2}\sigma_0\tau_0 - \frac{1}{2}\sigma_y\tau_y\\
    \rho^{0}_{f, \downarrow}(\nu) &= \begin{cases}
    \frac{1}{2}\sigma_0\tau_0 - \frac{1}{2}\sigma_y\tau_y &  \text{for } \nu = 0\\
    \text{diag}([1,0,0,0])& \text{for } \nu=-1\\
    \text{diag}([0,0,0,0])&  \text{for } \nu=-2\\
    \end{cases}\\
    \rho^0_{c(\Gamma_3)}(\nu)  =\rho^0_{c(\Gamma_1\oplus\Gamma_2)}(\nu)  &= \frac{1}{2} \sigma_0\tau_0\zeta_0
\end{aligned}
\end{equation}
where $\sigma_\mu, \tau_\mu, \zeta_\mu$, are Pauli matrices in orbital, valley and spin space, with $\mu = 0,x,y,z$, and the subscripts $(f, c(\ldots))$, $(\uparrow, \downarrow)$ signify projections to the electron species or spin sector.

The polarizing field $\hat{H}_{pol}$ used to bias the system towards a given ordered state is defined by the parent state density matrices 
\begin{align}
    \hat{H}_{pol} = -\lambda_{pol}\Psi^{\dagger}\left(\rho^0 -\rho^{0,symm}\right)\Psi
\end{align}
where $\Psi$ is a vector of $c$ and $f$ operators in the same order as in $\rho_0$, and $\rho^{0,symm}$ are the analog of the parent state density matrices for the symmetric phase:
\begin{align}
    \rho^{0,symm}_{c(\Gamma_3)}  =\rho^{0, symm}_{c(\Gamma_1\oplus\Gamma_2)}  &= \frac{1}{2} \sigma_0\tau_0\zeta_0,\\
    \rho^{0,symm}_{f}  &= \left(\frac{1}{2} + \frac{\nu}{8}\right)\sigma_0\tau_0\zeta_0,
\end{align}
where $\nu = 0,-1,-2$.
 
We chose the strength of the polarizing field $\lambda_{pol}$ to be $3$~meV. We found it sufficient to turn off the polarizing field after 10 iterations.

\subsection{Basis rotation and symmetrization}\label{app:breaking_symmetries:sym_proc}
In order to minimize the sign problem in the QMC simulations, we specify the impurity problem in the natural basis of the parent state density matrix, in which the hybridization function (the dynamical mean-field that describes electrons hopping in and out of the bath) is (almost diagonal) diagonal in the (ordered) symmetric phases. In the symmetric phase or when a spin sector is valley-polarized, the natural basis coincides with the original heavy Fermion basis, so no rotation is needed. When a spin sector has K-IVC order, a basis rotation is required. For instance, in the K-IVC+VP parent state at $\nu=-1$ described by the density matrix in \eqref{eq:parent-state-rho}, the corresponding wave function is written as
\begin{align}
    \ket{\text{K-IVC+VP}^{\nu=-1}_0} =& \frac{1}{2} f^\dagger_{1+\downarrow}(f^\dagger_{1+\uparrow} + f^\dagger_{2-\uparrow})\nonumber\\
    &\times(-f^\dagger_{1-\uparrow} + f^\dagger_{2+\uparrow})\ket{FS},
\end{align}
where $\ket{FS}$ is the Fermi sea of all the lower $c$ bands occupied. In this case, the natural basis in the up spin sector corresponds to the operators $\tilde{f}_{\tilde{\alpha}\tilde{\eta} \uparrow}$ defined by
\begin{align}
    \tilde{f}_{1\pm s} &= \frac{1}{\sqrt{2}}(f^\dagger_{1+s} \pm f^\dagger_{2-s})\\
    \tilde{f}_{2\pm s} &= \frac{1}{\sqrt{2}}(\mp f^\dagger_{1-s} + f^\dagger_{2+s})
\end{align}
In the natural basis, the self-energy is diagonal. We symmetrize the self-energy output by the QMC solver after every iteration. For calculations in the symmetric phase, this takes the form $\Sigma_{x'x'} \to \frac{1}{8}\sum_{x}\Sigma_{xx}$, where the index $x$ represents combined orbital, valley, and spin indices. In the symmetry-broken phase, a reduced symmetrization is applicable. Based on the parent states in \eqref{eq:parent-state-rho}, the sets of degenerate elements are

\begin{align}\label{eq:degSigma}
    \{\tilde{f}_{1+\uparrow}, \tilde{f}_{2+\uparrow}, \tilde{f}_{1+\downarrow}, \tilde{f}_{2+\downarrow}\},  \{\tilde{f}_{1-\uparrow}, \tilde{f}_{2-\uparrow}, \tilde{f}_{1-\downarrow}, \tilde{f}_{2-\downarrow}\} \quad\quad& \text{K-IVC at }\nu=0-\delta \nonumber\\
    \{\tilde{f}_{1+\uparrow}, \tilde{f}_{2+\uparrow}\},  \{\tilde{f}_{1-\uparrow}, \tilde{f}_{2-\uparrow}\}, \{f_{1-\downarrow}, f_{2-\downarrow}\}, \{f_{1+\downarrow}\}, \{f_{2+\downarrow}\} \quad\quad& \text{K-IVC+VP at }\nu=-1\mp \delta\\
    \{\tilde{f}_{1+\uparrow}, \tilde{f}_{2+\uparrow}\},  \{\tilde{f}_{1-\uparrow}, \tilde{f}_{2-\uparrow}\},\{{f}_{1+\downarrow}, {f}_{2+\downarrow}, {f}_{1-\downarrow}, {f}_{2-\downarrow}\}  \quad\quad& \text{K-IVC at }\nu=-2\mp \delta \nonumber
\end{align}

The self-energy is symmetrized within each set of degenerate orbitals. This symmetrization procedure is connected to the symmetries of the THF in Sec.~\ref{app:sec:hartree_fock}.

\subsection{Order parameter}
In the ordered state, we find the solution at an arbitrary filling $\nu$ and temperature $T$ by gradually doping the solution away from a reference integer fillings $\nu_{ref}=0,-1$ or $-2$ and temperature $T$. As we dope away from the integer filling, the system may or may not stay in the ordered state. We quantify this using the order parameter defined as follows. 
For an $f$-projected density matrix $\rho_f^{\nu_{ref}}(\nu, T)$ corresponding to the self-consistent solution at filling $\nu$ and temperature $T$ continued from the symmetry-broken self-consistent solution at $\nu_{ref} = 0, -1,$ or $-2$ and temperature $T$, the order parameter $O^{\nu_{ref}}(\nu, T)$ is given by
\begin{equation}
\begin{split}
   &O^{\nu_{ref}}(\nu, T) = \\
   &\left\langle\left(\rho^{0}_f(\nu_{ref}) - \left(\frac{1}{2}+\frac{\nu_{ref}}{8}\right)\mathbf{Id}\right),\left(\rho^{\nu_{ref}}_f(\nu, T)\right)\right\rangle,
   \end{split}
\end{equation}
where $\langle{A,B}\rangle = \text{trace}(\bar{A}^T \cdot B)$ is the matrix inner product, $\mathbf{Id}$ is the identity, and $\rho^{0}_f(\nu_{ref})$ is the density matrix of the parent state at $\nu_{ref}$. This definition has the advantage that all types of order (spin polarization, IVC, VP, etc.) are treated on the same footing, so we can plot all three panels in Fig.~\ref{fig:stability}(a) of the main text together. The order parameter is constrained to $[0,1]$ under the reasonable assumption that the self-consistent solution does not flip the sign of the polarization of the parent state.

\section{Finite-temperature Hartree-Fock theory for the THF model}\label{app:sec:hartree_fock}

This appendix provides a concise overview of the finite-temperature Hartree-Fock theory as it applies to the THF model, elaborating on the comprehensive discussion found in Ref.~\protect\cite{calugaru_ipt}. Initially, we revisit the notation set forth in Ref.~\protect\cite{calugaru_ipt}, which treats the $f$- and $c$-electrons on equal footing. Subsequently, we derive the Hartree-Fock Hamiltonian for a generic symmetry-broken ground state. In the final section, we adapt the self-consistent Hartree-Fock algorithm to accommodate the partial symmetrization methods delineated in \ref{app:breaking_symmetries:sym_proc}. This adjustment enables a direct comparison between DMFT and finite-temperature Hartree-Fock simulations, as both now incorporate a similar symmetrization approach.

\subsection{Generic notation}\label{app:sec:hartree_fock:notation}

Within the THF Hamiltonian from Eq.~\eqref{app:eqn:sp_ham_thf}, the $c$-fermions are only defined within a limited region around the $\Gamma_M$ point. Following Ref.~\protect\cite{calugaru_ipt}, we will extend the $\Lambda_{c}$ cutoff so that the $\cre{c}{\vk,a,\eta,\sigma}$ fermion covers exactly one Brillouin zone. Next, we introduce the $\cre{\gamma}{\vk,\eta,i,\sigma}$ fermions (for $1 \leq i \leq 6$), which are defined by
\begin{equation}
\label{app:eqn:shorthand_gamma_not}
\cre{\gamma}{\vk,\eta,i,\sigma} \equiv \begin{cases}
	\cre{c}{\vk,\eta,i,\sigma}, & \qq{for} 1 \leq i \leq 4 \\
	\cre{f}{\vk,\eta,i-4,\sigma}, & \qq{for} 5 \leq i \leq 6
\end{cases}.
\end{equation}
The corresponding single-particle Hamiltonian matrix, whose blocks are given by Eq.~\eqref{app:eqn:H_c} and Eq.~\eqref{app:eqn:H_fc}, reads as
\begin{equation}
\label{app:eqn:full_THF_Hamiltonian_gamma_bas}
h^{\eta} \left( \vk \right) = \begin{pmatrix}
	H^{(c,\eta)} \left( \vk \right) & H^{\dagger (fc,\eta)} \left( \vk \right) \\
	H^{(fc,\eta)} \left( \vk \right) & 0_{2 \times 2} 
\end{pmatrix}.
\end{equation}

With the new notation at hand, the density matrix of the system is given by
\begin{equation}
	\label{app:eqn:def_rho_HF}
	\varrho_{i \eta \sigma; i' \eta' \sigma'} \left(\vk \right) = \expec{ \normord{\cre{\gamma}{\vk, i, \eta, \sigma} \des{\gamma}{\vk, i' \eta' \sigma'}}},
\end{equation}
where $\expec{\hat{\mathcal{O}}}$ denotes the expectation value of the operator $\hat{\mathcal{O}}$ in the grand canonical ensemble of the system (at temperature $T$ and chemical potential $\mu$). Throughout this work, we restrict ourselves to states that preserve moir\'e translation symmetry. The total filling of the $f$- and $c$-electrons are related to the traces of the corresponding diagonal blocks of the density matrix,
\begin{equation}
	\nu_c = \frac{1}{N_M} \sum_{i=1}^{4} \sum_{\vk,\eta,\sigma} \varrho_{i \eta \sigma; i \eta \sigma} \left(\vk \right) \qq{and}
	\nu_f = \frac{1}{N_M} \sum_{i=5}^{6} \sum_{\vk,\eta,\sigma} \varrho_{i \eta \sigma; i \eta \sigma} \left(\vk \right),
\end{equation} 
while the total electron filling is $\nu = \nu_c + \nu_f$. 

\subsection{Hartree-Fock Hamiltonian}\label{app:sec:hartree_fock:hamiltonian}

In this section, we briefly review the finite-temperature Hartree-Fock theory of the THF model~\protect\cite{calugaru_ipt}. For a given state of the system characterized by the density matrix $\varrho_{i \eta \sigma; i' \eta' \sigma'} \left(\vk \right)$, the interaction Hartree-Fock Hamiltonian of the model can be obtained via the standard decoupling procedure~\protect\cite{calugaru_ipt} and is given by
\begin{equation}
	\label{app:eqn:hartree_fock_interaction}
	H_{I,\tMF} = \sum_{\substack{i,\eta,\sigma \\ i',\eta',\sigma'}} h^{I,\tMF}_{i \eta \sigma; i' \eta' \sigma'} \left( \vk \right) \cre{\gamma}{\vk,i,\eta,\sigma} \des{\gamma}{\vk,i',\eta',\sigma'} + E_0,
\end{equation}
where the corresponding Hartree-Fock Hamiltonian matrix reads as 
\begin{align}
	h^{I,\tMF}_{i \eta \sigma; i' \eta' \sigma'} \left( \vk \right) &= 
	\sum_{\alpha=1}^{2} \left( U\nu_f + \frac{1}{N_M} \sum_{a=1}^{4} W_a \sum_{\vk'} \sum_{\eta'',\sigma''} \varrho_{a \eta'' \sigma''; a \eta'' \sigma''} \left( \vk' \right) \right) \delta_{(\alpha+4) i} \delta_{(\alpha+4) i'} \delta_{\eta \eta'} \delta_{\sigma \sigma'} \nonumber \\
	&+ \sum_{a=1}^{4} \left( V \nu_c + W_a \nu_f \right) \delta_{a i} \delta_{a i'} \delta_{\eta \eta'} \delta_{\sigma \sigma'}  \nonumber \\
	&- \frac{U}{N_M} \sum_{\alpha,\alpha'=1}^{2} \delta_{(\alpha+4) i} \delta_{(\alpha'+4) i'}\sum_{\vk'}  \varrho_{i' \eta' \sigma'; i \eta \sigma} \left( \vk' \right) \nonumber \\
	&- \sum_{\alpha=1}^{2} \sum_{a'=1}^{4} \frac{W_{a'}}{N_M} \left[ \delta_{(\alpha+4) i} \delta_{a' i'} + \delta_{(\alpha+4) i'} \delta_{a' i} \right] \sum_{\vk'} \varrho_{i' \eta' \sigma'; i \eta \sigma} \left( \vk' \right) \nonumber \\
	& + \frac{1}{N_M} \sum_{\substack{\iesC{1} \\ \iesC{2}}} \left( \mathcal{J}_{\ies{1};\ies{2}; i \eta \sigma; i' \eta' \sigma'} + \mathcal{J}_{i \eta \sigma; i' \eta' \sigma'; \ies{1};\ies{2}} \right)  \sum_{\vk'}  \varrho_{\ies{1}; \ies{2}} \left( \vk' \right) \nonumber \\
	& - \frac{1}{N_M} \sum_{\substack{\iesC{1} \\ \iesC{2}}} \left( \mathcal{J}_{\ies{1}; i' \eta' \sigma' ; i \eta \sigma; \ies{2}} + \mathcal{J}_{i \eta \sigma; \ies{2}; \ies{1} i' \eta' \sigma'} \right)  \sum_{\vk'}  \varrho_{\ies{1}; \ies{2}} \left( \vk' \right),	\label{app:eqn:hartree_fock_interaction_matrix}
\end{align}

\begin{align}
    E_0 &=  -N_M \frac{U \nu_f^2}{2} 
    +\frac{U}{2 N_m} \sum_{\vk,\vk'} \sum_{\alpha,\alpha'=1}^2 \sum_{\substack{\eta, \sigma \\ \eta', \sigma'}} \varrho_{(\alpha'+4) \eta' \sigma'; (\alpha + 4) \eta \sigma}\left( \vk' \right) \varrho_{(\alpha + 4) \eta \sigma; (\alpha' + 4) \eta' \sigma'}\left( \vk \right)  \nonumber\\ 
    &-\sum_{a=1}^4 W_a \nu_f\sum_{\vk,\eta, \sigma}\varrho_{a\eta \sigma; a\eta \sigma } \left( \vk \right) 
    +\frac{1}{N_M}\sum_{\vk',\vk}\sum_{a=1}^4 W_a  \varrho_{a'\eta'\sigma';(a+4)\eta\sigma}(\vk')
     \varrho_{(a+4)\eta \sigma; a'\eta'\sigma'}(\vk)
    -V \frac{\nu_c^2}{2} \nonumber\\ 
    & -\frac{1}{2N_M} \sum_{\substack{\iesC{1} \\ \iesC{2}}} \left( \mathcal{J}_{\ies{1};\ies{2}; i \eta \sigma; i' \eta' \sigma'} + \mathcal{J}_{i \eta \sigma; i' \eta' \sigma'; \ies{1};\ies{2}} \right)  \sum_{\vk'}  \varrho_{\ies{1}; \ies{2}} \left( \vk' \right)
    \sum_{\vk}\rho_{i\eta \sigma;i'\eta'\sigma'} \left(\vk \right)
    \nonumber \\
	& + \frac{1}{2N_M} \sum_{\substack{\iesC{1} \\ \iesC{2}}} \left( \mathcal{J}_{\ies{1}; i' \eta' \sigma' ; i \eta \sigma; \ies{2}} + \mathcal{J}_{i \eta \sigma; \ies{2}; \ies{1} i' \eta' \sigma'} \right)  \sum_{\vk'}  \varrho_{\ies{1}; \ies{2}} \left( \vk' \right)\sum_\vk \rho_{i\eta \sigma;i'\eta'\sigma'} \left(\vk \right). \label{app:eqn:hartree_fock_ct_term}
\end{align}
In Eq.~\eqref{app:eqn:hartree_fock_interaction_matrix} and Eq.~\eqref{app:eqn:hartree_fock_ct_term}, we have introduced the following tensor 
\begin{align}
	&\mathcal{J}_{\ies{1};\ies{2};\ies{3};\ies{4}} = - \frac{J}{2} \sum_{\substack{\alpha,\alpha'=1 \\ \eta,\eta'}}^{2} \left[ \eta \eta' + \left( -1 \right)^{\alpha+\alpha'} \right] \delta_{(\alpha+4)i_1} \delta_{(\alpha'+4)i_2} \delta_{(\alpha'+2)i_3} \delta_{(\alpha+2)i_4} \delta_{\eta \eta_1} \delta_{\eta' \eta_2} \delta_{\eta' \eta_3} \delta_{\eta \eta_4} \delta_{\sigma_1 \sigma_4}  \delta_{\sigma_2 \sigma_3} \nonumber \\
	%%%%%%%%%%%%%%%%%%%%%%%%%%%%%%%%%%%%%%%%%%%%%%%%%%%
	-& \frac{J}{4} \sum_{\substack{\alpha,\alpha'=1 \\ \eta,\eta'}}^{2} \left[ \eta \eta' - \left( -1 \right)^{\alpha+\alpha'} \right] \delta_{(\alpha+4)i_1} \delta_{(\alpha'+2)i_2} \delta_{(\alpha'+4)i_3} \delta_{(\alpha+2)i_4} \delta_{\eta \eta_1} \delta_{\eta' \eta_2} \delta_{\eta' \eta_3} \delta_{\eta \eta_4} \delta_{\sigma_1 \sigma_4}  \delta_{\sigma_2 \sigma_3} \nonumber \\
	%%%%%%%%%%%%%%%%%%%%%%%%%%%%%%%%%%%%%%%%%%%%%%%%%%%
	-& \frac{J}{4} \sum_{\substack{\alpha,\alpha'=1 \\ \eta,\eta'}}^{2} \left[ \eta \eta' - \left( -1 \right)^{\alpha+\alpha'} \right] \delta_{(\alpha+4)i_2} \delta_{(\alpha'+2)i_1} \delta_{(\alpha'+4)i_4} \delta_{(\alpha+2)i_3} \delta_{\eta \eta_1} \delta_{\eta' \eta_2} \delta_{\eta' \eta_3} \delta_{\eta \eta_4} \delta_{\sigma_1 \sigma_4}  \delta_{\sigma_2 \sigma_3}, \label{app:eqn:j_tensor}
\end{align}
which encodes the $\hat{H}_{J}$ and $\hat{H}_{\tilde{J}}$ interaction terms of the THF Hamiltonian~\protect\cite{songMagicAngleTwistedBilayer2022}. The mean field Hamiltonian from Eq.~\eqref{app:eqn:hartree_fock_interaction} corresponds to the Hartree-Fock decoupling of the $\hat{H}_{U}$, $\hat{H}_{W}$, $\hat{H}_{J}$, and $\hat{H}_{\tilde{J}}$, interaction terms of the THF Hamiltonian (wherein the nearest-neighbor $f$-electron repulsion $U_2$ is ignored), as well as to the $\hat{H}_{V}$ term decoupled only the in Hartree channel. The exclusion of the Fock channel in $\hat{H}_{V}$ aligns with both Ref.~\protect\cite{songMagicAngleTwistedBilayer2022} and the methodology used in the DMFT simulations. Finally, the constant term in Eq.~\ref{app:eqn:hartree_fock_ct_term} simply results in a shift of the Hartree-Fock charge excitation bands. Therefore, it can be disregarded in our subsequent discussions

The \emph{total} mean field Hamiltonian is obtained by summing the single-particle Hamiltonian and the Hartree-Fock interaction one, $H_{\tMF} = \hat{H}_{0} + H_{I,\tMF}$, with the corresponding matrix reading as
\begin{equation}
	\label{app:eqn:TBG_HF_Hamiltonian}
	h^{\tMF}_{i \eta \sigma; i' \eta' \sigma'} \left( \vk \right) = h^{\eta}_{i i'} \left(\vk \right) \delta_{\eta \eta'} \delta_{\sigma \sigma'}+  h^{I,\tMF}_{i \eta \sigma; i' \eta' \sigma'} \left( \vk \right),
\end{equation}
such that 
\begin{equation}
	H_{\tMF} = \sum_{\vk} \sum_{\substack{i, \eta, \sigma \\ i', \eta', \sigma'}} h^{\tMF}_{i \eta \sigma; i' \eta' \sigma'} \left( \vk \right) \cre{\gamma}{\vk,i,\eta,s} \des{\gamma}{\vk,i',\eta',s'}.
\end{equation}
Finally, we note that at self-consistency, the Hartree-Fock Hamiltonian and the density matrix are related by~\protect\cite{calugaru_ipt}
\begin{equation}
	\label{app:eqn:hf_scf_fin_temp}
	\varrho^{T} \left( \vk \right) = \left\lbrace \exp \left[  \beta \left( h^{\tMF} \left( \vk \right) - \mu \one \right) \right] + \one \right\rbrace ^{-1} - \frac{1}{2} \one,
\end{equation}
where $\one$ denotes the identity matrix, $\beta = 1/T$ is the inverse temperature, and $\mu$ is the chemical potential of the system. 

\subsection{Density matrix symmetrization}\label{app:sec:hartree_fock:symmetrization}

In this section, we are interested in obtaining the phase diagrams of the correlated insulators discussed in Eq.~\eqref{eq:parent-state-rho} at finite temperature and doping. Similarly to the DMFT simulations, the density matrix away from integer filling can be obtained by gradually doping the integer-filled self-consistent solution in small increments, as will be explained in section \ref{app:sec:hartree_fock:sc_solution}. To ensure the consistency with the partial symmetrization employed in the DMFT simulations, as explained in section \ref{app:breaking_symmetries:sym_proc}, the density matrices of both the doped and undoped correlated insulators will be symmetrized according to the symmetries of the corresponding integer-filled parent states, as will be explained in detail below. In section \ref{app:breaking_symmetries:sym_proc}, a natural basis was constructed for each of the three correlated insulators from Eq.~\eqref{eq:parent-state-rho}. When computing the self-energy of the doped symmetry-broken phases, a reduced symmetrization was performed on the self-energy, such that the latter is diagonal in the natural basis of the corresponding parent state. In this section, we also offer a justification for this \textit{ad-hoc} symmetrization procedure, and show that it arises naturally if one requires that the doped correlated phases obey the symmetries of the undoped parent states. 

We start by analyzing the symmetries of the correlated insulators from Eq.~\eqref{eq:parent-state-rho}. Before doing so, it is useful to define a number of symmetry operators. The generators of the spin-$\SUN{2}$ symmetry of the system are given by 
\begin{equation}
	\hat{S}_{\mu} = \sum_{\vk} \sum_{\substack{\iesC{1} \\ \iesC{2}}} \left[ D \left( \hat{S}_{\mu} \right) \right]_{\ies{1}; \ies{2}} \cre{\gamma}{\vk,\iesC{1}} \des{\gamma}{\vk,\iesC{2}}, \qq{for} \mu = x,y,z,
\end{equation}
with $D \left( \hat{S}_{\mu} \right) = D^{c} \left( \hat{S}_{\mu} \right) \oplus D^{f} \left( \hat{S}_{\mu} \right)$ and
\begin{equation}
	D^{c} \left( \hat{S}_{\mu} \right) =\sigma_0 \tau_0 \zeta_{\mu} \oplus \sigma_0 \tau_0 \zeta_{\mu}, \quad D^{f} \left( \hat{S}_{\mu} \right) =\sigma_0 \tau_0 \zeta_{\mu}, \qq{for} \mu=x,y,z.
\end{equation}
Similarly, the valley $\UN{1}$ rotation is generated by
\begin{equation}
	\hat{V}_{\UN{1}} = \sum_{\vk} \sum_{\substack{\iesC{1} \\ \iesC{2}}} \left[ D \left( \hat{V}_{\UN{1}} \right) \right]_{\ies{1}; \ies{2}} \cre{\gamma}{\vk,\iesC{1}} \des{\gamma}{\vk,\iesC{2}},
\end{equation}
where $D \left( \hat{V}_{\UN{1}} \right) = D^{c} \left( \hat{V}_{\UN{1}} \right) \oplus D^{f} \left( \hat{V}_{\UN{1}} \right)$ and
\begin{equation}
	D^{c} \left( \hat{V}_{\UN{1}} \right) =\sigma_0 \tau_0 \zeta_{\mu} \oplus \sigma_0 \tau_z \zeta_0, \quad D^{f} \left(  \hat{V}_{\UN{1}} \right) =\sigma_0 \tau_z \zeta_{0}.
\end{equation}
We also define the (spinless) unitary threefold rotation $C_{3z}$, twofold rotation $C_{2z}$, as well as the (spinless) antiunitary time-reversal $T$ and Kramers $K$ symmetry operators, such that 
\begin{equation}
	g \cre{\gamma}{\vk,i,\eta,\sigma} g^{-1} = \sum_{i', \eta', \sigma'} \cre{\gamma}{g\vk,i',\eta',\sigma'} \left[ D \left( g \right) \right]_{i' \eta' \sigma';i \eta \sigma}, \qq{for} g=C_{3z}, C_{2z}, T, K, 
\end{equation}
for which the representation matrices are given by $D \left( g \right) = D^c \left( g \right) \oplus D^f \left( g \right)$, with
\begin{alignat}{4}
	D^{c} \left( C_{3z} \right) &&=& e^{\frac{2 \pi i}{3} \sigma_z \tau_z \zeta_0} \oplus \sigma_0 \tau_0 \zeta_0, \quad &D^{f} \left( C_{3z} \right) &&=& e^{\frac{2 \pi i}{3} \sigma_z \tau_z \zeta_0} \\
	D^{c} \left( C_{2z} \right) &&=& \sigma_x \tau_x \zeta_0 \oplus \sigma_x \tau_x \zeta_0, \quad &D^{f} \left( C_{2z} \right) &&=& \sigma_x \tau_x \zeta_0 \\
	D^{c} \left( T \right) &&=& \sigma_0 \tau_x \zeta_0 \oplus \sigma_0 \tau_x \zeta_0, \quad &D^{f} \left( T \right) &&=& \sigma_0 \tau_x \zeta_0 \\
	D^{c} \left( K \right) &&=& -i  \left(\sigma_0 \tau_y \zeta_0 \oplus \sigma_0 \tau_y \zeta_0 \right), \quad &D^{f} \left( K \right) &&=& -i \sigma_0 \tau_y \zeta_0.
\end{alignat}

Finally, we introduce five symmetry operations which act in specific spin-valley flavors. We define a Kramers and a twofold rotation symmetry operator acting only in the spin $\sigma = \uparrow$ sector, a time-reversal symmetry operator acting in spin $\sigma = \downarrow$ sector, as well as a $C_{2z}T$ symmetry operator acting only in the spin $\sigma = \downarrow$ and valley $\eta=-$ sector, whose actions on the $\gamma$-fermions are given, respectively, by
\begin{align}
	K_{\uparrow} \cre{\gamma}{\vk,i,\eta,\sigma} K^{-1}_{\uparrow} &= \begin{cases}
		\sum_{i', \eta'} \cre{\gamma}{-\vk,i',\eta',\uparrow} \left[ D \left( K \right) \right]_{i' \eta' \uparrow;i \eta \uparrow}, & \qq{if} \sigma=\uparrow \\
		\cre{\gamma}{\vk,i,\eta,\sigma}, & \qq{if} \sigma=\downarrow
	\end{cases},\\
	\left[C_{2z} \right]_{\uparrow} \cre{\gamma}{\vk,i,\eta,\sigma} \left[C_{2z} \right]^{-1}_{\uparrow} &= \begin{cases}
		\sum_{i', \eta'} \cre{\gamma}{-\vk,i',\eta',\uparrow} \left[ D \left( C_{2z} \right) \right]_{i' \eta' \uparrow;i \eta \uparrow}, & \qq{if} \sigma=\uparrow \\
		\cre{\gamma}{\vk,i,\eta,\sigma}, & \qq{if} \sigma=\downarrow
	\end{cases},\\
	\left[T \right]_{\downarrow} \cre{\gamma}{\vk,i,\eta,\sigma} \left[T \right]^{-1}_{\downarrow} &= \begin{cases}
	\sum_{i', \eta'} \cre{\gamma}{-\vk,i',\eta',\downarrow} \left[ D \left( T \right) \right]_{i' \eta' \downarrow;i \eta \downarrow}, & \qq{if} \sigma=\downarrow \\
	\cre{\gamma}{\vk,i,\eta,\sigma}, & \qq{if} \sigma=\uparrow
	\end{cases},\\
	\left[C_{2z}T \right]_{\downarrow,-} \cre{\gamma}{\vk,i,\eta,\sigma} \left[C_{2z}T \right]^{-1}_{\downarrow,-} &= \begin{cases}
		\sum_{i'} \cre{\gamma}{\vk,i',-,\downarrow} \left[ D \left( C_{2z}T \right) \right]_{i' - \downarrow;i - \downarrow}, & \qq{if} \sigma=\downarrow \qq{and} \eta=- \\
		\cre{\gamma}{\vk,i,\eta,\sigma}, & \qq{otherwise}
	\end{cases},
\end{align}
where $D \left( C_{2z}T \right) = D \left( C_{2z} \right) D \left( T \right)$. Additionally, we define a valley $\UN{1}$ rotation operator acting only on the spin $\sigma = \downarrow$ sector 
\begin{equation}
	\left[ \hat{V}_{\UN{1}} \right]_{\downarrow} = \sum_{\vk} \sum_{\substack{\iesCDn{1} \\ \iesCDn{2}}} \left[ D \left( \hat{V}_{\UN{1}} \right) \right]_{\iesDn{1}; \iesDn{2}} \cre{\gamma}{\vk,\iesCDn{1},\downarrow} \des{\gamma}{\vk,\iesCDn{2},\downarrow}.
\end{equation}

With the symmetry operators at hand, we now consider the symmetry-broken phases of the THF model at the Hartree-Fock level. In analogy with the DMFT simulations, in our finite-temperature Hartree-Fock simulations, we will ignore any correlations between $f$-electrons belonging to different lattice sites. As a result, for a given correlated phase of the THF model characterized by the density matrix $\varrho_{i \eta \sigma; i' \eta' \sigma'} \left( \vk \right)$, the $f$-electron block will be $\vk$-independent and given by 
\begin{equation}
	\varrho_{(\alpha + 4) \eta \sigma; (\alpha' + 4) \eta' \sigma'} \left( \vk \right) = O^{f}_{\alpha \eta \sigma; \alpha' \eta' \sigma'}, \qq{for} 1 \leq \alpha, \alpha' \leq 2.
\end{equation}
Moreover, if we assume that the corresponding correlated phase is symmetric under the crystalline symmetry $g$ or is invariant under a continuous symmetry generated by $g$, then its density matrix will obey the following constraint
\begin{equation}
	\label{app:eqn:density_sym}
	\varrho_{\ies{1};\ies{2}} \left( \vk \right) = \sum_{\substack{\iesC{3} \\ \iesC{4}}} \left[ D \left( g \right) \right]_{\ies{3};\ies{1}} \varrho^{(*)}_{\ies{3}; \ies{4}} \left( g \vk \right) \left[ D \left( g \right) \right]^{*}_{\ies{4};\ies{2}},
\end{equation}
where $g \vk = \vk$ if $g$ is a continuous symmetry and ${}^{(*)}$ denotes complex conjugation whenever $g$ is antiunitary. 

As will be explained in section \ref{app:sec:hartree_fock:sc_solution}, the correlated phases at finite doping can be obtained by doping into the charge-one excitation bands of an integer-filled insulator, then recalculating the self-consistent solution for that particular non-integer filling. We note that each integer-filled correlated insulator state breaks some of the symmetries of the system while retaining others. In principle, as the integer-filled solution is gradually doped, the system can undergo further symmetry-breaking transitions at fractional fillings. To prevent this scenario, our calculations for the doped correlated insulators ensure that the remaining symmetries of the corresponding integer-filled state \emph{persist} even in the doped phase. Practically, this is achieved by imposing the symmetry constraint from Eq.~\eqref{app:eqn:density_sym} for every symmetry of the undoped insulator at the level of the $f$-electron density matrix of the doped phase.

In what follows, we will individually discuss the unbroken symmetries for each of the correlated insulators from Eq.~\eqref{eq:parent-state-rho}. We will also show that by imposing these symmetries in the corresponding doped phases, the $f$-electron block of the density matrix both \emph{at} and \emph{away} from integer filling becomes diagonal in the natural bases introduced in section \ref{app:breaking_symmetries:sym_proc}.  

\subsubsection{The $\nu = 0$ K-IVC state}\label{app:sec:hartree_fock:symmetrization:0}

The $\nu=0$ K-IVC correlated insulator state from Eq.~\eqref{eq:parent-state-rho} is symmetric under the $C_{3z}$, $C_{2z}$, spin $\SUN{2}$, and $K$ symmetries. As a result, through Eq.~\eqref{app:eqn:density_sym}, we will require that the correlated phases obtained by doping it away from integer filling have the following $f$-electron block in their density matrices
\begin{equation}
	\label{app:eqn:paramet_of_KIVC}
	O^{f} = a_1 \sigma_0 \tau_0 \zeta_0  + a_2 \sigma_y \tau_y \zeta_0, \qq{with} a_1,a_2 \in \mathbb{R}.
\end{equation}
We note that if we cast the $f$-electron density matrix in the natural basis of the $\nu = 0$ K-IVC state introduced in section \ref{app:breaking_symmetries:sym_proc}, 
\begin{equation}
	\left( \cre{\tilde{f}}{\vk,1,+,\uparrow}, \cre{\tilde{f}}{\vk,2,+,\uparrow}, \cre{\tilde{f}}{\vk,1,+,\downarrow}, \cre{\tilde{f}}{\vk,2,+,\downarrow}, \cre{\tilde{f}}{\vk,1,-,\uparrow}, \cre{\tilde{f}}{\vk,2,-,\uparrow}, \cre{\tilde{f}}{\vk,1,-,\downarrow}, \cre{\tilde{f}}{\vk,2,-,\downarrow} \right),
\end{equation}
the resulting density matrix has the following form
\begin{equation}
	\tilde{O}^{f} =  \text{diag}\left( \left[ a'_1, a'_1, a'_1, a'_1, a'_2, a'_2, a'_2, a'_2 \right] \right), \qq{with} a'_1,a'_2 \in \mathbb{R}.
 \label{eq:breaking_symmetries:sym_rhof_nu_0}
\end{equation} 
By transforming to the natural basis, we enforce the density matrix of the $f$-electrons to take the form given in Eq.~\ref{eq:breaking_symmetries:sym_rhof_nu_0}, thus ensuring that all the remaining symmetries preserved by the parent $\nu=0$ K-IVC insulator are also preserved in the non-integer correlated phases obtained by doping it. 

\subsubsection{The $\nu = -1$ K-IVC+VP state}\label{app:sec:hartree_fock:symmetrization:1}

The case of the $\nu=-1$ K-IVC+VP state requires a slightly more involved analysis. We note that the parent state (the $\nu = -1$ K-IVC+VP state) we consider here still has $C_{3z}$ and spin $\hat{S}_z$ symmetries. When computing the self-consistent solutions of the correlated phases obtained by doping it away from integer filling, we will impose these symmetries ($C_{3z}$ and $\hat{S}_z$) to the corresponding density matrices. In doing so, we ensure that the unbroken symmetries of the integer-filled $\nu = -1$ K-IVC+VP insulator persist in the correlated phases obtained by doping it. This then leads to two constraints on the density matrix according to Eq.~\eqref{app:eqn:density_sym} and also implies that the density matrix is spin-diagonal. Additionally, the spin $\sigma = \uparrow$ sector has $K$ and $C_{2z}$ symmetry (meaning that the correlated insulator as a whole has $K_{\uparrow}$ and $\left[C_{2z} \right]_{\uparrow}$ symmetries), which requires that its density matrix obeys 
\begin{align}
	\varrho_{\iesUp{1};\iesUp{2}} \left( \vk \right) &= \sum_{\substack{\iesCUp{3} \\ \iesCUp{4}}} \left[ D \left( K \right) \right]_{\iesUp{3};\iesUp{1}} \varrho^{*}_{\iesUp{3}; \iesUp{4}} \left( - \vk \right) \left[ D \left( K \right) \right]^{*}_{\iesUp{4};\iesUp{2}}, \label{app:eqn:density_sym_k_up_1} \\
	\varrho_{\iesUp{1};\iesUp{2}} \left( \vk \right) &= \sum_{\substack{\iesCUp{3} \\ \iesCUp{4}}} \left[ D \left( C_{2z} \right) \right]_{\iesUp{3};\iesUp{1}} \varrho_{\iesUp{3}; \iesUp{4}} \left( - \vk \right) \left[ D \left( C_{2z} \right) \right]^{*}_{\iesUp{4};\iesUp{2}}. \label{app:eqn:density_sym_k_up_2}
\end{align}
The spin $\sigma  = \downarrow$ sector features valley $\UN{1}$ symmetry, which implies that the correlated insulator is symmetric under the continuous symmetry generated by the $\left[ \hat{V}_{\UN{1}} \right]_{\downarrow}$ operator and that its density matrix obeys
\begin{equation}
	\varrho_{\iesDn{1};\iesDn{2}} \left( \vk \right) = \sum_{\substack{\iesCDn{3} \\ \iesCDn{4}}} \left[ D \left( \left[ \hat{V}_{\UN{1}} \right]_{\downarrow} \right) \right]_{\iesDn{3};\iesDn{1}} \varrho_{\iesDn{3}; \iesDn{4}} \left( \vk \right) \left[ D \left(  \left[ \hat{V}_{\UN{1}} \right]_{\downarrow} \right) \right]^{*}_{\iesDn{4};\iesDn{2}}, \label{app:eqn:density_sym_k_up_3}
\end{equation}
thus being valley-diagonal in the spin $\sigma = \downarrow$ sector. Finally, the spin-valley sector corresponding to $\sigma= \downarrow$ and $\eta = -$ also features $C_{2z}T$ symmetry. The $\nu=-1$ K-IVC+VP correlated insulator is thus symmetric under $\left[C_{2z}T\right]_{\downarrow,-}$, which implies that its density matrix obeys
\begin{equation}
	\varrho_{\iesDnMin{1};\iesDnMin{2}} \left( \vk \right) = \sum_{i_3,i_4} \left[ D \left( \left[C_{2z}T\right]_{\downarrow,-} \right) \right]_{\iesDnMin{3};\iesDnMin{1}} \varrho^{*}_{\iesDnMin{3}; \iesDnMin{4}} \left( \vk \right) \left[ D \left(  \left[C_{2z}T\right]_{\downarrow,-} \right) \right]^{*}_{\iesDnMin{4};\iesDnMin{2}}. \label{app:eqn:density_sym_k_up_4}
\end{equation}
In practice, we require that the correlated phases obtained by doping the $\nu = -1$ K-IVC+VP insulator preserve the remaining symmetries of the corresponding integer-filled state. This implies that the constraints from Eq.~\eqref{app:eqn:density_sym_k_up_1}, Eq.~\eqref{app:eqn:density_sym_k_up_2}, Eq.~\eqref{app:eqn:density_sym_k_up_3}, and Eq.~\eqref{app:eqn:density_sym_k_up_4} must hold for the density matrices of the doped phases, which will then assume the following form in the $f$-block
\begin{align}
	O^{f}_{\alpha \eta \uparrow; \alpha' \eta' \downarrow} &= O^{f}_{\alpha \eta \downarrow; \alpha' \eta' \uparrow} = 0, \\
	O^{f}_{\alpha \eta \uparrow; \alpha' \eta' \uparrow} &= \left[ a_1 \sigma_0 \tau_0  + a_2 \sigma_y \tau_y \right]_{\alpha \eta; \alpha' \eta'}, \\
	O^{f}_{\alpha + \downarrow; \alpha' + \downarrow} &= \left[ a_3 \sigma_0 + a_4 \sigma_z \right]_{\alpha \eta; \alpha' \eta'}, \\
	O^{f}_{\alpha - \downarrow; \alpha' - \downarrow} &= a_5 \left[  \sigma_0  \right]_{\alpha \eta; \alpha' \eta'}, \\
	O^{f}_{\alpha + \downarrow; \alpha' - \downarrow} &= O^{f}_{\alpha - \downarrow; \alpha' + \downarrow} = 0 ,
\end{align}
where $a_i \in \mathbb{R}$, ($1 \leq i \leq 5$). In the natural basis of the $\nu=-1$ K-IVC+VP state introduced in section \ref{app:breaking_symmetries:sym_proc}, 
\begin{equation}
	\left( \cre{\tilde{f}}{\vk,1,+,\uparrow}, \cre{\tilde{f}}{\vk,2,+,\uparrow}, \cre{\tilde{f}}{\vk,1,-,\uparrow}, \cre{\tilde{f}}{\vk,2,-,\uparrow}, \cre{f}{\vk,1,+,\downarrow}, \cre{f}{\vk,2,+,\downarrow}, \cre{f}{\vk,1,-,\downarrow}, \cre{f}{\vk,2,-,\downarrow} \right),
\end{equation}
the $f$-electron density matrix of both the $\nu=-1$ K-IVC+VP insulator and of the phases obtained by doping it will have the following form
\begin{equation}
	\tilde{O}^{f} =  \text{diag}\left( \left[ a'_1, a'_1, a'_2, a'_2, a'_3, a'_4, a'_5, a'_5 \right] \right), \qq{with} a'_i \in \mathbb{R}, \qq{for} 1 \leq i \leq 5.
\end{equation} 

\subsubsection{The $\nu = -2$ K-IVC+VP state}\label{app:sec:hartree_fock:symmetrization:2}

Similarly to the $\nu=-1$ K-IVC+VP state, the $\nu=-2$ K-IVC state features $C_{3z}$ and spin-$\hat{S}_z$ symmetries, thus being spin-diagonal. The latter also features $C_{2z}$ symmetry in both spin sectors. These three symmetries constrain the density matrix according to Eq.~\eqref{app:eqn:density_sym}. Additionally, the $\nu=-2$ K-IVC state features $\left[ \hat{V}_{\UN{1}} \right]_{\downarrow}$ and $T_{\downarrow}$ symmetries which imply that 
\begin{align}
	\varrho_{\iesDn{1};\iesDn{2}} \left( \vk \right) &= \sum_{\substack{\iesCUp{3} \\ \iesCUp{4}}} \left[ D \left( \left[ \hat{V}_{\UN{1}} \right]_{\downarrow} \right) \right]_{\iesDn{3};\iesDn{1}} \varrho_{\iesDn{3}; \iesDn{4}} \left(  \vk \right) \left[ D \left( \left[ \hat{V}_{\UN{1}} \right]_{\downarrow} \right) \right]^{*}_{\iesDn{4};\iesDn{2}}, \label{app:eqn:density_sym_k_up_5}\\
	\varrho_{\iesDn{1};\iesDn{2}} \left( \vk \right) &= \sum_{\substack{\iesCDn{3} \\ \iesCDn{4}}} \left[ D \left( T_{\downarrow}  \right) \right]_{\iesDn{3};\iesDn{1}} \varrho^{*}_{\iesDn{3}; \iesDn{4}} \left( - \vk \right) \left[ D \left( T_{\downarrow} \right) \right]^{*}_{\iesDn{4};\iesDn{2}}. \label{app:eqn:density_sym_k_up_6}
\end{align}
As a result, the density matrix of the $\nu=-2$ K-IVC state, as well as of the correlated phases obtained by doping it away from integer fillings must have the following parameterization of their $f$-electron block
\begin{align}
	O^{f}_{\alpha \eta \uparrow; \alpha' \eta' \downarrow} &= O^{f}_{\alpha \eta \downarrow; \alpha' \eta' \uparrow} = 0, \\
	O^{f}_{\alpha \eta \uparrow; \alpha' \eta' \uparrow} &= \left[ a_1 \sigma_0 \tau_0  + a_2 \sigma_y \tau_y \right]_{\alpha \eta; \alpha' \eta'}, \\
	O^{f}_{\alpha \eta \downarrow; \alpha' \eta' \downarrow} &= a_3 \left[ \sigma_0 \tau_0 \right]_{\alpha \eta; \alpha' \eta'},
\end{align}
for $a_1,a_2,a_3 \in \mathbb{R}$. In the natural basis of the $\nu=-2$ K-IVC state from \ref{app:breaking_symmetries:sym_proc},
\begin{equation}
	\left( \cre{\tilde{f}}{\vk,1,+,\uparrow}, \cre{\tilde{f}}{\vk,2,+,\uparrow}, \cre{\tilde{f}}{\vk,1,-,\uparrow}, \cre{\tilde{f}}{\vk,2,-,\uparrow}, \cre{f}{\vk,1,+,\downarrow}, \cre{f}{\vk,2,+,\downarrow}, \cre{f}{\vk,1,-,\downarrow}, \cre{f}{\vk,2,-,\downarrow} \right),
\end{equation}
the $f$-electron density matrix is given by 
\begin{equation}
	\tilde{O}^{f} = \text{diag}\left( \left[ a'_1, a'_1, a'_2, a'_2, a'_3, a'_3, a'_3, a'_3 \right] \right), \qq{with} a'_1, a'_2, a'_3 \in \mathbb{R}.
\end{equation}

\subsection{Self-consistent solution}\label{app:sec:hartree_fock:sc_solution}

To obtain the self-consistent density matrix for a given correlated insulator at integer filling $\nu_0$, we start from an initial guess of the density matrix 
\begin{equation}
	\varrho^{(0)}_{i \eta \sigma; i' \eta' \sigma'} \left( \vk \right) = \begin{cases}
		\rho^{0}_{(i-4) \eta \sigma; (i'-4) \eta' \sigma' } - \frac{1}{2} \delta_{i i'} \delta_{\eta \eta'} \delta_{\sigma \sigma'} & \qq{if} 5 \leq i, i' \leq 6 \\ 
		0 & \qq{otherwise} \\ 
	\end{cases}.
\end{equation}
For the correlated phases obtained by doping the integer-filled insulator, we use a pre-computed density matrix at a nearby integer filling, as will be explained below. Starting from the initial condition, we iterate the following Hartree-Fock algorithm whereby a new density matrix $\varrho^{(n+1)}_{i \eta \sigma; i' \eta' \sigma'} \left( \vk \right)$ can be obtained from the density matrix at the $n$-th step, $\varrho^{(n)}_{i \eta \sigma; i' \eta' \sigma'} \left( \vk \right)$. The Hartree-Fock self-consistent loop can be summarized as follows:
\begin{enumerate}
	\item Using $\varrho^{(n)}_{i \eta \sigma; i' \eta' \sigma'} \left( \vk \right)$, we construct the Hartree-Fock Hamiltonian at the $n$-th step according to Eq.~\eqref{app:eqn:TBG_HF_Hamiltonian}.
	
	\item The Hartree-Fock Hamiltonian is diagonalized as 
	\begin{equation}
		\label{app:eqn:diag_hf_ham}
		\sum_{i',\eta',\sigma'} h^{\tMF}_{i \eta \sigma; i' \eta' \sigma'} 
		\left( \vk \right) \varphi_{m;i' \eta' \sigma'} \left( \vk \right) = \epsilon_m \left( \vk \right) \varphi_{m;i \eta \sigma} \left( \vk \right),
	\end{equation}
	with $\epsilon_m \left( \vk \right)$ and $\varphi_{m;i' \eta' \sigma'} \left( \vk \right)$ denoting the $m$-th Hartree-Fock energy and eigenvector. A new density matrix is then constructed according to 
	\begin{equation}
		\label{app:eqn:assemble_rho_hf}
		\varrho'_{i \eta s; i' \eta' s'} \left(\vk \right) = \sum_{m} \frac{1}{e^{\beta \left(\epsilon_m \left( \vk \right) - \mu \right)}+1}\varphi^{*}_{m;i \eta s} \left( \vk \right) \varphi_{m;i' \eta' s'} \left( \vk \right),
	\end{equation}
	where the chemical potential $\mu$ is fixed by requiring that the total filling of the system $\nu$ equals the desired value.
	
	\item The new density matrix $\varrho'_{i \eta s; i' \eta' s'} \left(\vk \right)$ will not generically have a $\vk$-independent $f$-electron block with the correct parameterization, as derived in section \ref{app:sec:hartree_fock:symmetrization}. This is because the $f$-$c$ hybridization of the single-particle THF Hamiltonian is $\vk$-dependent. We seek solutions where the $f$-electrons at different sites are uncorrelated, meaning that the $f$-electron block of the corresponding density matrix is $\vk$-independent. To drop the $\vk$-dependency of the density matrix in the $f$-block, we now replace the $f$-electron density matrix at different $\vk$ points by its average across the BZ 
	\begin{align}
		O^{\prime f}_{\alpha \eta \sigma; \alpha' \eta' \sigma'} &\equiv \left[ \frac{1}{N_M} \sum_{\vk'} \varrho'_{(\alpha+4) \eta s; (\alpha'+4) \eta' s'} \left(\vk' \right) \right], \qq{for} 1 \leq \alpha, \alpha' \leq 2 \\
		\varrho'_{(\alpha + 4) \eta \sigma; (\alpha' + 4) \eta' \sigma'} \left( \vk \right) &\to O^{\prime f}_{\alpha \eta \sigma; \alpha' \eta' \sigma'}, \qq{for} 1 \leq \alpha, \alpha' \leq 2, \qq{and any} \vk.
	\end{align}
	
	\item Although $\vk$-independent, the $f$-electron block of the new density matrix, $O^{\prime f}_{\alpha \eta \sigma; \alpha' \eta' \sigma'}$, will not generically obey the remaining symmetries of the correlated phase from which it was obtained. For example, the correlated phase at $\nu=-0.1$ is obtained by doping the $\nu = 0$ K-IVC state and, per our assumptions around Eq.~\eqref{app:eqn:density_sym}, should obey the remaining symmetries of the K-IVC state discussed in section \ref{app:sec:hartree_fock:symmetrization:0}. With infinite precision arithmetic, if the density matrix at the beginning of the self-consistent step $\varrho^{(n)} \left( \vk \right)$ obeys some of the symmetries of the THF model, then the corresponding mean field Hamiltonian $h^{\tMF} 
	\left( \vk \right)$ and, consequently, the new density matrix $\varrho' \left( \vk \right)$ should obey the same symmetries. However, in finite precision arithmetic, $\varrho' \left( \vk \right)$  will only obey these symmetries \emph{approximately}. Over the course of many iterations, these small errors would grow leading to the self-consistent solution for $\nu = -0.1$ \emph{not} obeying the symmetries of the $\nu = 0$ K-IVC state. 
	
	To prevent these numerical instabilities, we will require that $O^{\prime f}_{\alpha \eta \sigma; \alpha' \eta' \sigma'}$ has the correct symmetry-enforced parameterization at every Hartree-Fock step. For example, when computing the order parameter of the $\nu=-0.1$ correlated phase, we impose the parameterization in Eq.~\eqref{app:eqn:paramet_of_KIVC} as follows
	\begin{align}
		&a_1 \equiv \frac{1}{8} \Tr \left[ \sigma_0 \tau_0 \zeta_0 O^{\prime f}  \right], \quad a_2 \equiv \frac{1}{8} \Tr \left[ \sigma_y \tau_y \zeta_0 O^{\prime f}  \right], \\
		&O^{\prime f} \to a_1 \sigma_0 \tau_0 \zeta_0 + a_2 \sigma_y \tau_y \zeta_0,
	\end{align}
	which then leads to the new density matrix 
	\begin{equation}
		\varrho^{(n+1)}_{i \eta \sigma; i' \eta' \sigma'} \left( \vk \right) = \begin{cases}
			O^{\prime f}_{\alpha \eta \sigma; \alpha' \eta' \sigma'} & \qq{if} 5 \leq i, i' \leq 6 \\ 
			\varrho'_{i \eta s; i' \eta' s'} \left(\vk \right) & \qq{otherwise} \\ 
		\end{cases}.
	\end{equation}
	Similar types of symmetrization are performed for the correlated phases obtained by doping the $\nu = -1$ K-IVC+VP and $\nu= -2$ K-IVC correlated insulator by imposing the corresponding parameterizations derived in sections \ref{app:sec:hartree_fock:symmetrization:1} and \ref{app:sec:hartree_fock:symmetrization:2}, respectively.
\end{enumerate}
To mitigate fluctuations in the self-consistent iterative algorithm, we use a two-pronged approach for improved convergence. Initially, we employ linear mixing to smooth out oscillations when the algorithm is far from convergence (implying that $\rho^{(n+1)} \left( \vk \right) \to \frac{1}{2} \rho^{(n)} \left( \vk \right) + \frac{1}{2} \rho^{(n+1)} \left( \vk \right)$ is performed at the end of each self-consistent step). As the density matrix changes become smaller, we switch to the DIIS convergence acceleration method, as described in Ref.~\protect\cite{PUL80}. Convergence is determined by the criterion that the Hartree-Fock Hamiltonian and the density matrix should commute (i.e., $\commutator{\varrho^{T} \left( \vk \right)}{h^{\tMF} \left( \vk \right)} = 0$, up to numerical accuracy), a condition implied by Eq.~\eqref{app:eqn:hf_scf_fin_temp}.

To generate the Hartree-Fock phase diagrams depicted in Fig.~\ref{fig:stability} in the main text, we first calculate the self-consistent density matrix for integer-filled correlated insulators at a specific temperature $T$. Once the self-consistent solution is obtained for an insulator at an integer filling $\nu_0$ and temperature $T$, we can obtain the symmetry-broken phases in the vicinity of $\nu_0$ by incrementally doping the system and requiring the doped states have the same symmetries as the state at integer filling $\nu_0$. Specifically, we examine fillings of $\nu_0 + n \delta\nu $, where $n\in \mathbb{Z}$ and $\abs{n} \leq \frac{1}{2 \delta\nu}$. This means that we incrementally dope the integer-filled correlated insulator by at most $1/2$, using $\delta \nu$ as the step size. To solve for each new filling $\nu_0 + n \delta \nu$, we use the previously computed self-consistent density matrix $\varrho \left( \vk \right)$ at $\nu_0 + (n-1) \delta \nu$ (if $n>0$) or $\nu_0 + (n+1) \delta \nu$ (if $n<0$) as initial conditions.
% Moreover, we also enforce the symmetries of the doped phase to be the same as the ones of the corresponding correlated-insulator state at integer filling.
In our calculations, we set $\delta \nu = 1/50$.

Finally, we note that for each iteration of the algorithm outlined above, we enforce that the $f$-electron density matrix, when expressed in the natural basis of the doped correlated insulator, must be proportional to the identity matrix within each set of degenerate elements. These sets are defined as per Eq.~\eqref{eq:degSigma}. This procedure is equivalent to enforcing that the self-consistent solutions have the same symmetry as our initial parent state or, in other words, have the same symmetry as the correlated state at the corresponding integer filling $\nu_0$. Our numerical checks also confirm that this constraint leads to a self-consistent density matrix that respects the symmetries of its correlated insulator parent state. This, according to Eq.~\eqref{app:eqn:hf_scf_fin_temp}, ensures that the Hartree-Fock Hamiltonian also adheres to these symmetries.

Specifically, the $f$-electron component of the Hartree-Fock Hamiltonian, when expressed in the natural basis of the parent correlated insulator, will also be proportional to the identity matrix within each degenerate set, as outlined in Eq.~\eqref{eq:degSigma}. Furthermore, Eq.~\eqref{app:eqn:hartree_fock_interaction_matrix} dictates that this Hartree-Fock Hamiltonian will exhibit $\vk$-independence in the $f$-electron block. Consequently, the $f$-electron block of the self-energy is $\vk$-independent, and, when expressed in the natural basis of the parent correlated insulator, is proportional to the identity matrix within each set of degenerate elements. Since the Hartree-Fock Hamiltonian is precisely the static part of the electron self-energy, this implies that within the natural basis, the $f$-electron static self-energy is proportional to the identity matrix within each degenerate set. This mirrors the partial symmetrization used in the DMFT calculations and outlined in section \ref{app:breaking_symmetries:sym_proc} which requires that the entire $f$-electron self-energy (i.e., both the \emph{dynamic} and the \emph{static} parts) are diagonal in the natural basis and proportional to the identity within each set of degenerate elements.

\section{Quasiparticle analysis}\label{app:QP}
The first step in the quasiparticle analysis is to find the zeros of $\hat{H}_{aux}$ defined in the main text as \eqref{eq:Hqp}. We do this using an iterative inflation scheme. We start by evaluating the eigenvalues of $\hat{H}_{aux}$ on the regular centered sampling of the first Brillouin zone as defined in \ref{app:implementation} with \texttt{sample\_len=20}. We filter the $k$-points by their smallest eigenvalue, i.e. we sort the $k$-points by their smallest eigenvalue and keep a fraction of the them. We determine this fraction heuristically. Then we inflate the $k$-mesh around the selected $k$-points. This new set of $k$-points is a finer sampling of a subset of the Brillouin zone within which the zeros are contained. We perform the filtering and inflation procedure iteratively until the smallest eigenvalue of all the filtered $k$-points falls below a threshold, which we chose to be 0.1~meV. 

\subsection{Quasiparticle weight, scattering rates, and transport}

In the DMFT approximation, the scattering rate is solely determined by the self-energy. More precisely, the inverse quasi particle life-time is given by $\Gamma = -Z \Im \Sigma(0)$, while the transport scattering rate rate is $\Im \Sigma(0)$---at least in a single band system.

\begin{figure}
    \centering
    \includegraphics[width=0.4\textwidth]{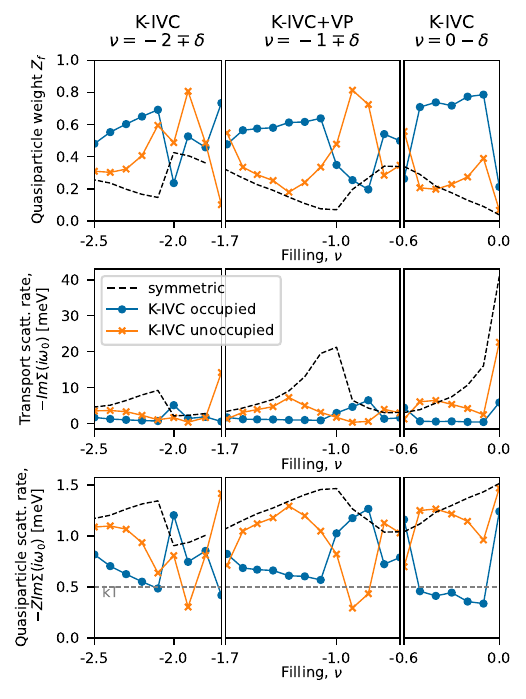}
    \caption{The quasiparticle weight, and the transport and quasiparticle scattering rates: a comparison between the symmetric and ordered state as a function of filling at $5.8$~K}
    \label{fig:Scattering rates}
\end{figure}

Here, the situation is complicated by the existence of the two types of delocalized and localized carriers $c$ and $f$. While $Z$ and $\Gamma$ are straightforwardly-defined and available for $f$-orbitals, disentangling the contributions of the two types of orbitals towards transport is currently a work in progress. Within the scope of the present work, in Fig.~4(h), we have compared the quasiparticle weight of the localized orbitals in the symmetric and symmetry-broken phases at the same temperature, and we conclude that since the quasiparticle weight is much lower in symmetric phase vs. the symmetry-broken phase, transport in the symmetry-broken phase is more coherent. The same conclusion is borne out when comparing the quasiparticle or transport scattering rates (see. Fig.~\ref{fig:Scattering rates}). The scattering rates in the symmetry-broken phase are below those for the symmetric states for most fillings. The $c$-electrons are not included in this analysis, and therefore only qualitative statements can be made regarding the resulting transport properties. %A more quantitative treatment is outside the scope of the current work.

\begin{figure}
    \centering
    \includegraphics[width=\textwidth]{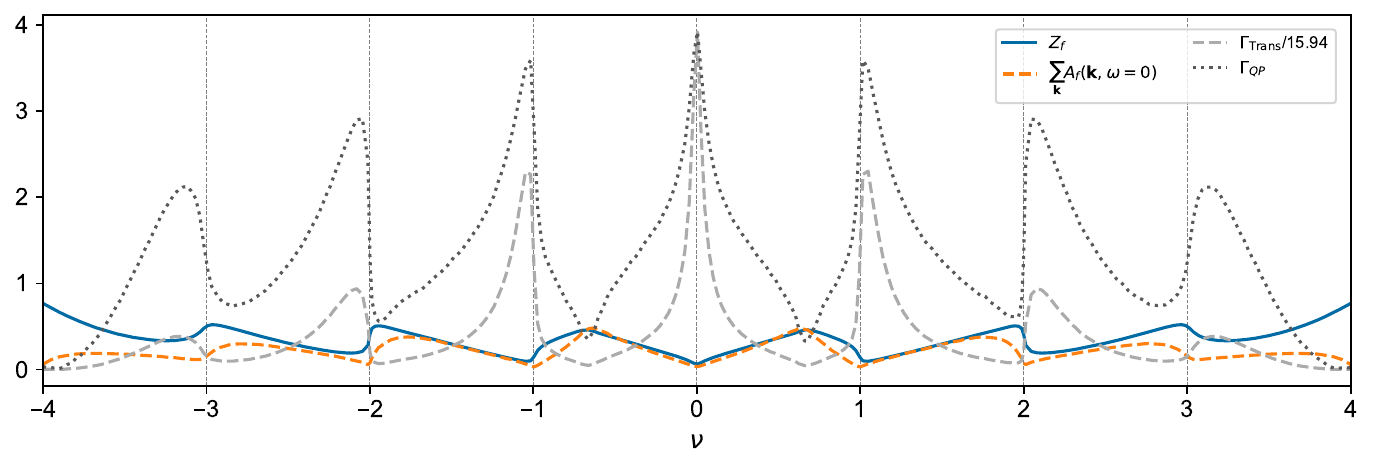}
    \caption{Comparison between different correlation estimators at T=11.6K: quasiparticle weight, spectral weight at the Fermi level extracted from $\sum_{\mathbf{k}}A_{f}(\mathbf{k},\omega=0)\approx G_{f}^{\mathrm{QMC}}(\tau=\beta/2)$, quasiparticle scattering amplitude $\Gamma_{QP}=-Z\mathrm{Im}\Sigma(\omega_{0})$ and transport scattering amplitude $\Gamma_{\mathrm{Trans}}=-\mathrm{Im}\Sigma(\omega_{0})$  (normalized to the max of the former).}
    \label{fig:Scattering rates symmetric}
\end{figure}
In Fig.~\ref{fig:Scattering rates symmetric}, we show the Quasiparticle weight $Z_f$, the quasiparticle scattering rate, the transprot scattering rate, and the spectral weight at the Fermi level for the symmetric state at the Fermi level. All indicators show that correlations are strongest close to integer fillings. While the two scattering reach local maxima close to integer fillings, the quasiparticle weight and the spectral weight at the Fermi level reach their local minima near integer fillings. 

\section{Chemical potential definition and geometric capacitance}\label{app:capacitance}
\begin{figure}
    \centering
    \includegraphics[width = 0.5\textwidth]{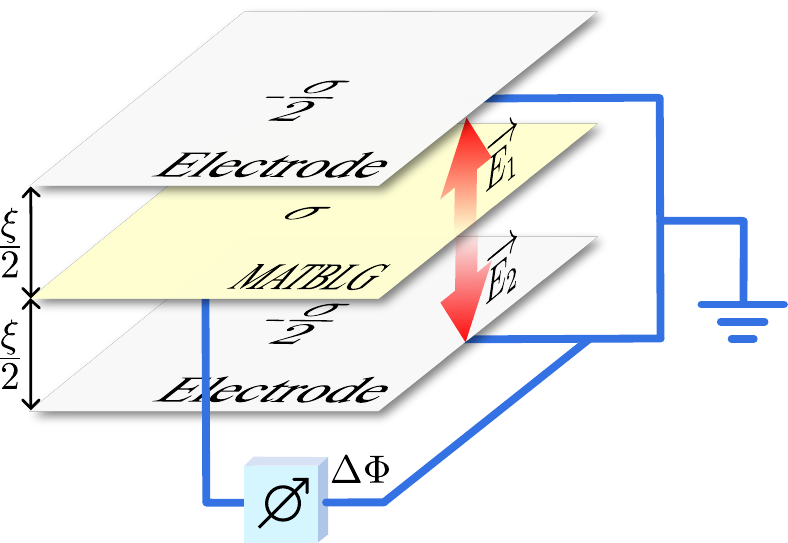}
    \caption{Sketch of the MATBLG realization assumed in the Song-Bernevig model, assumed MATBLG in between two capacitor plates.}
    \label{fig:geomcap}
\end{figure}

\begin{figure}
    \centering
    \includegraphics[width=\textwidth]{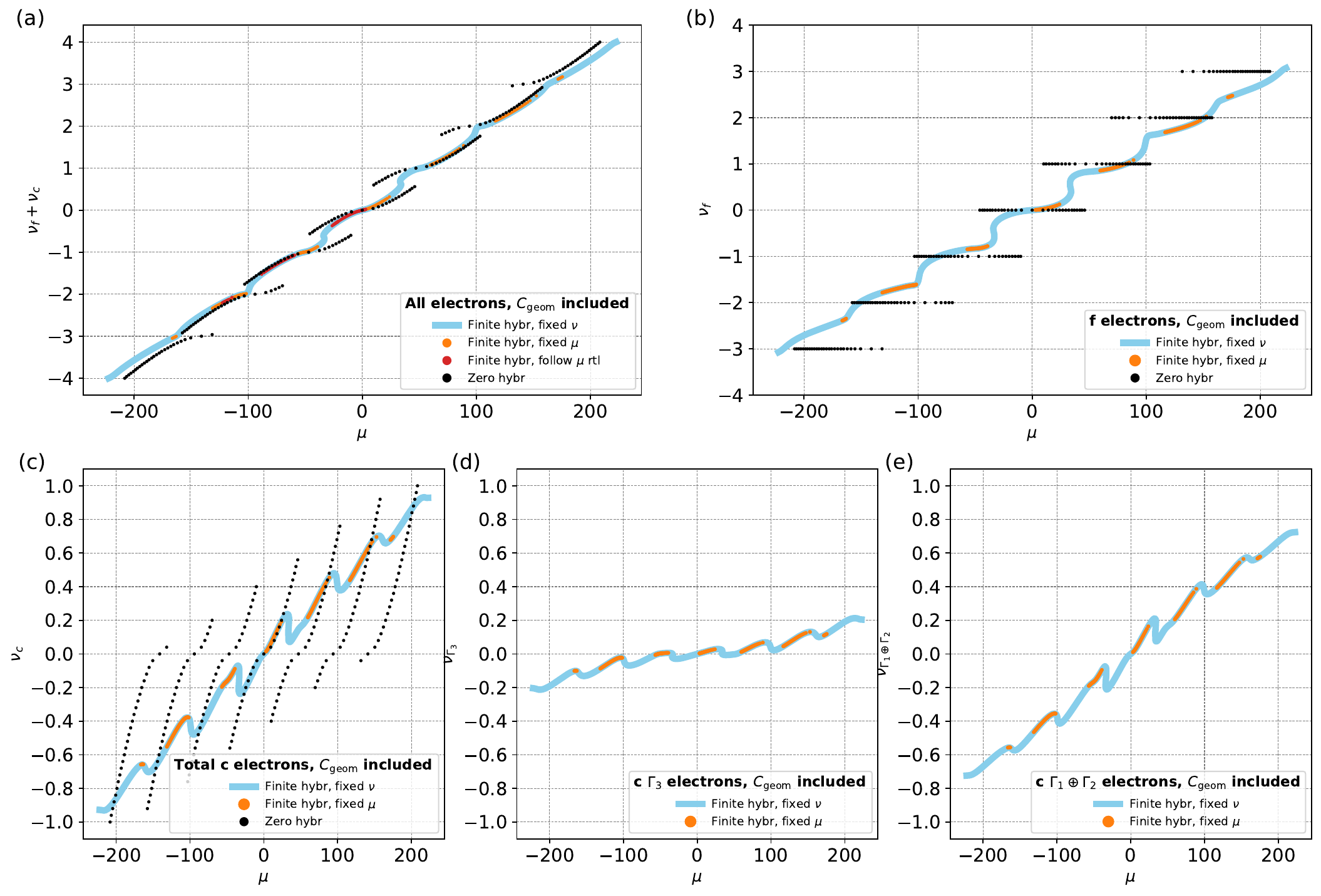}
    \caption{Occupation with respect to CNP as a function of the chemical potential, total value (a) and orbital-valley resolved (b-c), for T=11.6 K. The blue curves refer to DMFT simulations tuning the chemical potential to obtain the requested (total) occupation value. The orange curves refer to DMFT simulations performed at fixed chemical potential. The red curve follows the fixed-chemical-potential solution for decreasing value of $\mu$ (right-to-left). The black data refer to the zero-hybridization model, solved in ~\protect\cite{huKondoLatticeModel2023c}. In these plots, the geometrical capacitance contribution is \emph{included} in the value of $\mu$, contrary to Fig.~\ref{fig:nvsmu} of the main text. It is clear how, the presence of this geometrical term was the cause of the wide negative compressibility regions displayed by the $\nu_f+\nu_c$ curve 
    %, proving that effectively the experimental setup is globally stable 
    throughout most of the parameter range. Remarkably, a small region of negative compressibility is however still present around $\pm 0.6$.
    }
    \label{fig:nvsmuwithgeom}
\end{figure}

\begin{figure}
    \centering
    \includegraphics[width=\textwidth]{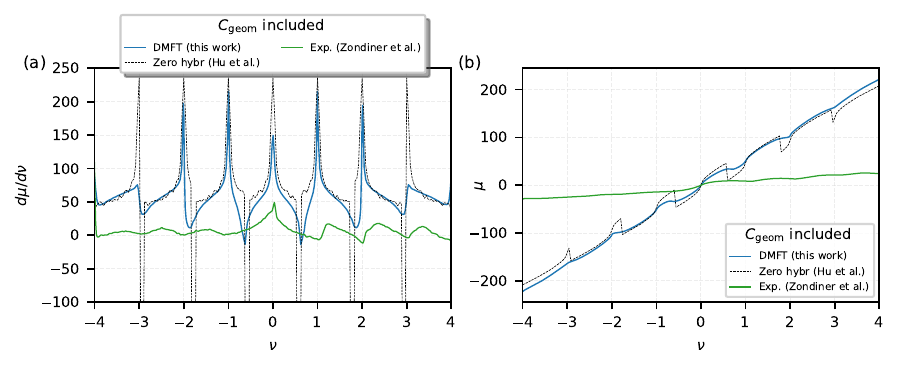}
    \caption{(a) Inverse compressibility obtained from the fixed-occupation DMFT simulations for $T=11.6 K$, compared with the experimental data of~\protect\cite{zondinerCascadePhaseTransitions2020}, for $B_{||}=0$, $T=4K$ and $\theta=1.13^\circ$. Here, the geometric capacitance contribution has been taken into account, causing a positive shift of the compressibility values. While the negative spikes of the zero-hybridization solution are still present, such shift is enough make $d\mu/d\nu$ positive except for the small regions around $\pm 0.6$. (b) Chemical potential for DMFT simulations at $T=11.6 K$ compared with ~\protect\cite{zondinerCascadePhaseTransitions2020} data. Again, this scenario is relative to the convention according to which the contributions of the geometric capacitance of the plane-plate capacitor have not been subtracted out. This bad agreement with the green curve from experiments confirms the choice made in the main text with the geometric capacitance contributions to the chemical potential.}
    \label{fig:expwithgeom}
\end{figure}

Interacting electron models can contain terms which relate to what is referred to as \textit{geometric capacitance} in experimental contexts. (See for instance the capacitor circuit model detailed in the SI of \protect\cite{zondinerCascadePhaseTransitions2020}.) Hence, special care has to be taken when comparing chemical potentials, compressibilities or related quantities between different theoretical models and different experimental setups, since geometric capacitance effects have to be consistently treated or consistently cancelled. We explain in the following how to avoid 
the geometric capacitance contributions in our calculations. 

We base our calculations on the Song-Bernevig model from Ref. \protect\cite{songMagicAngleTwistedBilayer2022}. 
The interaction terms in the Song-Bernevig model derive from Coulomb integrals, where TBLG is assumed to be placed in a double gate structure (see Sec.~3A in the SI of \protect\cite{songMagicAngleTwistedBilayer2022}), which also contains a geometric capacitance contribution. Specifically,
%In the following, we will use a simple picture to motivate that subtracting the geometric capacitance contribution is essentially equivalent to using the local chemical potential $\mu_{loc}$ as described above.
the Coulomb integrals correspond to a geometry, where the TBLG layer is sandwiched by two capacitor plates (see Fig.~\ref{fig:geomcap}) and where the total system is charge neutral. The two plates are assumed to be equivalent and equally spaced from the TBLG layer by a distance of $\xi/2$. If the arbitrarily doped TBLG layer has surface charge $\sigma$, each plate possesses charge $-\sigma/2$. The surface charge density corresponding to the filling $\nu_{\mathrm{tot}}$ is $\frac{\nu_{\mathrm{tot}} e}{\Omega_0}$, where $\Omega_0$ is the area of the moir\'e unit cell. The voltage between the top plate and the TBLG layer $\Delta\Phi$ is given by $\frac{\sigma}{2\epsilon_0\epsilon_r}\frac{\xi}{2} = \nu_{\mathrm{tot}} \frac{e\xi}{4\Omega_0 \epsilon_0\epsilon_r}$ where $\epsilon_0$ and $\epsilon_r$ are the vacuum permittivity and the relative permittivity. Ref.~\protect\cite{songMagicAngleTwistedBilayer2022} assumes $\epsilon_r=6$ stemming from the hBN encapsulation and $\xi=10$~nm. The contribution from the geometric capacitance to be subtracted is then

\begin{equation}
e\Delta\Phi = \nu_{\mathrm{tot}} \frac{e^2\xi}{4\Omega_0 \epsilon_0 \epsilon_r} = \nu_{\mathrm{tot}}\cdot 47\,\rm meV.
\label{eq:geom_capa_SB}
\end{equation}

%The definition of the chemical potential of the heavy fermion model is of crucial importance, and a nontrivial task in itself. The best choice is arguably to define it coherently with experimental setups, as detailed for example in \protect\cite{zondinerCascadePhaseTransitions2020}, where the negative inverse compressibility is equated to the electrostatic potential of the TBLG layer in a setup that is effectively a capacitor circuit (see SI of \protect\cite{zondinerCascadePhaseTransitions2020}). An important feature of this setup is the renormalization of the compressibility by the geometric capacitance, which is the capacitance of the plane plate capacitor formed by the TBLG layer and the back gate electrode.

We explain in the following how we have approximately cancelled the geometric contribution according to Eq. \ref{eq:geom_capa_SB} in our theoretical framework. In order to apply DMFT to solve the Song-Bernevig model \protect\cite{songMagicAngleTwistedBilayer2022}, we have split the interaction terms in three categories: the first ($H_{U1}$ term) is dynamically treated, the second ($H_{W}$, $H_{V}$, $H_{J}$ terms) is accounted for at the mean-field level and the third ($H_{U2}$ term) is small enough to be neglected.
Let us for simplicity restrict ourselves to the symmetric case, and treat $H_{\mathrm{MF}}=H_{W}+H_{V}$ at the Hartree level: since $H_{W}$ and $H_{V}$ are density-density terms, apart from constant energy shifts and omitting the indices,
\begin{equation}
    H_{W}= W \big[\nu_{f}\big(\sum c^{\dagger}c\big) + \nu_{c}\big(\sum f^{\dagger}f\big)  \big]
    \label{eq:H_w}
\end{equation}
\begin{equation}
    H_{V}= V \big[\nu_{c}\big(\sum c^{\dagger}c\big) \big]
    \label{eq:H_v}
\end{equation}

$H_{\mathrm{MF}}$ can equivalently be decomposed in the two following terms

\begin{equation}
    H_{\mathrm{tot}} = \big(W \nu_{f} + V \nu_{c}\big)\big(\sum f^{\dagger}f + \sum c^{\dagger}c\big) = \big(W \nu_{f} + V \nu_{c}\big)\hat{N} 
    \label{eq:H_tot}
\end{equation}
\begin{equation}
    H_{\mathrm{f}}= \big(W (\nu_{f} - \nu_{c}) - V \nu_{c}\big)\big(\sum f^{\dagger}f\big) = \big(W (\nu_{f} - \nu_{c}) - V \nu_{c}\big)\hat{n}_{f} 
    \label{eq:H_f}
\end{equation}
where $\hat{N}$ and $\hat{n}_{f}$ are the total and f-subspace occupation number operators respectively.

Remembering that $W \approx V \approx 47 \pm 1$~meV and comparing $H_{\mathrm{tot}}$ to the geometric capacitance contribution from Eq. \ref{eq:geom_capa_SB}, we notice that discarding $H_{\mathrm{tot}}$ is equivalent, within 1meV (and hence generating an error smaller than, for example, the exclusion of the $H_{U2}$ interaction term) to cancelling the geometrical capacitance contribution. Hence, we discard $H_{\mathrm{tot}}$ and impose $H_{\mathrm{MF}}=H_{\mathrm{f}}$. We refer to the chemical potential derived within this convention as $\mu$. All data show in the manuscript refer to this definition of the chemical potential $\mu$, which can also be directly compared to experiments like \protect\cite{zondinerCascadePhaseTransitions2020}.

The addition of symmetry-breaking terms such as $H_{J}$ does not alter this picture, since the total and $f$-subspace shifts only couple to diagonal entries in the system density matrix.

We could relate $\mu$ to the chemical potential $\mu_1$, which would be obtained from treating the Song-Bernevig model without geometric capacitance subtraction. In this case, $H_{\mathrm{tot}}$ is to be included leading to \begin{equation}
    \mu_{1}=\mu + W \nu_{f} + V \nu_{c}.
    \label{eq:mu_conventions}
\end{equation}

When the DMFT simulations are performed at fixed density, the chemical potential $\mu$ is not set \textit{a priori} as a parameter, but iteratively adjusted as to obtain the desired total occupation. 

As stated above, the expression of the mean-field interaction terms via Eq.~\ref{eq:H_tot} and \ref{eq:H_f} has, as a first benefit, the peculiarity of isolating a term ($H_{\mathrm{tot}}$) which is to all intents and purposes equivalent to the contribution coming from the geometric capacitance $e\Delta\Phi$. 
It also comes with a relevant computational advantage: instead of dealing with two different energy shifts for the $f$- and $c$-subspaces, to be self-consistently determined at every DMFT loop, we now effectively consider only one shift restricted to the $f$-subspace, and hence analogous to the usual double-counting corrections typical of DFT+DMFT calculations~\protect\cite{Anisimov1991,Solovyev1994,Czyzik1994}. Moreover, the overall energy shift is calculated in one single operation upon self-consistently adjusting $\mu$, without calculating $H_{\mathrm{tot}}$. This avoids convergence instabilities due to the oscillations of $\nu_{f}$ and $\nu_{c}$.\\

It is instructive to look at the filling-vs-$\mu$ and inverse compressibility data when the geometrical capacitance contribution is not subtracted out (see Figs.~\ref{fig:nvsmuwithgeom} and~\ref{fig:expwithgeom}). 
Regarding the inverse compressibilities (cfr. Fig.\ref{fig:expcomparison} in the main text), while the huge negative spikes of the zero-hybridization solution are obviously hardly affected, the negative inverse compressibility regions in the DMFT curve have almost disappeared.
Relatedly, the filling data of Fig.~\ref{fig:nvsmuwithgeom} acquire a sizable horizontal offset, with the value of $\mu$ now varying in a range roughly ten times bigger than the one in Fig.~\ref{fig:nvsmu} in the main text.
The $f$ orbitals get filled up in an almost-staircase fashion, and the $c$ orbitals display a clearer saw-tooth behavior, as shown in Fig.~\ref{fig:nvsmuwithgeom}(c). It is interesting to note how the zero-hybridization data, which in Fig.~\ref{fig:nvsmu}(c) travel multiple times along the same curve, here move along a set of shifted disconnected branches, each one corresponding to a $f$-electron plateau of Fig.~\ref{fig:nvsmuwithgeom}(b). 

The data points in orange and red refer to two other sets of DMFT simulations run using an alternative procedure: instead of fixing the overall occupation, we required the converged solution to have a specific chemical potential, letting the occupation value freely adjust as a consequence. The orange data thus obtained are in agreement with a portion the positive-slope branches of the total filling curve in Fig.~\ref{fig:nvsmu}(a) of the main text. We can access a larger section of the fixed-$\nu$ curve (red data) by converging each $\mu$-point and using the resulting self-energy as the starting point of the next simulation, in a decreasing $\mu$ direction. In both cases, however, the negative-slope branches of Fig.~\ref{fig:nvsmu}(a) remain out of reach. 
%\LC{Fixed-$\mu$ DMFT simulations were recently employed in a paper by Datta et al.~\protect\cite{dattaHeavyQuasiparticlesCascades2023a} using a multi-orbital model adapted from the low-energy band structure by Carr et al.~\protect\cite{CarrPrr2019}, which also features the peculiar saw-tooth behavior for inverse compressibility depicted in Fig.~\ref{fig:expcomparison} in the main text, but no negative values. The overall evolution of the relative occupation of the flat and dispersive bands, as well as the coherent/incoherent sequences, are instead in fairly good agreement.}

Even including the geometric contribution, a small region of negative compressibility still persists in the range $\nu_{f}+\nu_c\in [0,\pm 1]$, as apparent from Fig.~\ref{fig:expwithgeom}, suggesting perhaps the tendency of the full TBLG system towards intrinsic (electronic) phase separation around $\nu$ values close to $\pm0.6$. A larger separation between the screened Coulomb interaction gates of the interaction Hamiltonian would however strech out the $\nu_{f}+\nu_c$ vs. $\mu$ curves further and stabilize the system eventually at all dopings.

\section{Experimental data comparison}\label{app:exp_data_comparison}

The experimental data of Fig.\ref{fig:expcomparison} in the main text have been extracted from work by Zondiner et al. (in the following denoted as Z,~\protect\cite{zondinerCascadePhaseTransitions2020}), Pierce et al. (P,~\protect\cite{pierceUnconventionalSequenceCorrelated2021}) and Saito et al. (S,~\protect\cite{saitoIsospinPomeranchukEffect2021}). In the two panels of the figure, the plotted quantities are related by derivative/integration. Hence, for each experimental source only one set of data has been considered, and the other has been numerically obtained. For Z, this is the chemical potential, while for P and S it is the inverse compressibility.
An important point to notice is the difference between charge carrier density (denoted in Z and P as $n$) and filling factor $\nu$, which is analogous to the one used in our work. The two are related via $\nu=n\Omega_{0}$, where $\Omega_{0}$ is the area of the moir\'e  unit cell, and depends on the twist angle $\theta$ via $\Omega_{0}=8\pi^2/3\sqrt{3}k_{\theta}^2$ where $k_{\theta}=2|\mathbf{K}|\sin \theta/2$ and $\mathbf{K}$ is the Dirac point of one graphene layer. This conversion factor, dependent on the twist angle, has to be applied to data from Z (which provides $\mu$ as a function of $n$) and P (which provides $d\mu/dn$ as a function of $\nu$). When the $\mu(\nu)$ data are obtained through integration, an offset given by the value $\mu(0)$ is subtracted.

\section{Temperature dependence of local moments and charge fluctuations}\label{app:highT}

\begin{figure}
    \centering
    \includegraphics[width=0.7\textwidth]{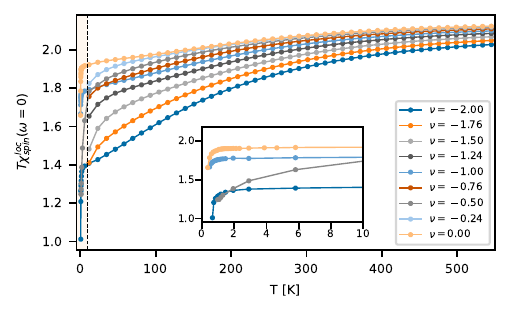}
    \caption{Local static spin susceptibility in the symmetric phase as a function of temperature. The shaded region of the main panel, delimited by the black dashed line and enlarged in the inset, represents the Curie-Weiss susceptibility plateau, associated to the presence of local moments at integer fillings. }
    \label{fig:susz_high_T}
\end{figure}

\begin{figure}
    \centering
    \includegraphics[width=\textwidth]{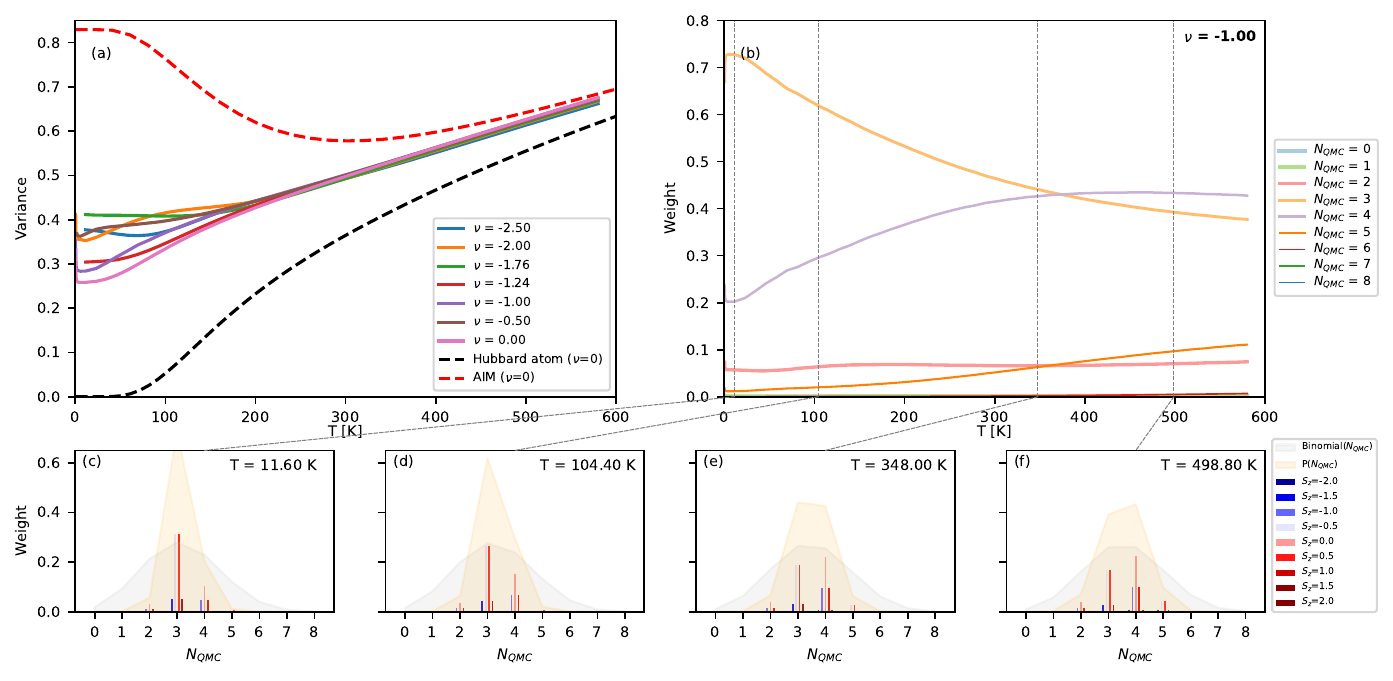}
    \caption{(a) Variance of the spectral weight distribution from our continuous-time quantum Monte Carlo solver for occupation values of the narrow orbitals of TBLG at different dopings, compared with that of a 4-orbital Hubbard atom and a 4+4 Anderson Impurity Model with $U,\Delta$=$U_{1},\gamma$. (b) Evolution of Monte Carlo spectral weight for each charge sector as a function of temperature for $\nu=-1$. (c-f) Snapshots of spectral weight distribution at different $T$, resolved by spin and charge. The grey shaded area shows the corresponding binomial distribution for the given $f$-occupation value. }
    \label{fig:composite}
\end{figure}

In an interacting system, local moments are formed when charge fluctuations are frozen due to high Coulomb energy penalties, and the relevant electronic degrees of freedom can be assumed to be only those associated with spin ($=1/2$ in the simplest one-orbital realization of a Kondo system).
%For the paradigmatic single-orbital case, 
This reflects in the Curie-like behavior of the local spin susceptibility $\chi_\text{spin}^\text{loc}(T)$ whose $1/T$-behavior measures the strength of the (square) fluctuating moment \protect\cite{Hausoel2017}. Inspecting how $\chi_\text{spin}^\text{loc}(T)$ deviates from the Curie law, one can identify two temperature scales: $T_{\mathrm{K}}$, below which Kondo screening takes place, and $T_{\mathrm{Moments}}$, above which thermally activated charge fluctuations set in. In the simplest case of a one-orbital model, charge fluctuations lead to the doubly-occupied and empty sectors contributing significantly to the partition function.
This range of very high temperatures, known as \textit{free orbital regime}~\protect\cite{KrishnaMurthy1980,Bulla2008,Drouin2021}, is associated with a decrease in the $1/T$-coefficient of $\chi_\text{spin}^\text{loc}(T)$ and an increase in entropy from $\about\log 2$ to $\about\log 4$, reflecting the higher number of contributing local eigenstates.

In twisted bilayer graphene, we have the SU(4)-flavor structure of the narrow $f$-manifold in addition to the SU(2)-spin degree of freedom which enriches this picture. Further, the absence of Hund's rule favors low-spin configurations resulting in a relatively small value of the effective local moment even in the strong local-moment regime. 
%features deviations from this picture in both the low and high temperature limit. This is a consequence of the multi-orbital structure of the narrow bands manifold combined with the absence of an interaction term enforcing the Hund's rules, thereby equally penalizing all double occupations.
In Fig.~\ref{fig:susz_high_T} we show the behavior of the quantity $T\cdot\chi_\text{spin}^\text{loc}(\omega=0)$ in the symmetric phase and across a fairly extended temperature range. As it can be seen in the inset, Curie-like behavior (i.e. where the curve is horizontal) persists all the way down to 1-2 K for integer filling.
This allows us to quantify the fluctuating local moment at the integer fillings. As discussed in Ref.~\protect\cite{huSymmetricKondoLattice2023a}, due to the absence of a Hund's coupling, these values are considerably smaller than the nominal ones for an SU(4)-atom.
At the same time, this analysis also reveals that Kondo screening at low temperatures will be preempted in TBG by the transition to the symmetry-breaking phase which, as shown in the main text, sets in at around $T\sim 10$~K. 

Let us now turn to the estimate of $T_{\mathrm{Moments}}$. % from the behavior of local susceptibility.
%in the case of TBG requires a little more care. 
%alone there is no indication of the other relevant temperature scale . 
Upon increasing $T$, the value of $T\cdot\chi_\text{spin}^\text{loc}(\omega=0)$ slowly increases and eventually saturates at $T$ well above room temperature. This is yet another manifestation of the absence of Hund's coupling: since both high- and low-spin configurations have equal energy cost, the value corresponding to the free-orbital regime is not necessarily smaller than the low-temperature local moment. Hence, the local susceptibility alone is not an indicator of the other relevant temperature scale in the case of TBG. 

Based on the previous discussion, we estimate $T_{\mathrm{Moments}}$ considering the statistical weight of %all orbital configurations of 
 different $f$-electron charge sectors, which are readily accessible from our numerical Monte Carlo simulations. 
The Boltzmann weight of all occupation sectors, from empty ($N=0$) to completely full ($N=8$) $f$-electron manifold is plotted in Fig.~\ref{fig:composite}(b) for $\nu=-1$. In the bottom panels (c-f) the sector weights are resolved by both occupation number and spin eigenvalue. At low temperatures, one sector is dominant over all the others and charge fluctuations are massively suppressed. Increasing $T$, the occupation of neighboring charge sectors steadily increases. At $T=300$~K, the weight of the dominant sector is reduced to less than half the total, and continues to decrease upon increasing $T$.
It is also worth noticing that low-spin (e.g. $S_z=0$ for even values of $N$) configurations have the largest weight at each $N$-sector, in agreement with $J_{\mathrm{Hund}}$ being zero. 

The spread of the sector occupation distribution can be estimated by the statistical variance 
\begin{equation}
\sigma^{2} = \sum_i{w_{i}\big(N_{i}-\overline{N}\big)^{2}}  
\end{equation}
where $\overline{N}$ is the average value of $N_{i}$ and $w_{i}$ is the relative weight ($\sum_{i}w_{i}=1$). This quantity is shown in Fig.~\ref{fig:composite}(a) for various dopings, along with that of exactly solvable theoretical models at half-filling, provided for reference.
The black dashed line refers to a SU(4) Hubbard atom at half-filling, while the red one corresponds to the correlated orbitals of an Anderson Impurity Model (AIM). The AIM is constructed by setting $\mathbf{k}=0$ in Eqn.~\ref{app:eqn:H_c} and ~\ref{app:eqn:H_fc}, and considering only the $\Gamma_{3}$ c-bands since the others are decoupled from the f-subspace at $\Gamma_{M}$. In both cases, we chose the interaction parameters to be comparable to those of the THF model.

%The variance of the AIM shows three distinct behaviors. First, at low temperatures, it is rather constant, reflecting the quantum fluctuations of the model in its ground state, which is essentially a Kondo system. Then, when local moments develop (at a temperature dependent on the single-particle and interaction parameters, for this specific toy model at around 50 K), the variance decreases as the relative weight of the half-filled impurity sector grows. Finally, thermal fluctuations take over and the variance shows the same increasing behavior as the other systems we considered.

The variance of the AIM shows three distinct behaviors. At low temperatures, where the ground state is Kondo-screened, the variance is high and relatively constant as a consequence of quantum charge fluctuations between correlated orbitals (impurity) and uncorrelated ones (bath) within the ground state. The variance reflects the relative weights of the various occupation configurations for a \textit{globally} (impurity+bath) half-filled system, which depend on the model parameters (in our specific case, on the relative strength of $U$ and $\gamma$). As temperature increases, the variance decreases due to the growing relative weight of the half-filled \textit{impurity} sector. This signals the development of a local moment, which sets in at around 50 K in this specific AIM. Finally, thermal fluctuations take over and the variance shows the same ``universal" increasing behavior of the other curves. 

With regards to the TBLG model, the solid lines in Fig.~\ref{fig:composite}(a) illustrate that for temperatures below 100-200K, the variance exhibits different behavior depending on the filling.
 For integer filling, there is initially a sharp decrease, coinciding with the formation of a well-defined local moment at around $2\;K$, as per the inset of  Fig.~\ref{fig:susz_high_T}.
For fractional filling, the variance stays rather constant, reflecting the fact that the system is in an intermediate-valence regime, with two sectors contributing more than the others to the partition function.

\begin{figure}
    \centering
    \includegraphics[width=\textwidth]{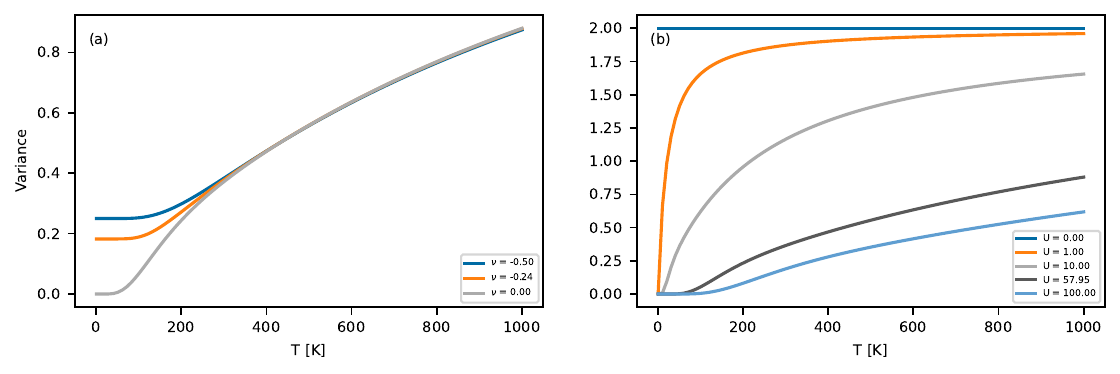}
    \caption{(a) Variance of sector occupation distribution for a 4-orbital Hubbard model, with $U=U'=U_{1}$, $J_{H}=0$ and at varying occupation, as a function of temperature. (b) Variance of sector occupation distribution for a half-filled 4-orbital Hubbard model at varying U, as a function of temperature.}
    \label{fig:only_hubbard}
\end{figure}

Above 100-200 K, the variance does not appreciably differentiate between occupations anymore. Heuristically, this happens when the system starts to consistently explore neighboring occupation sectors due to thermal fluctuations.
%, no matter the original weight distribution. 
This trend is persistent across systems, as shown for the Hubbard atom at various occupations in Fig.~\ref{fig:only_hubbard}(a). 
Hence, we estimate the temperature $T_{\mathrm{Moments}}$ characterizing the crossover from a behavior driven by Coulomb correlations to one with thermally activated charge fluctuations to be on the order of 100~K to 200~K. 

This links directly to the Hartree-Fock analysis: the energy needed to explore different charge sectors is on the order of the splitting between upper and lower Hubbard bands in Hartree-Fock calculations. Thus, thermal effects will suppress (iso)spin-polarized Hartree-Fock solutions in the very same temperature range where thermally activated charge fluctuations set in DMFT. This is why $T_{\mathrm{Moments}}$ corresponds to the Hartree-Fock ordering temperature.

%, and signals the onset of a regime where the $N$ distribution is dominated by statistics. 

%\LC{This is unfortunately just a qualitative description. We have no real explanation on why the figures at varying doping have such a large region where the variance is almost equal independently of filling. We tried for the Hubbard atom to put this in relation to the value of T where $T\chi$ ceases to be constant, but unfortunately this is not the case. We are still thinking on it.}
%\LC{Luca's comment: the "slope" of the universal region is dependent on U. The "universal" region is the region when the occupation distribution start to be "bell-shaped" around the dominant sector, but has not reached yet the "border" sectors. This would be the same slope in the infinite-orbital limit.}

One can ask what happens at even higher temperatures: $T\gg T_{\mathrm{Moments}}$. Clearly, at some very high temperature, the system should not only explore neighboring charge sectors as for $T\gtrsim T_{\mathrm{Moments}}$ but explore all charge sectors. In the limit of infinite temperature, the occupation distribution at a given $n_{f}$ is given by the binomial formula
\begin{equation}
    p(N)=\binom{8}{N}\bigg(\dfrac{n_{f}}{8}\bigg)^{N}\bigg(1-\dfrac{n_{f}}{8}\bigg)^{8-N}
\end{equation}
with constant variance
\begin{equation}
    \sigma^{2}=8\cdot\bigg(\dfrac{n_{f}}{8}\bigg)\cdot\bigg(1-\dfrac{n_{f}}{8}\bigg),
\end{equation}
as shown in Fig.~\ref{fig:only_hubbard}(b) for $U=0$ in the simple case of an $SU(4)$ Hubbard atom. This is counteracted by the Coulomb interaction, the effect of which is narrow down the range of thermally accessibly configurations (intermediate temperatures) and ultimately to stabilize a single configuration (low temperatures) with respect to all the others.

%
% ****** End of file apssamp.tex ******

\end{document}